**Title:** Single-molecule techniques in biophysics: a review of the progress in methods and applications

Authors: Helen Miller[1], Zhaokun Zhou[1], Jack Shepherd, Adam J. M. Wollman and Mark C. Leake[2]
Affiliations: Biological Physical Sciences Institute (BPSI), University of York, York, YO10 5DD, U.K.
[1] These authors contributed equally
[2] Correspondence to be spent to mark.leake@york.ac.uk

**Abstract**
Single-molecule biophysics has transformed our understanding of biology, but also of the physics of life. More exotic than simple soft matter, biomatter lives far from thermal equilibrium, covering multiple lengths from the nanoscale of single molecules to up several orders of magnitude to higher in cells, tissues and organisms. Biomolecules are often characterized by underlying instability: multiple metastable free energy states exist, separated by levels of just a few multiples of the thermal energy scale $k_BT$, where $k_B$ is the Boltzmann constant and $T$ absolute temperature, implying complex inter-conversion kinetics in the relatively hot, wet environment of active biological matter. A key benefit of single-molecule biophysics techniques is their ability to probe heterogeneity of free energy states across a molecular population, too challenging in general for conventional ensemble average approaches. Parallel developments in experimental and computational techniques have catalysed the birth of multiplexed, correlative techniques to tackle previously intractable biological questions. Experimentally, progress has been driven by improvements in sensitivity and speed of detectors, and the stability and efficiency of light sources, probes and microfluidics. We discuss the motivation and requirements for these recent experiments, including the underpinning mathematics. These methods are broadly divided into tools which detect molecules and those which manipulate them. For the former we discuss progress of super-resolution microscopy, transformative for addressing many longstanding questions in the life sciences, and for the latter we include progress in 'force spectroscopy' techniques that mechanically perturb molecules. We also consider *in silico* progress of single-molecule computational physics, and how simulation and experimentation may be drawn together to give a more complete understanding. Increasingly, combinatorial techniques are now used, including correlative atomic force microscopy and fluorescence imaging, to probe questions closer to native physiological behaviour. We identify the trade-offs, limitations and applications of these techniques, and discuss exciting new directions.

1. Introduction

Ensemble biophysics methods have produced a wealth of information, but in life, more so than in classical condensed matter, each member of a population is an individual. Single-molecule techniques allow researchers to probe the heterogeneity one molecule at a time to reveal a highly complex bigger picture of biological systems [1,2], general soft condensed matter [3], and biological physics[4], as well as intricate nanoscale biomolecular machines [5] and a range of single molecule behaviours in native living cells [6,7]. We begin with a brief review of the key experiments enabling the observation, perturbation and prediction of the behaviour of single molecules and a discussion of the nanometre length scale single-molecule environment.

1.1. What is 'single-molecule biophysics'?

There are many ensemble average techniques in biophysics that are able to provide valuable information about the mean average state of a biological system. However, by utilizing single-molecule biophysics technique we are able to explore energetically metastable, heterogeneous states one molecule at a time, which is not possible with ensemble average methods.

In biology we study the processes and components of 'life'. 'Life' as a philosophical concept has no unique agreed definition, nevertheless, from a physics perspective we can say that the processes involved are typically far from thermodynamic equilibrium. Typically, in studying life we can capture information at a sufficiently meaningful level by investigating properties of single molecules rather than their atomic or sub-atomic constituents: this is typically the smallest length scale at which we can probe to understand biological processes at the level of apparent 'biological function' in a relevant organism [8].

The type of 'molecular heterogeneity' we encounter will vary depending on the system. In solid systems we might have a static heterogeneity, for example, arising from the defects in a crystal, but often heterogeneity is temporal; in particular when a biomolecule undergoes conformational changes related to its biological role; or spatial if the molecule has interactions over the nanometre length scale with other biomolecules.

Single-molecule biophysics techniques allow us to look at biological features of interest, one molecule at a time, and build up a picture of the underlying molecular heterogeneity in the system. Within the relatively small volume of one cell there can be vast differences in environment, for example, in viscosity or in the local concentration of a biomolecule. Single molecule studies aid us in understanding the mechanisms behind the properties we are investigating, as the average may not necessarily correspond to a real, achievable state of the system. To explain this notion we can use an analogy of the average speed of a group of swimmers who join a swimming session at a swimming pool but in separate lanes. We imagine the whole swimming pool to be the biological system we are interested in and the separate lanes as sub-regions of the system which we can sample, with our swimmers as individual biomolecules. If we first average the speed of each swimmer in the whole pool we will find an average swimming speed that might not correspond with the swimming speed of any one person. If we look at the swimmers in each lane one at a time we may also observe spatial heterogeneity; a different average swimming speed in each lane, seen as a variation in swimming speed from one side of the pool to the other. If we were not looking at swimming humans, but individual molecules, we might infer a gradient in viscosity or temperature causing this spatial change in speed, however the underlying cause is instead that of molecular heterogeneity.

Our analogy of swimming here emphasizes the key aspect of heterogeneity when taking averages over a population. However, it should also be noted that the role of Brownian fluctuations and thermal forces are important factors in regards to molecular heterogeneity. This is the main source of noise and variability in single-molecule experiments. In many circumstances it is difficult to discriminate molecular heterogeneity from just stochastic noise due to thermal forces.

Both temporal and spatial heterogeneity can be present in a system, and single molecule studies allow us to observe these differences. Where ensemble methods provide us with just a mean average value, single molecule methods can, ultimately, generate the average value but also produce a probabilistic distribution of values either side of that mean. A distribution with distinct clusters of measurements separated by gaps might indicate different energetic or conformational states, and the location of the average relative to two distinct clusters might indicate a preference for one state over the other. Probability distributions can highlight deviations from mean average behaviour and are able, with appropriate biological and physical insight, to suggest potential mechanisms for the observed behaviour, far beyond what can be inferred from a simple mean average value obtained by an ensemble, population-level technique.

With the potential to glean so much information from single-molecule biophysics there come also extreme technical challenges. For example, as we strive to collect data at higher temporal and spatial resolutions we must aim to reduce the background level for detection to achieve a signal-to-noise ratio greater than one; we must maximise collection efficiency and, perhaps most importantly, perform checks to ensure that what we believe are single molecules, are in fact just that and not multiples. The suite of biophysical techniques that have been developed to study single molecules are not inherently high throughput, but for many of the techniques, recent advances are being made towards multiplexing measurements.

In the following pages this review will cover a brief history of the key experiments that have enabled single molecule studies, before summarising the most recent progress in techniques which detect, manipulate, or simulate single molecules. We conclude by looking at the increasing number of correlative techniques which combine at least two single molecule methods and the challenges faced by the field as techniques develop from proof of principle to the study of real, complex biological questions.

1.2. Key historical experiments

We briefly review some of the key experiments that have carved the current research landscape for the field of single-molecule biophysics, and laid the way for the new developments discussed in detail later. Here we consider single molecules in the condensed phase. Single molecule/atom/ion traps in a vacuum/gaseous state have a longer history that we are do not consider in the context of biophysics.

1.2.1. Detection of single molecules using electron microscopy

Since the 1950s many experiments have contributed advances to reach the point where we can perform single-molecule experiments on biologically relevant molecules in cells, rather than there being one clear and unique example of a 'pioneer' experiment. The full story includes experiments on non-biological specimens and low temperature studies; here, a few studies of particular historical relevance to current work in the field of single-molecule biophysics are outlined.

The first images of single biological molecules were taken using transmission electron microscopy (EM) of single filamentous molecules including both deoxyribonucleic acid (DNA) molecules and proteins such as collagen, taken by Hall in 1956 [9]. Small biological molecules are difficult to image directly in electron microscopy because their constituent atoms, consisting predominantly of low atomic number atoms such as carbon, do not generally have high enough electron density to strongly scatter electrons. To visualise biological samples a process known as shadow casting [10] is often used. Here, a thin coating of gold is deposited on the sample from an angle. The non-coated areas in the shadow are easily seen in the EM images. In EM the sample is prepared on mica substrate; this means that the images may not be of the molecules in their native state as mica has a net electrical positive charge on its surface and many biological molecules, including DNA, also have net electrical charges. Further, most biological samples are usually solvated by water and the drying of the sample in EM has since been shown to lead to non-native conformations.

The first indirect measure of single biological molecules in aqueous solution was performed by Rotman in 1961 [11]. Small droplets containing a low concentration of enzyme and a high concentration of substrate that becomes fluorescent once it has been acted on by the enzyme were produced by atomisation. After incubation for a given time, the fluorescence of droplets of similar volume was measured, and was found to be zero, a given amount of fluorescence, or multiples thereof. This implies that the droplets contained zero, one, two etc. enzyme molecule each, and further the distribution of the number of droplets containing each number of fluorophores was found to be consistent with Poisson statistics for the starting concentrations, thus he could measure the activity of single biological molecules *indirectly*.

In 1976 Thomas Hirschfeld detected single globulin protein molecules in aqueous solution which had each been labelled with hundreds of fluorescent organic dye molecules [12]. In 1982 Barack and Webb [13] tracked single lipid molecules labelled with multiple fluorescent tags diffusing on a cell membrane bleb. They used a crude method of intensity centroid localisation on a frame by frame basis to find the centre of molecules using tracing paper overlaid on photographic papers. Centroid localisation and subsequent Gaussian fitting is now a key stage in many force transduction and optical super-resolution techniques. In 1988 Gelles et al. [14] found the centre of a plastic bead being rotated by a single kinesin motor to a precision of a few nanometres by digitising computer images and processing them. Building on these advances the first localisation of a single fluorescent molecule was performed on a single rhodamine labelled lipid in 1996 by Schmidt [15], achieving ca. 30 nm resolution, and Sako et al. [16] performed the first direct single-molecule imaging (i.e. which utilises just a single fluorescent dye tag) on live cells in 2000.

1.2.2.   'Super-resolved fluorescence'

The Nobel Prize in chemistry 2014 was awarded to Eric Betzig, Stefan Hell and William Moerner (Nobel lectures: [17–19]) for their key roles in establishing the field of 'super-resolution' imaging, namely imaging with a an effective spatial precision which is better than the standard optical-resolution limit of a few hundred nm for visible light microscopy, what is now the cutting-edge of light microscopy after its ca. 300 years history [20]. In less than 20 years microscopic imaging in the life sciences was transformed from being diffraction-limited to achieving resolutions of a few tens of nanometres. Imaging of single molecules via electron microscopy has already been discussed but for the life sciences, detection in solid and aqueous *native* environments was key to moving towards imaging in live cells, since all proteins look the same to electrons – light microscopy allows us to add distinct markers to our molecules of interest. Moerner was the first to detect single molecules in the solid phase at cryogenic temperatures [21], and Orrit et al. [22] used fluorescence to improve the signal-to-noise ratio of the measurement. Betzig worked to develop near-field scanning microscopy (NSOM) [23,24] and was the first to achieve super-resolution imaging in cells [25], and then single fluorescent molecules in a monolayer at room temperature [26]. But, in a cell the fluorophores were too densely packed to achieve single-molecule imaging. Hell and Betzig independently published the idea that what was required was a method to have only one molecule 'on' in the diffraction limited spot at a time in 1994 and 1995 [27,28]. Betzig solved this problem using modified versions of fluorescent proteins and activating only a random subset of them for each imaging acquisition frame; well-separated molecules could be localised by fluorescence intensity centroid fitting, before turning them off in effect and exciting a different subset until a full image was formed, a technique called photoactivated localization microscopy (PALM) [29]. Hell experimentally developed stimulated emission depletion microscopy (STED), the theoretical concept for which appears to have been independently formulated in the 1980s by the Okhonin who patented the first super-resolution microscope based on stimulated emission [30]. STED acts by reducing the volume from which a fluorescent particle can emit until it can only contain one particle. By scanning this small volume over the sample the single molecules can be located one at a time [31]. The most recent developments in super-resolution fluorescence microscopy are discussed in section 2.1 below.

### 1.2.3. Molecular motor dynamics

Molecular motors are a class of, primarily, proteins that transform chemical potential energy released from adenosine triphosphate (ATP) hydrolysis or from proton pumping into kinetic energy in the form of movement and rotation. This mechanical output drives a wide range of biological activities, from the transportation of molecular cargos inside cells, through to the locomotion of the cells themselves, to the movement of whole muscles. An individual motor moves in steps of a few nm (i.e. $10^{-9}$ m) with forces of a few pN (i.e. $10^{-12}$ N). The structures of these motors can be interrogated with EM, nuclear magnetic resonance and X-ray crystallography but the dynamics and motions can be studied with spectroscopy tools, the details of which will be discussed in section 3. In this section here we introduce milestones in molecular motor experiments performed with spectroscopy. Other biological molecules are suitably probed with spectroscopy too, such as the viscoelasticity nucleic acid polymers in water solution, also the strength of receptor binding; these are discussed in many excellent reviews [32–35].

Kinesin is responsible for the transportation of cellular cargos along microtubules [36]. Block et al. [37] used optical tweezers (OT) to hold a kinesin-coated bead and position it onto microtubule. OT, which are discussed in specific detail later in this review, use a focussed laser beam to create a potential energy well in the vicinity of refractile particles, such as a latex or glass beads of diameter around a micron, which results in an optical trapping force field of effective diameter close to the wavelength of the laser, due to combination of scattering and refraction effects from the laser on the particle. These result in a net force directed approximately towards the centre of the laser focus. This OT approach increases the effective experimental efficiency compared to waiting for kinesin to attach to the microtubule by diffusion alone, since we can physically move a trapped bead coated in kinesin to its points of action at a microtubule. Once attached, the OT can counteract the Brownian motion of the bead in the surrounding aqueous media (a 'pH buffer' which can chemically stabilise the pH of the solution to within reasonably narrow limits) and apply a controllable manipulation of the bead's position to allow the force dependence of the bead motion to be interrogated. Figure 1 (a) shows a general schematic diagram of an OT trapping a kinesin-coated bead, which translocates along a microtubule track. Two models are proposed for the movement: the stroke-release model and hand-over-hand model, the difference being in the continuous or brief attachment to the microtubule (see figure 1 (b) for a general schematic of this). Block et al. determined that the latter model better explains the data from kinesin experiments. With 1 kinesin molecule per bead, the bead moves on average 1.4 µm before detachment takes place.

In a later paper, Svoboda and Block [38] used an improved OT with differential interference contrast optics and dual quadrant photodiodes (QPDs) to increase tracking precision to sub-nm levels. This precision facilitated measurement of the kinesin movement, indicating that its velocity decreases linearly with load up to forces of 5-6 pN. They also compared the force-velocity curves at high and low ATP concentrations to deduce that the movement per catalysed ATP decreases at higher load.

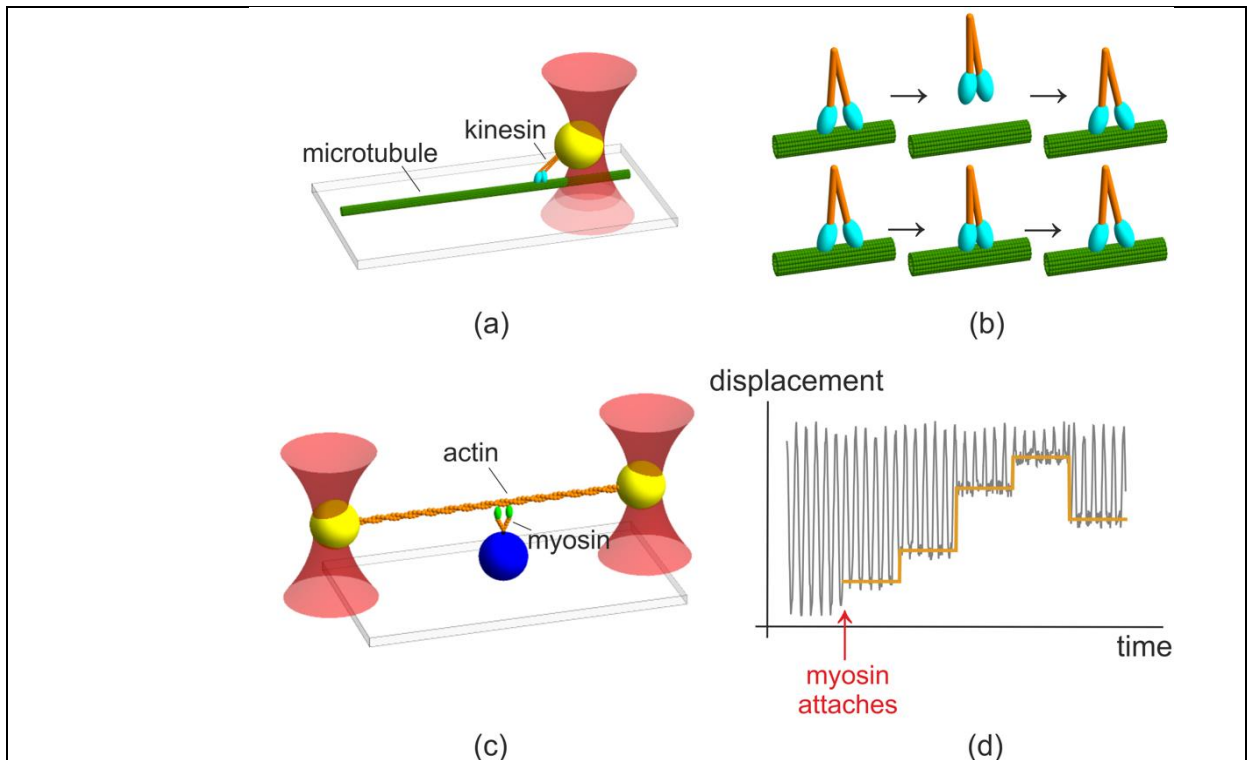

Figure 1. Schematics of a general assay for molecular motor type single-molecule force spectroscopy experiment setups and illustrations of results and conclusions. (a) Optical tweezers can position the kinesin-coated bead onto microtubule, which hugely increases the experimental yield. Optical tweezers apply a controllable force to the bead to investigate the effects of force on kinesin movement. (b) General schematic showing two possible models of kinesin motion. Top panel: stroke-release model, in which the molecule briefly detaches from the fibre and diffuses back to carry on the movement; bottom panel: hand-over-hand model where the molecule stays attached to the fibre for multiple cycles. (c) Dumbbell optical tweezers setup used to determine myosin-V stepping on actin. Here we show a schematic of a general assay. This type of assay was first reported in reference [39]. The blue bead is functionalised with myosin and is immobilised on the coverslip. While the myosin moves along the actin, the optical traps can pick up the step size. The trap can also apply controlled forces to monitor myosin stepping behaviours at various force levels. (d) Schematic showing representative idealized displacement of the bead relative to the trap centre. Initially the trap is set to oscillate triangularly. At the red arrow, myosin attaches to the actin and starts pulling, and the amplitude of the oscillations is reduced. This allows the determination of the myosin displacement along the actin (yellow trace).

Myosins are a family of molecular motors that transport organelles along actin filaments. Biochemical studies suggested that myosin-V is a processive motor in the sense that it undergoes multiple catalytic cycles, or walks multiple steps, before detaching from the actin (the alternative theory is that myosin walks one step before detachment). Mehta et al. [39] used optical tweezers to confirm the processive motor hypothesis and to determine that the step size is 36 nm. Figure 1 (c) shows general schematic of a 'dumbbell' optical tweezers configuration used in the experiment. One trap is set to oscillate with a triangular displacement-time curve. When the myosin moves along the actin, it pulls the actin taught so the detected bead oscillation would have part of its oscillatory peaks removed (figure 1 (d) for hypothetical 'idealized' displacements of a bead). The removed amplitude indicates the amount by which the myosin has moved. This method reveals step size, translocation speed and direction of motion of an individual myosin-V in real time. The application of forces at physiological levels and simultaneous measurement of bead movement are currently possible with single-molecule force transduction devices, most important of which are optical tweezers and magnetic tweezers. Most high resolution optical tracking techniques use dual traps where instrumental drift is strongly suppressed for dumbbells suspended in water [40–43].

1.2.4. <u>Computational biophysics</u>

Computational biophysics has its origins in the 1950s, when modest simulations of the motion of hard [44] and subsequently elastic [45] spheres were conducted over short time scales, demonstrating the potential of molecular dynamics techniques to study dynamical properties of systems and generate trajectories. The possibilities this afforded for biological physics did not go unnoticed and with increased computational power, higher quality data, and general molecular dynamics force fields enough progress was made in a decade to allow refinement of protein structures derived from X-ray crystallography [46], and shortly thereafter detailed analysis of the dynamic behaviour of interior atoms in folded proteins [47]. Larger scale protein dynamics with biological function was also within reach, for example key sites of bending of the molecular structure [48]. However, these molecular simulations all had one challenge in common - that of solvation.

Water was a difficult proposition at the time - not only are thousands of molecules necessary to solvate any but the most trivial systems, but the interactions water has with other molecules are generally extremely complex. Simple water molecules such as the TIP3P model [49] were created in the early 1980s, but due to computational limitations early studies of solvation concentrated largely on the geometry of the solvation shell around biological molecules as these include only a small number of additional residues - for example 72 in the solvation shell of a B-DNA dodecamer [50]. Through time, it became clear that improving classical force fields would be of little benefit without a robust and computationally tractable solvation scheme which could be used in large simulations. To this end, the Generalised Born solvation model was developed [51] and implemented in a variety of molecular dynamics (MD) software, improving the accuracy of simulations. Explicit water models remained out of reach in many cases until the development of the particle mesh Ewald technique [52], which allows a cut-off in calculating the effect of solvent at long distances, and therefore cuts the computational complexity of a simulation significantly. This step also allowed fully solvated simulations of ribonucleic acid (RNA), which had previously been elusive.

Classical MD simulations are not the only outfit in town when it comes to computation. Working with atomic length scales and femtosecond time scales means that the quantum landscape is always in view. Expensive and complex though it is, some work has been done on *ab initio* density functional theory based molecular simulations [53], made possible by the work on computational quantum mechanics of Car and Parrinello [54]. However, these simulations were limited again by computational power, complexity of the system, and the large amount of memory needed to store and work on a molecular wave function.

At the other end of the scale, coarse-graining techniques to reduce computational complexity to its bare minimum were also underway. Using classic statistical mechanics considerations such as the wormlike chain, predictions about the force-extension behaviour of DNA were made and compared with experiment [55]. Computational models of coarse-grained molecules were developed in parallel with atomistic methods, and met with some success [56]. But, these simulations were accurate only under specific conditions, and it is only recently that more generalised coarse-graining has been possible for nucleic acids, as we shall see in section 4.

1.3. <u>Length, time, force and energy scales</u>

The environment experienced by single molecules is vastly different to that experienced by humans in terms of energy, time, force and distance. For reference some typical magnitudes of these quantities are shown in figure 2.

The length scales of biological molecules are mostly smaller than the wavelength of visible light and so cannot be imaged directly using conventional light microscopy techniques. Super-resolution fluorescence microscopy can be used to achieve localisation precisions of a few tens of nanometres, whilst atomic force microscopy (AFM) can attain atomic level spatial precision.

The time scales for biological processes cover many orders of magnitude, at least from nanosecond fluorescence lifetimes to animal, or even ecosystem, lifetimes. In optical detection the localisation precision is related to the number of photons collected, so single-molecule super-resolution imaging is limited to around millisecond time scales, capable at best of imaging molecular conformational changes which happen on a millisecond timescale, as is exploited in Förster resonance energy transfer (FRET) to be discussed later in this review. However, quadrant photodiodes used for back focal plane (BFP) detection in optical and magnetic tweezers can record at least one thousand times faster than this [57–59].

Biomolecular forces tend to lie in the pN range and molecular displacements are usually nanoscale. It is unsurprisingly that the energy scale is roughly $10^{-21}$ J, i.e. the energy required to move 1nm through a force of 1pN. More precisely, the energy scale is that due to thermal fluctuations, i.e $k_B T$, since biomolecules are immersed in a thermal reservoir of water molecules. This energy level is roughly the energy transferred in thermal collisions, with a value of around $4.1 \times 10^{-21}$ J, or equivalently 4.1 pN.nm at room temperature.

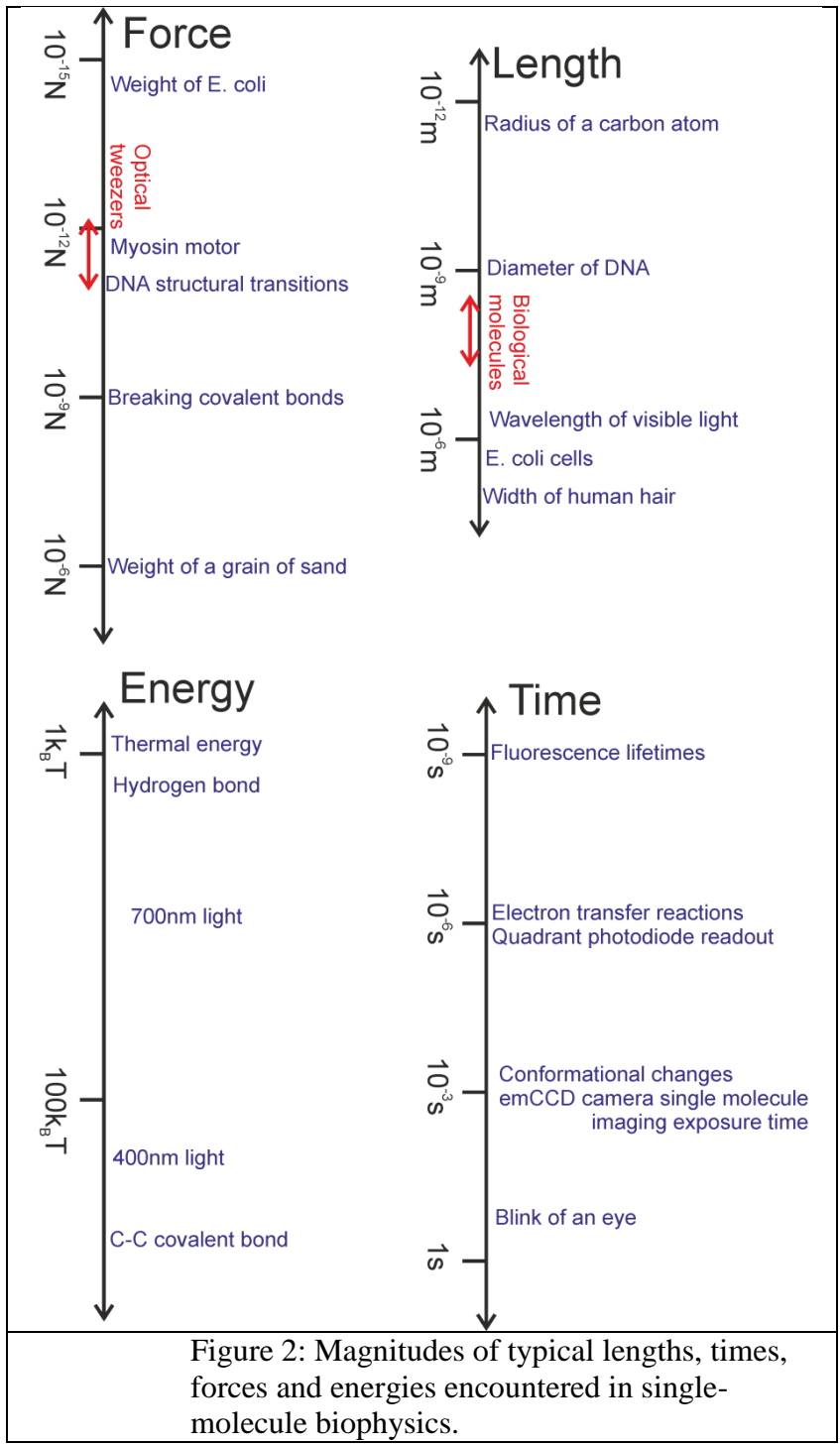

Figure 2: Magnitudes of typical lengths, times, forces and energies encountered in single-molecule biophysics.

The biological environment is in water, but for micron sized objects such as microbeads the environment is of low Reynolds number, i.e. inertial forces are much smaller than viscous drag forces, both for particles in opto-magnetotraps and for *in vivo* studies. This means that a particle will stop moving when external forces cease to act on it.

The very low magnitudes of energies, length, time and force that biological molecules experience means that there are compromises to be made in the design of single-molecule experiments. Temporal resolution may be sacrificed for spatial localisation, or *vice versa*, depending on the biological question and the method used to address it.

1.4. The influence of noise

At high temporal resolution noise in the system poses the greatest limit of the precision that is attainable. In visible light super-resolution fluorescence microscopy the localisation precision of a fluorophore is a function of the number of detected photons, $N$, and is given by [60]:

$$< (\Delta x)^2 > = \frac{s^2 + a^2/12}{N} + \frac{4\sqrt{\pi} s^3 b^2}{aN^2} \tag{1}$$

where $\Delta x$ is the error in localisation, $s$ is the standard deviation of the point spread function (PSF), $a$ is the size of the camera pixel edge length and $b$ is background noise. To reduce localisation error, one can increase the number of collected photons and/or decrease background noise (camera improvement is discussed in section 2.2.3). One particular strength of fluorescence imaging is that the emission wavelength differs from that of the excitation wavelength so an emission filter can be used to block out all but a ~50 nm window of light. This eliminates noise in all the rest of the camera-sensitive spectrum, particularly the bright excitation light. Nevertheless, noise still poses hurdles to the extraction of useful information from fluorescent micrographs.

Many breakthroughs in the past decade are methods and technologies that have increased localisation precision to such an extent that the ultimate limiting factor on imaging precision now becomes the size of labelling probe and the labelling density. Partial illumination of the sample space leaves the rest of the imaging volume in the dark and vastly reduces the out of focus fluorescence. One example of this method is total internal reflection fluorescence (TIRF) microscopy [61]; here the excitation beam arrives at the sample at an angle where it is totally internally reflected: an evanescent wave travels into the sample, which can be detected by its ability to excite fluorophores, but since the evanescent wave decays exponentially only fluorophores within ~100-200 nm of the surface are excited. A second example is light sheet microscopy [62] where only a thin slice of the sample is illuminated at any one time. A similar approach involves confining the excitation beam size to a far smaller diameter so the intensity is many fold higher [63] to achieve higher levels of emission. The brightness and lifetime of fluorophores/dyes have been improved [64–67]. Denoising super-resolution image reconstruction algorithms have also been implemented [68,69] to extract maximum signal from the image data. Some of these techniques will be discussed in more detail below.

By contrast, in so-called 'force spectroscopy' noise is often the fundamental limit on measurement precision. Sources of noise such as air currents, mechanical vibration, thermal expansion and electrical noise in sensors, etc. have been the focus of innovation in instrumentation and protocols that effectively reduce them to levels below biological signals. Nevertheless, ultimately Brownian noise sets the ceiling on measurement resolution. Force spectroscopy measurement with nm or sub-nm resolution is routinely achieved in temperature-controlled ($\pm 0.2°C$), acoustically-isolated rooms [70] with controlling instruments such as laser drivers and PCs typically housed in a separate room. To reduce mechanical vibration, spectroscopy instruments, particularly ones that use quadrant photodiodes (QPDs) to track beads via interferometry, tend to have a bespoke reinforcement skeleton added to the setup. For example, the microscope condenser pillar was strengthened with an aluminium trapezoid in [70], or the condenser mount can be replaced with heavy-duty variants [71].

For measurement of protein motions within the frequency range of noise induced by air currents, optical components can be enclosed in sealed boxes filled with helium, which has lower refractive index than air, and thus any helium flow will cause less deflection of the laser beam. This method has been used to decrease noise spectral density 10-fold at 0.1 Hz so the power of noise was below 1 Å [72]. A more convenient way to suppress air currents not involving replacing air with helium is to simply enclose optical components in boxes.

In optical trapping, optical tweezers often use 1064 nm wavelength near infrared (NIR) lasers as the light source, which utilises the relative affordability of Nd:YAG laser sources, ideal in avoiding the high-absorption region of many proteins in the visible light spectrum. But, the water absorption coefficient at 1064 nm is approximately 1 $cm^{-1}$ (compared to $10^{-4}$ to $10^{-2}$ $cm^{-1}$ in the visible range) and microscope slides/coverslips do not always have minimised absorption at 1064 nm. In magnetic trapping, permanent-magnet magnetic tweezers do not generate significant heat but electromagnetic variants often do. Even though heat can be partially removed with water [73,74] or fan cooling [75], a temperature gradient is inevitably created. A 1°C gradient potentially causes mechanical drift of optical components on the order of ~100 nm [76]. To put this in context, the step size of kinesin is 8 nm [77], the motion of enzymes on DNA is measured in base-pair sized steps (0.34 nm) [78], and the unfolding events of protein domains are 20-30 nm [79,80]. The standard way to tackle measurement errors due to thermal-expansion is by measuring the drift with a marker bead tethered to the assay chamber and remove the drift from the experiment measurement [81]. Furthermore, balanced photodiodes such as these are currently the highest performance method to measure position of trapped beads in optical tweezers; they can measure the ballistic regime in colloidal motion at MHz sample rate. They remove the common mode quantum noise.

Figure 3 shows the typical arrangement of optical components for drift measurement and removal. A tracking laser (green) and QPD2 is used to monitor the position of the marker bead with interferometry. Its signal is sent to calculate a compensation displacement, which is executed with the motorised stage. The trapping laser (red) traps and traces the larger 'experiment' bead independently with a separate QPD (QPD1).

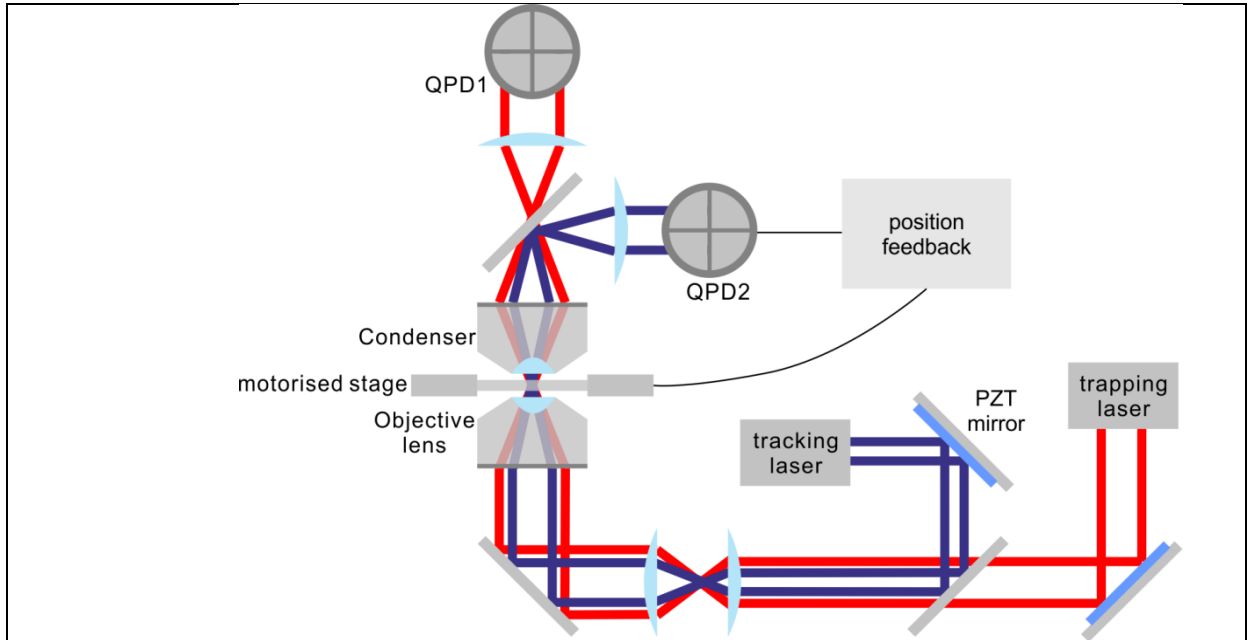

Figure 3. General schematic of a typical optical path diagram of a system that corrects for mechanical drift of optical components. The tracking laser tracks the position of a marker bead that is fixed on the assay chamber, which measures the movement of the assay chamber. The movement is supplied to a position feedback system that instructs the motorised nanostage to compensate the drift, resulting ~0.1 nm stabilisation of the system along all 3 spatial axes. This general type of 3D stabilization was first reported in reference [81], which is a useful source for additional practical information.

Brownian noise arises from the fact that the probe bead is immersed in a liquid solution. The bead can be modelled as attached to a Hookean spring undergoing damped simple harmonic oscillation. The equipartition equation along one spatial axis indicates:

$$\frac{1}{2} k <x^2> = \frac{1}{2} k_B T \qquad (2)$$

where $x$ is the position of the bead and the angle brackets mean averaging over a long period of time. This relation can be used to quantify the size of the thermal noise; $k$ is the spring constant, $k_B$ is Boltzmann constant and $T$ is the absolute temperature. Rearranging the equation, we get:

$$\Delta x = \sqrt{\frac{k_B T}{k}} \qquad (3)$$

where $\Delta x$ is the magnitude of the displacement noise. Since the power spectrum of the bead displacement fits a Lorentzian curve, noise other than Brownian noise can be filtered/discarded and the resulting noise becomes [76]:

$$x = \sqrt{\frac{4\beta k_B T \Delta f}{k^2}} \quad (4)$$

where $\beta$ is the hydrodynamic drag on the bead and $\Delta f$ is the frequency range in which the measurement is taken. Stiffness $k$ is set by the experiment, so to reduce noise, for example, one can use a smaller bead (so lower $\beta$) or reduce $\Delta f$.

2. Single-molecule detection techniques

Techniques which allow us to observe single molecules in their native state can potentially enable the elucidation of molecular interactions and complex dynamic behaviour. These techniques include light microscopy methods, structural investigation tools and electrical conductance measurements.

2.1. Light microscopy approaches

A suite of light microscopy tools has been developed due to the minimal perturbation they cause and the efficiency of labelling that is possible. Many of these techniques use fluorescence emission, a technique by which incident light of one wavelength is absorbed by a molecule and emitted at a longer wavelength, over time scales of ~ns (i.e. $10^{-9}$ s, see figure 4). The separation between the absorption and emission wavelengths allows filtering of the excitation wavelength using 'dichroic mirrors' to increase image contrast. Fluorescence emissions also occurs not just in single fluorophore molecules but also longer length scale material such as gold (plasmon resonance effects), as well as fluorescence in diamond and in quantum dots, but the physical processes are more complex than those we describe here.

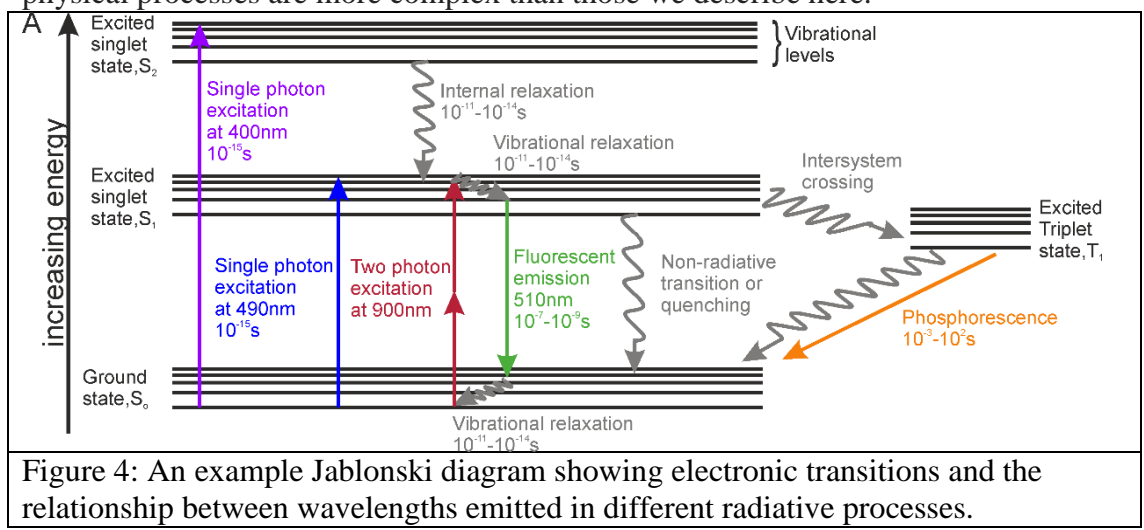

Figure 4: An example Jablonski diagram showing electronic transitions and the relationship between wavelengths emitted in different radiative processes.

2.1.1. Localization microscopy to overcome the optical resolution limit

Optical microscopy is a particularly appealing biophysical tool to study small molecules because it is less perturbing to their natural state than other methods [82]. However, the wave nature of light limits the diameter a beam can be focussed down to, equivalent to roughly half a wavelength, and is known as the diffraction limit. Many biological molecules of interest are approximately two orders of magnitude smaller than this, so we cannot study them by simply 'looking down the microscope'.

When two light emitting molecules are within a few nm of each other their diffraction-limited images overlap and the positions of each cannot be accurately determined. Nevertheless, the shape of the distribution of light from a single emitter can be modelled as Bessel functions (which in turn are reasonably well approximated by a Gaussian function when recorded on pixelated photon detector devices such as cameras). If we can create a low spatial density of emitters we can find the intensity centroid of a molecule to a few nm precision [60,83] by fitting its' intensity distribution to an approximation of the analytical Bessel/Gaussian intensity function (see figure 5).

For some samples a low spatial density is easily achieved, but for dense concentrations of emitters the concentration must be controlled [28,84]. There are a suite of methods to achieve this which will be discussed below, but mostly they utilise the photophysical properties of the fluorophore to achieve a low spatial density in any given frame of an image acquisition.

2.1.2. <u>Photoactivation, blinking and switching methods</u>

There are now a multitude of methods that achieve super-resolution by exploiting the photophysical properties of fluorescent molecules to turn on a random subset of molecules that densely label a biological structure. Photoactivated localization microscopy (PALM) [29,85] and Stochastic Optical Reconstruction Microscopy (STORM) [86] start with all the molecules either in a dark state for PALM or a different colour such as green in STORM, which are activated in PALM (or switched in STORM) to a different (typically red) colour by using an intense ultraviolet laser pulse. The short activation pulse results in stochastic activation/switching of a small random subset of the molecules to an emitting state, which is then read-out using a shorter wavelength. These methods are technically challenging to extend to two colour measurements partially due to the different chemical conditions (such as oxygen levels) required by the different fluorophores used, though there are now some good antioxidants available that work well with different coloured fluorophores in multi-colour STORM. A more major issue is with aligning the beams so they have the same illumination profile in the microscope. Binding-Activated Localization Microscopy (BALM) [87] and bleaching/blinking assisted localization microscopy (BaLM) [88] can be more simple to implement as only one illumination wavelength is required and images are acquired continuously instead of stroboscopically.

These technologies have been extended into multiple colours, 3D, living cells and living samples, although optimising imaging buffers for multiple colour experiments and resolving clusters of fluorophores remain significant challenges [89]. Notable recent developments are: the high localisation precision of 20 nm (xy) and 50 nm (z) achieved in a study of the 3D structure of chromatin [90] and 3D imaging in live cells applied to study the stoichiometry of DNA polymerase PolC in live bacteria *Bacillus subtilis* [91].

The increase in complexity of the imaging data being generated is driving advances in analysis software. Whilst early single-molecule super-resolution fluorescence experiments utilised simple software such as QuickPALM [92] which localises molecules from their centre of inertia, the drive for higher temporal resolution has led to data containing more closely spaced molecules and more complex software. RainSTORM [93], ThunderSTORM [94], and SRRF [95] are examples of popular reconstruction software utilising different methods to reconstruct data. A comprehensive resource for choosing an appropriate software for a given dataset has been compiled [96], comparing around 70 different software packages on simulated data.

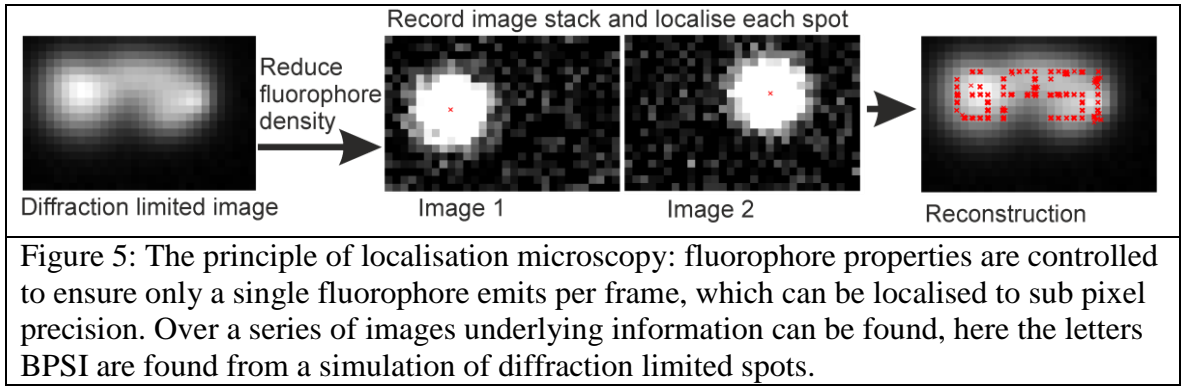

Figure 5: The principle of localisation microscopy: fluorophore properties are controlled to ensure only a single fluorophore emits per frame, which can be localised to sub pixel precision. Over a series of images underlying information can be found, here the letters BPSI are found from a simulation of diffraction limited spots.

2.1.3.  Stimulated emission depletion and related microscopies

In STED microscopy the fluorophores are excited with one beam, and de-excited via stimulated emission using a donut shaped intensity profile at a wavelength longer than their fluorescence emission. This reduces the area where fluorescence can occur such that any detected photons must have been generated in the small region towards the centre of the donut. By scanning the beams across a sample the locations of single molecules can be determined one molecule at a time [31,84].

Reversible saturable optical linear fluorescence transitions (RESOLFT) microscopy [97,98] uses a longer-lived conformational state change to de-excite molecules outside of the donut centre. This requires much lower laser intensity, allowing the use of less power, or the creation of an array of low power beams. An array of over one hundred thousand beams has been demonstrated [99], allowing areas greater than 100μm by 100μm in living cells to be scanned in less than one second. The method has also been demonstrated for time lapse imaging in live Drosophila melanogaster larvae [100].

In MINFLUX imaging [101] the donut shaped beam is used for excitation, and no de-excitation beam is required. Fluorophores at the centre of the beam are not excited, and the position where intensity is lowest is the most likely position of the particle. By repeating the measurement at three positions forming an equilateral triangle around the found position and comparing the intensities, nm localisation is achieved at much lower photon numbers than are required for intensity maxima fitting [83], but the fluorophores must be well separated for this method to work (see figure 6 (a)).

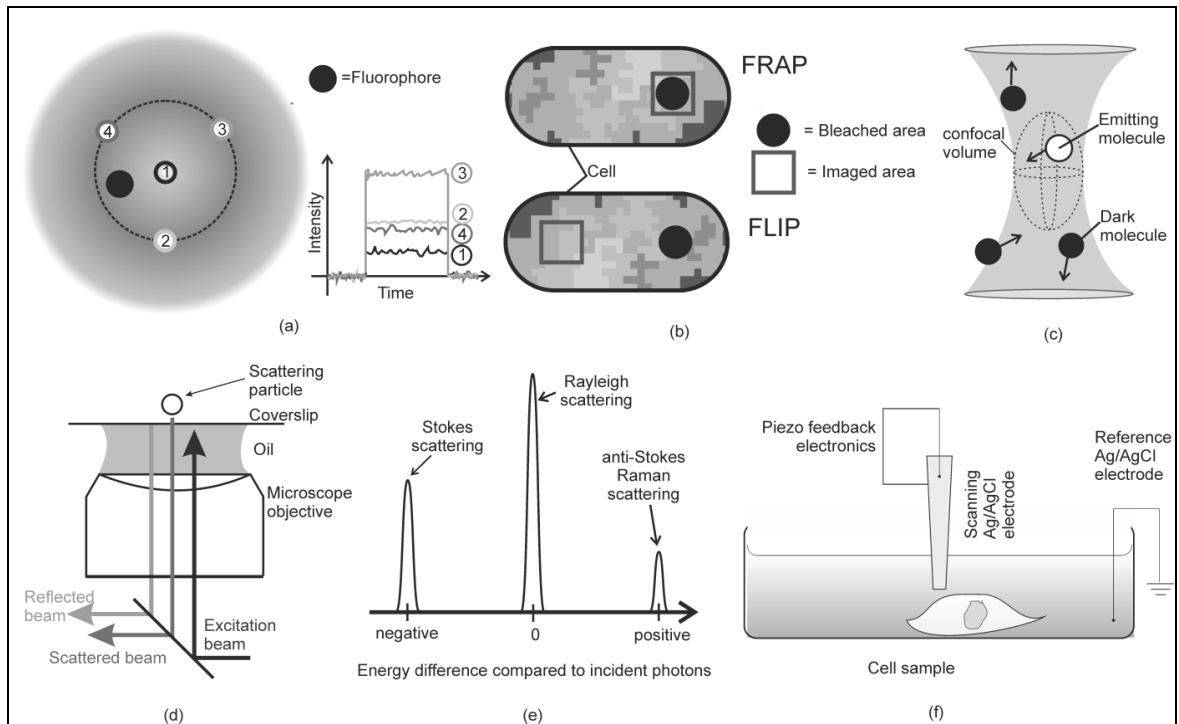

Figure 6: Schematic diagrams of selected microscopy techniques. (a) Schematic diagram of MINFLUX for localisation. The excitation donut beam (shown in blue) is centred sequentially at each of the positions 1,2,3,4 (shown centred on 1) to localise the fluorophore (yellow). The measured intensity for each position is lower the closer the position is to the fluorophore (graph). (b) A schematic diagram of FRAP and FLIP microscopy, showing the relative locations of the bleached and imaged areas. (c) Cartoon of FCS confocal volume. The laser is focused to an oblate volume indicated with dotted lines. Signal is recorded as a fluorescent molecule passes through the volume. (d) A schematic diagram of the illumination geometry for iSCAT microscopy. (e) Energy shifts in Raman Spectroscopy. The Stokes scattering peak is small compared to the Rayleigh scattering and must be enhanced. (f) Schematic of a typical scanning ion conductance setup, here used for generating topographical information of a live cell surface.

### 2.1.4. 3D localization microscopy tools

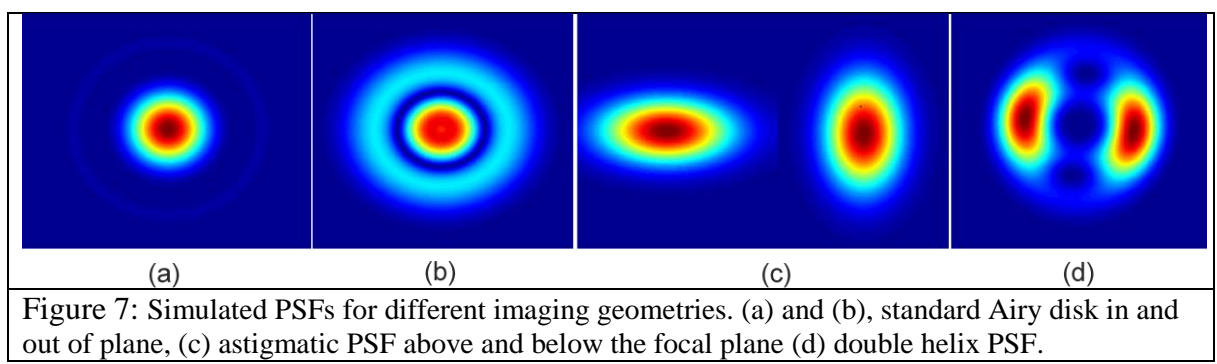

Figure 7: Simulated PSFs for different imaging geometries. (a) and (b), standard Airy disk in and out of plane, (c) astigmatic PSF above and below the focal plane (d) double helix PSF.

Localisation based microscopy approaches have been extended into 3D by fitting the PSF along the z dimension, perpendicular to the image plane. In a standard light microscope the width of its PSF changes relatively little above and below the focal plane over an axial (z) length scale of the depth of field (a few hundred nm for a high magnification microscope), but can still be used for 3D tracking by focusing on one surface of the sample or using biplane microscopy [102]. The latter utilises a second objective lens, opposite the first, focused on a second plane to sample the PSF of a single molecule at 2 different z heights to reconstruct the 3D position. The PSF itself can be engineered to better encode 3D information. The simplest method is astigmatism microscopy [103] which utilises a cylindrical lens in the microscope imaging path to slightly offset the focal plane along one lateral axis creating an elliptical distortion of the PSF dependent on z. Combined with STORM, this method has achieved 30 nm and 50 nm lateral and axial resolution respectively. [104] Utilisation of a phase mask or spatial light modulator in the imaging path, allows near direct engineering of the PSF to have desired properties. The most widely used engineered PSF is the double helix PSF [105] which contains two lobes which rotate in the image plane depending on the z-distance, achieving the highest resolution for 3D localisation techniques [106]. A whole zoo of PSFs have recently been designed with different operating ranges and resolutions including corkscrew [107], self-bending [108], saddle-point and tetrapod [109], some of these are shown in figure 7.

Also, an emerging method to measure z distances involves radiative decay engineering. Here, metal or graphene surfaces can increase or decrease the radiative decay rate from fluorophores. This process has a very sensitive distance dependence and so can be used as quantitative metric from height of the fluorophore from the surface.

### 2.1.5. Determining molecular interactions using Förster energy resonance transfer

Exquisitely sensitive measurements of the interaction between molecules are made possible by FRET. If two fluorophores with overlapping excitation and emission spectra are within ~10 nm of each other, energy can transfer non-radiatively from the shorter wavelength donor fluorophore to the longer wavelength acceptor [110]. The efficiency is inversely proportional to the sixth power of the distance between them allowing extremely sensitive distance measurements between ~1-10 nm [111]. Extra information is provided when combined with alternating laser excitation (ALEX) of the donor and acceptor excitation wavelengths, allowing sorting of the number and type of fluorophores present [112]. The first observation of FRET at a single-molecule level (smFRET) was made using a nearfield imaging approach, utilising scaning nearfield optical microscopy using a short DNA test construct to separate the FRET dye pairs [113]. ngle molecule FRET can be observed in diffusing molecular species inside a confocal volume. This geometry was used to observe initial transcription by RNA polymerase through the DNA-scrunching mechanism [114]. Immobilising molecules on a surface and combining with TIRF microscopy enables dynamic information to be obtained by observing FRET efficiently as a function of time. This has been used to detect mRNA as it exits RNA polymerase II in real time [115]. The technique has now been extended to use 4 fluorophores and 4 alternating lasers [116]. Recently, MD simulations have been used to map the possible orientations of labelling fluorophores on a single molecule to convert FRET efficiencies into Ångstrom resolution distances (i.e. $10^{-10}$ m, the diameter of a single hydrogen atom)[117], extending FRET to provide structural information rivalling crystallography, although only between pairs of points measured sequentially. FRET can also be used in conjunction with colocalisation analysis which uses single particle tracking of typically two different types of biomolecules tagged with different coloured dye molecules, to determine if their point spread function images overlap or not [118], to within the spatial precision of the tracking at least.

2.1.6. <u>Measuring molecular turnover</u>

Fluorescence recovery after photobleaching (FRAP) [119,120] and fluorescence loss in photobleaching (FLIP) [121,122] (see figure 6 (b)) have been used as ensemble techniques for decades, but they are increasingly used to measure the turnover of single molecules in live cell experiments.

FRAP performs a complete bleach of an area of interest and monitors the return of fluorescent molecules to the area, whilst in FLIP an area is continuously bleached and the loss of fluorescence from a different area is measured. Commonly determined quantities are diffusion coefficients and the relative abundance of the mobile and immobile population, and often FRAP and FLIP are used to complement each other [123]. Whilst most measurements from FRAP and FLIP are ensemble measures, the technique can be used to estimate stoichiometries and area densities as single-molecule differences are detected [124]. A technique has recently been developed that the authors call single-point single-molecule FRAP [125]. This technique has been used to determine the turnover of single molecules in a photobleached area of nuclear membrane in the form of transmembrane proteins returning to the bleached region as they are localised one molecule at a time, to build up a measure of the concentration ratio between the inner and outer nuclear membranes.

2.1.7. <u>Structured illumination methods</u>

Structured illumination microscopy (SIM) uses multiple images taken with patterned excitation light to improve spatial resolution [126]. By patterning the light at almost the optical resolution limit of the microscope the resulting light from the object contains Moiré fringes that encode the higher frequency information. By recording images with three different phases (for 2D), one can simply add the images together in reciprocal space, and inverse Fourier transform to produce a final image with twice the effective spatial resolution of conventional microscopy (and the method can be extended to 3D). If non-linear illumination is used the illumination pattern contains harmonics of the illumination frequency, which can in theory give infinite resolution. In reality, saturated structured illumination microscopy (SSIM) with pulsed lasers gives a resolution of around 50 nm [127].

Non-linear SIM can also be achieved using photoswitchable proteins to introduce the non-linearity [128]. This requires much lower laser powers than SSIM, making it biologically compatible, and achieves similar localisation precision to SSIM. "Instant SIM" develops this technique to 3D, and the usual post-processing performed in a computer on hardware, is instead performed via two microlens arrays, a pinhole array and a galvanometric mirror [129], allowing three-dimensional imaging at 100 Hz (although the spatial resolutions at this speed are ~350 nm axially and ~150nm laterally). Extending instant SIM to incorporate two-photon excitation increases the penetration depth and makes the method suitable for thick specimens [130,131].

The latest developments in SIM are mainly being driven by combination with other techniques, such as TIRF and patterned illumination - and these techniques themselves can be combined to achieve two colour imaging in live cells to achieve resolutions of 50-100nm [132].

2.1.8.  Fluorescence correlation spectroscopy
Fluorescence Correlation Spectroscopy (FCS) uses the transit of single fluorescent molecules across a femtolitre detection volume (i.e. $10^{-18}$ $m^3$, equivalent to an effective diameter of ~one micron, or $10^{-6}$ m), usually formed at the diffraction limited focus of a laser in confocal microscopy (see figure 6 (c)), to determine the 2D diffusion coefficient via the autocorrelation of the intensity *vs.* time trace [133–136]. Unlike most other single molecule fluorescence techniques, FCS was extended to two colours almost twenty years ago [137], but has had limited popularity due to the difficulty in getting focal volumes of the same size (requiring optical masking) to precisely overlap. Two photon excitation [138] removes some of the alignment difficulties of two colour but severely limits the probe choice.

Most recent progress has been in cell membrane applications or supported lipid bilayer mimics of cell membranes. Single cell FCS is challenging because the total number of fluorescent labels per cell is limited and they are bleached over time due to the laser illumination outside the confocal volume; and the cell can be highly heterogeneous, with very different properties in different spatial regions. Lower-phototoxicity techniques being developed include using total internal reflection microscopy or a light sheet to illuminate the sample [139,140] whilst combination with stimulated emission depletion microscopy allows a more accurate determination of spatial heterogeneity [141–145] and TIRF-STED-FCS has also been demonstrated [146]. Due to the heterogeneity many cells must be imaged to find meaningful statistics, and high throughput (sixty thousand measurements in ten thousand cells) FCS has recently been shown [147].

2.1.9. <u>Interferometric scattering</u>

In non-fluorescence light microscopy, images are formed by collating the scattering of light from the particle. The efficiency of a particle to absorb, scatter and extinguish can be calculated from the Q factors, derived from the complex refractive index (or complex dielectric function) using Mie Theory. The end result, of relevance to this section, is that the scattering signal is proportional to the sixth power of the diameter, *d* for Rayleigh regime scattering for which the length scale of the scatterer is much less than the wavelength of light. For nm length scale particles the scattering is too small to detect. Interferometric scattering techniques use an interference term to reduce the dependence on the diameter to the third power.

A basic microscopic arrangement is shown in figure 6 (d); spherical particles in a medium with a different dielectric constant which have a diameter much less than the wavelength of the illumination light scatter a spherical wave which is collimated by the objective lens. Some of the incident light is reflected directly from the coverslip and interferes with the scattered light giving three terms in the intensity at the detector– the scattering signal is negligible, background signal from the reflected unscattered beam, and the interference term which is the largest for very small particles [148].

By increasing the incident power, the number of photons collected in a given amount of time is increased, allowing imaging at high temporal resolution. Interferometric scattering microscopy (iSCAT) at 2 kHz sampling was applied to study the contentious topic of lipid microdomains, or 'rafts' [149], the existence of which had been under doubt due to difficulties in observation, and they were observed to have transient structural and mobility behaviour on a time scale on a few hundreds of milliseconds, faster than most scattering methods can image.

2.1.10. <u>Raman spectroscopy</u>

Raman microscopy uses inelastic photon scattering to determine the chemical bonds present in a molecule and therefore can be used to identify the specific type of molecule. Monochromatic photons are inelastically scattered with an energy shift caused by a change in the energy state of the molecule, for example in the rotational or vibrational energy levels. As the photon is scattered rather than absorbed, the shift can be seen with any input wavelength, so infrared lasers are usually used. The energy of the incident photon is usually reduced, which is known as a Stokes shift; if the energy is shifted up it is an anti-Stokes shift (see figure 6 (e)). Many of the incident photons are scattered elastically (Rayleigh scattering) and so notch or edge filters are often required to implement the technique.

The Stokes peak is too small for single molecules to be detected without enhancement. There are a number of techniques for this such as surface-enhanced Raman scattering (SERS) [150] in which the molecule of interest is adsorbed to a rough metal surface, such as gold or silver, and tip-enhanced Raman spectroscopy (TERS) in which a metal tip as used in atomic force microscopy is brought above the sample. Single molecule detection at room temperature has been available via SERS for twenty years [150], but work on biological samples is relatively recent, with work on a relatively large single mitochondria via TERS [151], and the detection of single dye labelled phospholipids in lipid membranes [152]. The detection of bending of individual molecules in lipid membranes [153] and observations of independent mobility of lipids and proteins in cell membranes [154] are exciting developments that will hopefully permit studies of structure and function, and the detection of small numbers of p53 (the expression levels of the protein p53 are altered in cancer) in human serum [155] suggest that single-molecule detection of cancer signatures by Raman spectroscopy may be possible.

Related techniques of Stimulated Raman Scattering (SRS) and Coherent Anti Raman (CARS) are now at, or very close to, single molecule sensitivity. CARS is a non-linear beam mixing process which involves a pump laser beam (wavelength in range ~700-1,000nm) and a lower frequency and a Stokes laser beam (usually the near infrared wavelength of 1064nm) producing an anti-Stokes interference signal of twice the pump frequency minus the Stokes frequency. If the beat frequency between pump and Stokes beams matches an active vibrational mode of a chemical bond in the sample then bond oscillations will be driven coherently, producing a signal several orders of magnitude stronger than conventional Raman. SRS microscopy has similarities with piump and Stokes beams, but here when the beat frequency matches a vibrational mode in the sample then stimulated excitation of this vibrational energy transition will occur. This results in minimal background for 'chemical imaging' compared to CARS, which is an advantage for single-molecule level sensitivity. Both techniques in particular have been used in conjunction with surface enhancement via nanoscale plasmonic antennas, and it is this approach that shows the most promise in regards to achieving single-molecule level detection.

## 2.2. Developments in light microscopy accessories

Developments in light microscopy have largely been driven by developments in the enabling technologies which continue to push the techniques to higher temporal and spatial resolutions. The most important of these are described below.

### 2.2.1. Probes

Since the first fluorescence experiments with fluorescent proteins much effort has been expended to extend the colour range (see figure 8), lifetime and photon budget of such dyes [156,157]. Similar advances have been made in the other classes of fluorescent probes [158,159], allowing faster image acquisition for longer times and thereby enabling the study of processes over different time scales.

The increasing number of correlative single-molecule techniques requires a new generation of probes that are optimised for multiple methods. In particular there are many probes being developed for correlative light- electron microscopy (CLEM). Click-EM [160] is a very elegant method for labelling non-protein biomolecules. Cells uptake a click chemistry substrate, and fluorescent dyes that produce singlet oxygen on fluorescence illumination are attached. During acquisition of the fluorescence image molecular oxygen is produced and causes a chemical reaction, the product of which can be seen in electron microscopy. To target proteins a fluorescent indicator and peroxidase for precipitation with EM resolution (FLIPPER) probe can be used [161]. In this technique a fluorescent protein attached to a peroxidase enzyme is genetically fused to the protein of interest, with the fluorescent protein providing contrast in optical imaging, and the peroxidase precipitating in electron microscopy. An example of a less specific biological probe is a fluorescent nanodiamond-gold nanoparticle (FND-Au) [162]; a ~10 nm diameter particle which is non-bleaching in fluorescence microscopy and has a high electron density for EM.

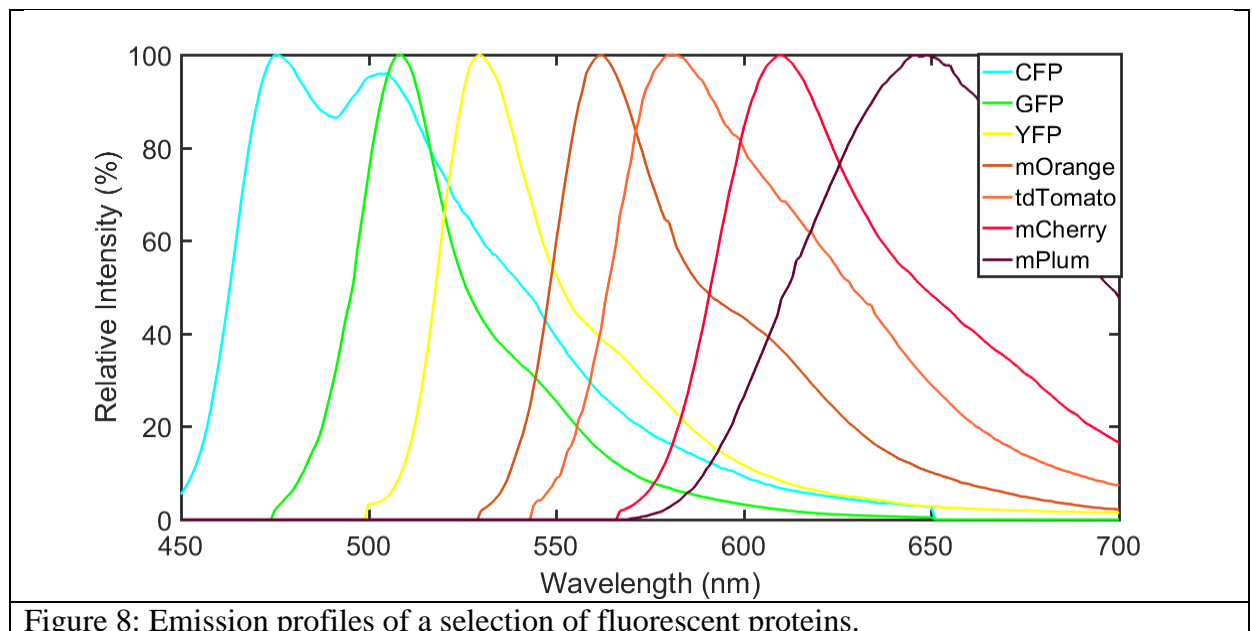

Figure 8: Emission profiles of a selection of fluorescent proteins.

2.2.2. <u>Microfluidics</u>

Microfluidics devices enable exquisite control of the fluid environment of single molecules and real time addition and observation of new species in an experiment [163]. Most microfluidics implementations are simple devices exploiting laminar flow, constructed from slides and tape [164] or moulded from polydimethylsiloxane (PDMS) resin to create multiple inlet channels. Fluid flow can be driven by gravity, a syringe pump or even passively using the surface tension of liquid droplets [165]. These types of devices have been used for ultra-fast mixing combined with FRET to see protein folding [166]. More complex devices trap molecules in small droplets by passing them through immiscible liquids [167]. Trapping reactants in droplets enables single-molecule observation of solution based reactions, including single polymerase chain reaction (PCR) reactions observed via fluorescent DNA probes [168]. Extremely intricate devices can be manufactured by multilayer soft lithography [169]. Here microfluidic valves are akin to transistors and are patterned into microfluidic geometries analogous to microprocessors. These arrangements have enabled very high throughput observation, via smFRET, of DNA reactions with different complementary strands at varying ionic strengths [170].

2.2.3. <u>Detectors</u>

Electron multiplying charge coupled device (emCCD) cameras have been the choice of imaging detector for single-molecule sensitivity for many years [171]. In CCD devices incident photons cause a build-up of charge on the detector elements, which is "shunted" between elements in readout. The shunt action can introduce errors, which are minimised by slow transfer times, but this limits the acquisition speed. Electron-multiplying cameras use on-chip amplification of the signal to detect single molecule events, but on-chip amplification also amplifies shot noise and therefore reduces the signal to noise compared to non-electron multiplying CCDs.

In the last 3-5 years scientific complementary metal-oxide semiconductor (sCMOS) sensors have achieved single-photon sensitivity via improvements in the semiconductor junctions [172,173]. In sCMOS cameras readout is directly from the detector element, and so there is no readout noise, meaning sCMOS cameras can achieve single-photon detection at low noise levels at fast frame rates. The choice of sCMOS or emCCD now depends on the specific details of the application [174].

Single photon avalanche photodiodes (SPADs) are based on semiconductor junctions in Geiger mode – that is a single photon will create an avalanche of signal. As a semiconductor based technology it is compatible with designs for CMOS cameras, although specialist configurations have been developed [175]. To date SPAD arrays are generally smaller than CCD arrays and are starting to be used for single-molecule experiments [176,177].

Balanced photodetectors, often standard photodiodes or avalanche photodiodes (APDs) are often used to increase the effective signal-to-noise ratio in signal molecule experiments, having some advantages to a simple amplification of the signal from a single photodetector. In particular, use of balanced photodetectors can reduce the overall "common mode noise" from laser sources. The simplest method to achieve balanced photodetection is to use two photodiodes connected such that their photocurrents cancel out. The effective signal from the balanced photodiode pair is thus zero until there is some difference in the intensity of one of the laser beams incident on the detectors.

2.2.4. Light sources

Lasers are the main illumination source in single-molecule fluorescence microscopy due to their single or narrow bandwidth and high illumination intensity. The propagation properties of highly collimated Gaussian profile beams have enabled many ingenious illumination and super-resolution techniques, such as in light sheet microscopy [178] or STED microscopy [84] mentioned previously. Light emitting diodes (LEDs) offer similar wavelength specificity to lasers at a lower power but, since they are inherently non-collimated, they are not suited for beam shaping techniques. LEDs do offer huge advantages in techniques where a large field of view is desired. The low power density of LED illumination is potentially less phototoxic than conventionally used arc lamps and provides uniform illumination; moreover, the intensity of the light source does not decay over time and the cost price of LED units is comparatively low. LED illumination has been shown to be suitable for video-enhanced differential interference contrast microscopy, a technique held back by the search for an incoherent light source with a narrow bandwidth [179]. LEDs are also suitable for inline holographic microscopy [180], a technique used for 3D single-cell tracking of motile species. LEDs are now available in every colour and their intensity is increasing to rival lasers.

Coherence is an important issue when comparing lasers with LEDs; the coherence length of LEDs is around a few tens of microns, compared to single-mode lasers which can be as low as ~20cm for cheap semiconductor laser and in excess of 100m for more expensive models, with some fibre lasers having coherence lengths of hundreds of km. Single molecule detection methods which rely on light interference are thus infeasible with LED sources, nevertheless there may be some benefit from incoherence of lasers in scrambling the effects of interference at the level of excitation of the sample in widefield fluorescence microscopy. LED illumination is also more difficult to implement in objective lens TIRF since the light source is spatially extended and so results in a large focal waist in the back aperture the objective lens which can thus couple light into non-TIRF illumination modes.

2.2.5. Software

There is no standard single-molecule microscopy software *per se*. Commercial microscopes are usually bundled with their own proprietary software packages, such as the Zeiss Zen suite. Camera manufacturers also often provide acquisition software such as the Andor Solis and iQ packages. Bespoke and some commercial instruments make use of the open source Micro-manager software (https://micro-manager.org/) which is based on the popular NIH ImageJ and is highly adaptable for many pieces of hardware. Many bespoke instruments also use bespoke software control, often written in LabView and the more cost effective Igor Pro. In terms of analysis of single-molecule data the super-resolution light microscopy community has developed various forms of software to reconstruct fluorescence images to better than the standard optical resolution limit, which are discussed elsewhere in this review. Similarly, the community of researchers engaged in molecular simulations have developed an extensive suite of software, which are discussed elsewhere in this review. Scientific languages such as MATLAB (Mathworks) and Mathematica are both heavily utilised in developing bespoke code for the analysis of single-molecule data, in particular in methods of step detection for the extraction of distinct single-molecule events above the level of background noise which is often comparable to the level of the background detector noise. Also, in methods used to analyse complex images such as live cell light microscopy data and to resolve subcellular compartments using automated image segmentation. MATLAB in particular has a thriving user community which encourages the sharing and discussion of code, which can of course facilitate developing new code for different applications.

2.2.6. Adaptive optics

Chromatic or spherical aberration is always present in a microscope, although at very low levels using high quality optics. The worst source is often the biological specimen itself, particularly with deeper observations into cells and tissues, due primarily to variation in the refractive index across the sample [20]. Adaptive optics attempts to correct for spherical aberration using spatial light modulators or deformable mirrors to alter specific parts of the light path. These techniques have yet to have a big impact in single-molecule imaging, partly due to the difficultly in optimising the adaptive optic with the stochastic bursts of photons received from single molecules, although a genetic algorithm has been developed to overcome this [181]. Adaptive optics have been used to extend STED deeper into tissues [182], and commercial systems now exist. The combination of adaptive optics with astigmatism imaging has been used to achieve axial localisation precisions of 20nm with fluorescent dyes and 40nm with fluorescent proteins [183].

- 2.2.7. Hydrodynamics and viscosity of single molecules
  There are a range of techniques that utilise hydrodynamics/viscosity properties of single molecules. For example, capillary electrophoresis (CE) can now achieve single-molecule precision to obtain high resolution separations of molecules based on differences in electrophoretic mobility due primarily to differences in frictional drag properties of the molecules. This example of hydrodynamic separation has also been combined with Cylindrical Illumination Confocal Spectroscopy (CICS) to separate DNA molecules of different lengths.

2.3. Single-molecule structural tools
  Structural tools have long been a workhorse of ensemble biophysics, but in the last few years great leaps have been made to enter the single-molecule precision regime.

- 2.3.1. Single-molecule crystallography
  Crystallography has long been an essential tool in the biophysicist's armoury. Able to atomically resolve the structure of suitably prepared molecules, great insight into the relationship between form and function on a biological level has been obtained through finding the structure of interacting biomolecules, and the crystal structures of proteins may be used as the basis for MD simulations to further probe dynamic and biological properties. Nevertheless, crystallography has historically been slow in tackling cellular membrane proteins, for which obtaining crystals of the requisite size represents a significant challenge. Recently, new methods have been developed which although not being explicitly 'single-molecule' are approaching exceptional levels of detection precision in requiring only nanocrystals while keeping molecular damage by X-rays at an acceptable level[184]. Nanocrystals of the target protein are fired in solution through a femtosecond X-ray beam and the resulting snapshots collated to form the final image and 3D atomic map. Further refinement of this technique, which has come to be known as serial femtosecond crystallography (SFX), makes use of existing lipidic cubic phase (LCP) techniques [185].

  The LCP injector combined with SFX imaging reduces the amount of sample required by making more efficient use of the microcrystals present. SFX technology promises to address the dearth of membrane crystal structures in the literature, and may open a new cellular realm for investigation. Progress made with the X-ray free electron lasers (XFELs) may also make possible imaging of individual proteins held coherently in the gas phase [186], eliminating the need for microcrystals altogether; experimental progress towards this has been rapid in the past decade. Similarly, rather than express molecules known to exist in eukaryotic cells in crystal form, it has been demonstrated that they may be investigated with X-ray crystallography *in situ* [187]. That being said, XFELs still have issues with poor temporal coherence. Nevertheless, methods have been developed to work round this issue which use the good temporal coherence of a longer wavelength source to act as a seed coherence at shorter harmonic wavelengths.

Traditional crystallographic techniques have also been meeting with success. The details of molecular growth and translocation are of fundamental importance to the cell, with implications for virtually all processes. Resolving such phenomena, which are by their nature dynamic, is a problem which will need to be addressed if full understanding of the cell is to be obtained. Work towards that was published in 2013 [188] in which the authors successfully crystallised membrane proteins together with a translocating cellulose molecule, and gave structures of the synthase itself as well as the cellulose channel formed. The structures found not only suggest the function of the individual sites of the proteins, but also help to elucidate the dynamic behaviour of the complexes.

### 2.3.2. Developments in electron microscopy

Operating below the diffraction limit of light microscopes, electron microscopy is able to resolve detail in the sample at the level of Ångströms, and is therefore an unparalleled tool for structure determination. But, as with X-ray crystallography, a snapshot of a biomolecule on its own cannot give the full picture. Interactions, movements, even phonon modes are important properties which deserve careful consideration: however, traditional electron microscopy tools do not permit these dynamic features to be measured directly. One method for probing this is known as ultrafast electron microscopy (4D UEM), a technology which is becoming more ubiquitous in the study of biomolecular behaviour. Through a variant of the original 4D UEM methodology, it has been possible to image in femtosecond resolution protein vesicles combined with nanometre precision spatially, and to image a full *E. coli* cell [189]. The detailed information which arises from this type of experiment is at the level necessary to truly understand the interactions between biomolecules.

Suffering as it does from the diffraction limit, optical microscopy in biophysics was energised by the development of fluorescent tags and concomitant super-resolution analysis techniques to obtain information which otherwise would be hidden. Much standard electron microscopy approaches do not traditionally need to tag proteins of interest through genetic means or otherwise, though immuno-EM methods do achieve this through the use of specific antibodies which have gold particles conjugated as an electron opaque contrast reagent. However, being able to perform an experiment of this type raises possibilities in the realm of correlated fluorescence and EM experiments *in vivo*. Demonstration of such a genetic tag was published in 2011 [190]. Here, the authors designed a tag which generated singlet oxygen, similar to the process described above for Click-EM, which catalysed a reaction that was visible in an EM image. This tag was shown to localise, and correlate well with simultaneously obtained fluorescence data, giving structural and localisation information that was previously unobtainable.

Other means of tagging molecules have also come about. Gold nanoparticles, used extensively in magnetic and optical tweezing, have also been used as an identifier for single molecules in cells imaged with EM in physiologically relevant aqueous solution [191] and imaging at a resolution of 4nm - well below many super-resolution imaging techniques.

### 2.4. Single-molecule electrical conductance measurements

Whilst conductance measurements have long been used to study cell membrane channel proteins, the technique is increasingly being applied to more diverse questions, such as DNA sequencing.

### 2.4.1 Patch Clamp

The patch clamp has been a workhorse of biophysics since the 1970s. In the most traditional form of single channel patch clamp a glass pipette containing an electrolyte solution is sealed onto a small area of cell membrane containing only one pore/channel, and the flow of ion currents through the opening can be recorded.

Single channel patch clamp recording has been used extensively to study ion channel proteins involved in electrical conduction in nerves and the transduction of electrical signals at neuromuscular junctions. More recent and diverse applications include biological processes such as those in insect olfactory receptors [192], those of direct relevance to impaired muscle contraction as occurs during atrial fibrillation [193], and the mechanisms of anion conduction by coupled glutamate receptors [194]. Importantly, it is also finding new applications in correlative techniques, as will be described in section 5.1.2.

### 2.4.2 Nanopore based detection

Nanopores enable single-molecule electrical measurements. Typically two reservoirs of ionic solution are separated by a nm length scale pore. The molecule of interest is introduced at one side of the pore and a voltage applied, forcing molecules through the pore via electrophoresis. As these molecules translocate and occlude the pore, characteristic drops in current are measured dependent on size and shape of the molecule [195]. The pores themselves can be solid state, silicon dioxide [196], graphene [197], or protein based, such as the S. aureus toxin α-haemolysin [198]. Recently pores have been constructed from DNA origami. [199]

Ever since the first applications of nanopores for detection of single molecules of nucleic acids [200] much of the pore research has focused on attempts to sequence DNA, with many believing this technology will cross the $1000 whole genome sequencing threshold. Oxford Nanopore Technologies has produced a USB-sized, chip-based nanopore sequencer with long read lengths which can produce several GB data/day although early results have been mixed in regards to high read-error rates [201].

Nanopores have also been used to interrogate a diverse range of single molecules. An α-haemolysin protein pore was used to detect mRNAs from lung cancer patients [202]. They have also been used to detect proteins directly [203] and protein interactions with nucleic acids [204]. More recently, the unfolding kinetics of single proteins has been measured [205] such that this technology might one day sequence or fingerprint proteins directly.

### 2.4.3 Scanning ion conductance microscopy

Scanning ion conductance microscopy (SICM) is a scanning probe microscopy technique which uses measurements of ion flux as a metric for the proximity of a surface of a sample due to an increase in probe electrical resistance as the surface is approached [206]. Here, the probe is a glass nanopipette, made by melting and drawing out the tip of a standard micropipette until the diameter is only ~10-30 nm. A small electrical potential difference is applied between the tip and a physiological ionic buffer (Figure 6 (f)). The technique is similar to the physiology method of patch clamping but includes additional scanning with the probe of the surface of the sample. As the tip is moved to within its own diameter from the biological sample being scanned then the ion flow is impeded (i.e. the measured electrical resistance through the probe increases).

With fast feedback electronics similar in nature to those used for atomic force microscopy this level of electrical resistance can be used to maintain a constant distance between the tip and the sample, and so can generate topographical information as the tip is laterally scanned over the surface. The spatial resolution is limited by the diameter of the nanopipette, which is the length scale of relatively large protein complexes on cellular surfaces. This is worse than atomic force microscopy but has an advantage of causing far less sample damage. The technique has also been adapted to be combined with single-molecule folding studies of fluorescent proteins: here the same nanopipette is used primarily to deliver a chemical denaturant to unfold, and therefore photobleach, a single fluorescent protein molecule, prior to their refolding and gaining photoactivity, and so this method can be used to study the kinetics of these processes [207].

### 2.4.4 Dielectric spectroscopy AFM

Dielectric spectroscopy AFM is a variant of AFM that measures the local electric dipole moment (i.e. the electrical permittivity) of a sample with an applied electric field. The field varies over a range of frequencies so the frequency response of the sample is probed. Dielectric spectroscopy measures molecular fluctuation directly and it has high enough sensitivity to precision measure thermal expansion [208].

Non-biological applications include fuel cell testing [209], microstructural characterisation [210], etc. Biological applications include molecular interaction [211], detecting cancerous cells [212] and more generic cells [213] including bacteria [214] and yeast [215]. Label-free microfluidic biosensors have also been contrived to apply dielectric spectroscopy [216].

## 3. Single-molecule manipulation techniques

Forces, torques and conformational changes underlie most cellular processes: myosin pulls against actin to cause muscle contraction [217]; kinesin [38,218] walks along microtubule tracks to transport cellular cargos; during DNA replication, positive supercoiling of the DNA builds up due to the translocation of DNA polymerase, which needs to be relaxed by a class of enzyme called a topoisomerase to a less stressed conformation for continued replication to then proceed [219,220]; ATP synthase creates ATP from adenosine diphosphate (ADP) by rotating its gamma subunit [221], as well as monitoring DNA topology, phase transitions and even knots at a single-molecule level. Force spectroscopy tools, notably optical tweezers, magnetic tweezers and atomic force microscopy, have the capability of measuring the forces and torques on single molecules [222–224]. They can also apply forces and/or torques of biologically relevant values to perturb innate systems. The basics of the mechanism and design of these instruments have been described extensively in literature [76,225–229] although a summary is provided here for convenience. In this section we intentionally do not discuss the many pioneering, seminal papers of force spectroscopy and general single-molecule biophysics, since we wish to emphasize the more recent progress of the past decade. These papers are reviewed in full elsewhere [230], but for clarity the reader is also steered towards the original seminal papers: single-molecule force spectroscopy to stretch single DNA molecules[231,232]; optical tweezers used for direct observation of a molecular motor (kinesin) translocating on its (tubulin) track[77]; AFM force spectroscopy measurements of adhesion forces between ligand and receptor pairs[233]; fluorescence imaging observation of rotation of single F1 rotary motors[234]. This chapter, instead, will focus on the innovative designs, combinations, and applications of these instruments that have emerged in the past decade or so.

3.1. Molecular manipulation using light

Photons carry linear and angular momenta. When photons hit an object, their momenta transfer to the object because they are either refracted, reflected or absorbed. Macroscopically this manifests as a force/torque applied on the object. These are the essential physics underlying optical tweezers.

3.1.1. Optical tweezers (OT)

Optical tweezers use a high numerical aperture (NA) objective lens to focus a collimated laser beam such that an object – usually a ~micron diameter sphere – close to the geometrical centre of the focus is at a potential energy minimum. Figure 9 (a) shows a basic laser tweezers setup. The bead is trapped by the laser so by moving the laser focus, one can move the bead. Near infrared (NIR) lasers serve as the light source as continuous-wave NIR lasers up to 10 W are readily available and this wavelength range causes less photodamage to biological molecules compared to visible wavelengths. A basic OT uses laser light with a Gaussian profile so that the intensity maximum is at the centre of its beam waist. The force transduction of the OT can be explained with ray optics as well as dipole approximation. In the ray optics description, the bead, having higher refractive index than the surrounding liquid, will experience gradient forces from the peripheral part of the beam which push the bead towards the focus where light intensity is at a maximum. The central part of the beam, being partially absorbed and reflected by the bead, pushes the bead forward, so the equilibrium position is slightly downstream of the laser waist. The scattering and the gradient contributions to the trapping force most come from light deflection (refraction) rather than light reflection or absorption from the bead. In fact, reflection coefficients of silica or polystyrene/latex beads are typically not larger than 5%.

The maximum optical force that the trap can apply to the bead along any of the three spatial directions can be expressed as [235]:

$$F = Q\frac{nP}{c} \quad (5)$$

where $Q$ is a constant between 0 and 1; $n$ is the refractive index of the surrounding liquid; $P$ is the power of the laser entering the trap, and $c$ is the speed of light in vacuum. Four example scenarios are depicted in figure 9 (b)-(e); note that the simplification and approximation of the sketches mean that they are suitable only for qualitative descriptions of the scattering and gradient forces; for quantification of forces, see [235,236]. The same three rays across the four scenarios are selected as representative to illustrate the gradient and scattering forces. In figure 9 (b), the bead is at the equilibrium position. The left and right rays refract at the bead boundaries to apply two gradient forces that balance out each other horizontally and add up vertically to result in a net downward force. The reflection of the two rays at the boundaries has been neglected. The ray shone on the bottom of the bead provides an upward scattering force by reflecting off the bottom boundary. The refraction of this ray is neglected. The overall resultant force is thus zero. Figure 9 (c) shows the bead at the centre of the focal plane. There is zero net gradient force so the only force from the laser is the upward scattering force. The overall resultant force pushes the bead upward. Figure 9 (d) shows the bead below the focal plane. Both scattering and gradient forces point upward so the net force is upward. In figure 9 (e) the bead is positioned off the optical axis. The net gradient force pulls the bead towards the left. The scattering force pushes the bead to the right but it is weaker than the gradient force. The overall net force is leftward. For a Gaussian profile beam, the scattering force partially balances out the axial gradient force, so $Q_{axial} < Q_{radial}$. Thus $Q_{axial}$ limits $Q$ of the trap. In the dipole approximation description, the bead is treated as a dielectric particle, which is polarised by the optical field in the laser trap. The optically induced dipole then interacts with the light field and tends to move along the field gradient, resulting in the trap applying a restoring force to the bead in all three dimensions.

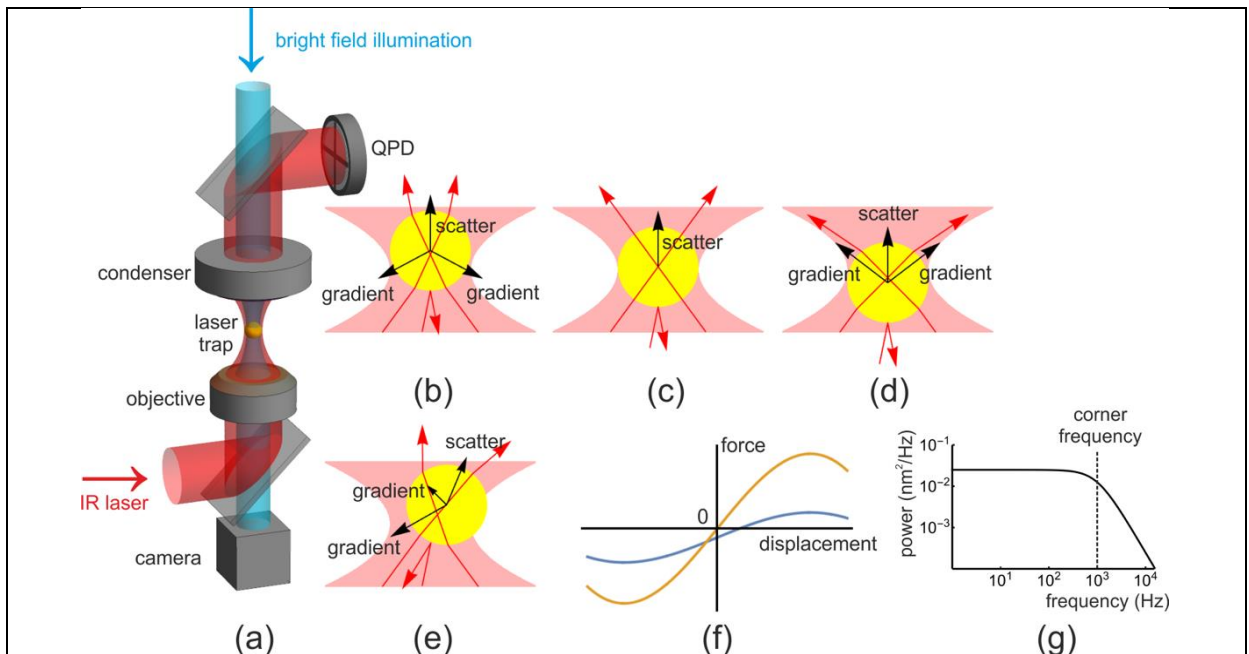

Figure 9: Schematics of optical tweezers and tweezing forces. (a) Key optical components in the OT. A collimated infrared laser beam enters the objective from below, then focuses to a diffraction limited spot, before being re-collimated by the condenser. The IR beam is then imaged on a quadrant photo diode, or QPD (the imaging lens between the condenser and the QPD has been omitted). (b) – (e) Describe forces applied on the bead in four scenarios in which the bead is (b) in the equilibrium position, (c) in the centre of the laser focus, (d) below the centre and (e) displaced to the right of the optical axis. (f) Shows two representative force-displacement curves for vertical (blue) and horizontal (orange) displacements. The curves are linear near equilibrium positions, suggesting a Hookean-spring force model. The gradient is the stiffness of the trap, which is limited by the axial stiffness. (g) Theoretical power spectrum of a trapped bead trace follows a Lorentzian shape. The corner frequency is approximately at the 'corner' of the log-log plot and is labelled with a dashed line.

The bead position is measured via interferometry. Either the trapping light itself or a separate tracking laser source is re-collimated after the trap by the condenser lens. This light is the mixture of the laser light that passes the bead unaffected by it and that which is scattered by it – an interference pattern at the back focal plane of the condenser – which then is imaged through an imaging lens onto a quadrant photodiode (QPD). A QPD is a 2 x 2 array of photodiodes arranged each in a quadrant, essentially a 4-pixel camera with high imaging frame rate (>10 kHz). A QPD does not take frames of images and its temporal resolution is characterised by the frequency range in which the diode has sufficiently high responsivity, or the bandwidth. Any movement of the trapped bead relative to the trap centre will cause shifts in the interference pattern, picked up by the QPD [237]. The QPD output signal is electric potentials that need to be translated into displacement with a pre-established look-up table. The trapped bead in a thermal reservoir of surrounding water molecules has its movement described by the Langevin equation:

$$m\ddot{x} + \beta\dot{x} + kx = F_{\text{thermal}} \qquad (6)$$

where $m$ is the mass of the bead, $\beta$ the hydrodynamic drag on the bead, $k$ the stiffness and $F_{\text{thermal}}$ the Brownian force. Upon Fourier transformation, equation 6 turns into a Lorentzian shaped curve in the frequency domain:

$$S(f) = \frac{k_{\text{B}}T}{\pi^2\beta(f^2 + f_0^2)} \qquad (7)$$

where $f_o$ is the corner frequency. Figure 9 (g) shows an example power spectrum plot. The corner frequency is where the power density drops to 50% of that at zero frequency and it is related to the stiffness of the trap by $k = 2\pi\beta f_o$. This can be used to obtain $k$. A QPD rather than a camera is required for this because the camera's frame rate is too low. Once $k$ is known, the force applied to the bead by the trap is easily obtained from Hooke's Law (an example of the linear response region between trapping force and displacement is shown in figure 9 (f)).

3.1.2. <u>Low axial-intensity beams</u>

The intensity profile of a laser beam can be characterized by its transverse electromagnetic (TEM) mode. The lowest order mode, $TEM_{0,0}$, has the same form as a Gaussian beam. For an optical trap formed with a $TEM_{0,0}$ Gaussian beam, the axial stiffness is lower than the radial stiffness. This is due to a combination of (i) the confocal volume being extended axially compared to radially/laterally by a factor of 2-3, and (ii) the effects of the scattering force from axial part of the beam profile. By contrast, a uniform intensity profile laser beam will have less axial scattering, and thus achieves a higher $Q_{axial}$:$Q_{radial}$ ratio. This ratio can be achieved by further decreasing the intensity of the axial part until a desired level is reached.

In practice the TEM$_{0,1}^*$ Laguerre-Gaussian (LG) mode beams have been used to yield a ~20% increase in $Q_{axial}$ [238], which is similar to the $Q_{axial}$ boost of a uniform intensity beam [235]. The field distribution of an LG mode with radial and angular mode of orders $p$ and $l$ is expressed as [239]:

$$U_{p,l}(r,\phi) = U_0 \rho^{\frac{|l|}{2}} L_p^{|l|}(\rho) \exp\left(-\frac{\rho}{2}\right) \exp(il\phi) \tag{8}$$

where $r$ and $\phi$ are the cylindrical coordinates, $\rho = 2r^2/w^2$, $w$ is the 1/e beam waist and $L_p^{|l|}(\rho)$ the Laguerre polynomial of order $p$ and index $l$. Figure 10 (a) and (b) show plots of the intensity distribution of TEM$_{0,0}$ and TEM$_{0,1}^*$ beams.

Such beams can be converted from Hermite-Gaussian mode beams by using a mode converter [240,241], which is essentially two cylindrical lenses, using the Gouy phase shift. The Hermite-Gaussian beam is produced by inserting metallic wires inside the laser cavity [242]. An alternative way to generate Laguerre-Gaussian beams is to use a computer generated hologram [243] or a spiral phase plate [244] to modulate a Gaussian beam. In the former case, a forked grating intercepts the Gaussian beam and the interference pattern has a specific $l$ but a range of various $p$. The blazed pattern of the hologram is given by [243]:

$$H(r,\phi) = \frac{1}{2\pi} \text{mod}(l\phi - \frac{2\pi}{\Lambda} r \cos\phi, 2\pi) \tag{9}$$

Where $\Lambda$ is the period of the grating. Figure 10 (c) shows a plot of the hologram. The transmittance function is then given by $T(r,\phi) = \exp[i\delta H(r,\phi)]$. In the case of spiral phase plate, a phase displacement, the amount of which depends on the azimuthal angle, is added to the beam, resulting in destructive interference that causes the characteristic ring shape. Figure 10 (d) shows a sketch of a spiral phase plate. One limit with this method is that higher order modes are generated, which reduces the purity of the beam.

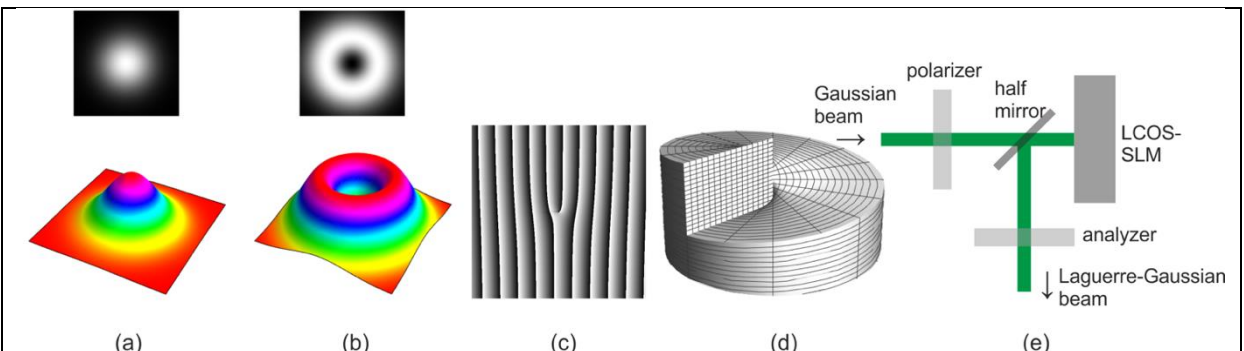

Figure 10: Laser profiles and laser mode conversion. Beam profiles for (a) TEM$_{0,0}$ mode Gaussian and (b) TEM$_{0,1}^*$ mode Laguerre-Gaussian beams. (c) Shows an example of blazed hologram for generating Laguerre-Gaussian

beams from Gaussian beams. The expression is given in equation 9. (d) An example shape of a phase plate. The thickness increases with the azimuthal angle to introduce phase displacement into the Gaussian beam to turn it into a Laguerre-Gaussian beam. (e) Optical diagram for converting Gaussian beams to Laguerre-Gaussian beams utilising a LCOS-SLM.

Another way to convert a Gaussian into an LG beam involves using a liquid-crystal-on-silicon spatial light modulator (LCOS-SLM) [245]. An LCOS-SLM comprises a 2D array of computer-controlled light modulators that can change the phase, polarisation or intensity of incoming light. The phase change needed to impose high order modes on a Gaussian beam can thus be applied to it at the position of each modulator. Unlike a lithographically created hologram, the pattern of which is fixed once the pattern is blazed, spatial light modulation enables dynamic control of the LG beam generation. Figure 10 (e) shows a sketch of the arrangement of optical components for such a method of beam conversion. The polariser aligns the polarisation of the incoming beam to the liquid crystal (LC) molecules that actuate the phase change. This is needed as light polarised in other directions will not be modulated by the LC molecules. The light then reflects off the LCOS-SLM, where its phase is modulated, before moving on to the downstream optics. The disadvantages of an LCOS-SLM include phase distortion during phase modulation that eventually manifests as a decrease of trapping stiffness but the distortion can be compensated. Similar variations of phase across a beam profile can be achieved using deformable micromirror arrays. In fact, deformable mirrors are used a lot for active optics in general and they have a number of big advantages over spatial light modulators e.g. much less light is lost which is important for experiments with low SNR, such as those with single-molecule microscopy

3.1.3. <u>Bessel traps</u>

Bessel beams have an intensity profile defined by a Bessel function of the first kind [246]:

$$I(r) = I_0 J_n(k_r r) \tag{10}$$

where $J_n$ is an $n$th-order Bessel function, $k_r$ is the radial wavevector and $r$ is the radial coordinate. Figures 11 (a) and (b) show a zeroth-order and a first-order Bessel beams, in both grayscale density plots and 3D plots. Apart from being non-diffractive, a true Bessel beam also has the peculiar property of self-reconstruction after being partially occluded with an obstruction.

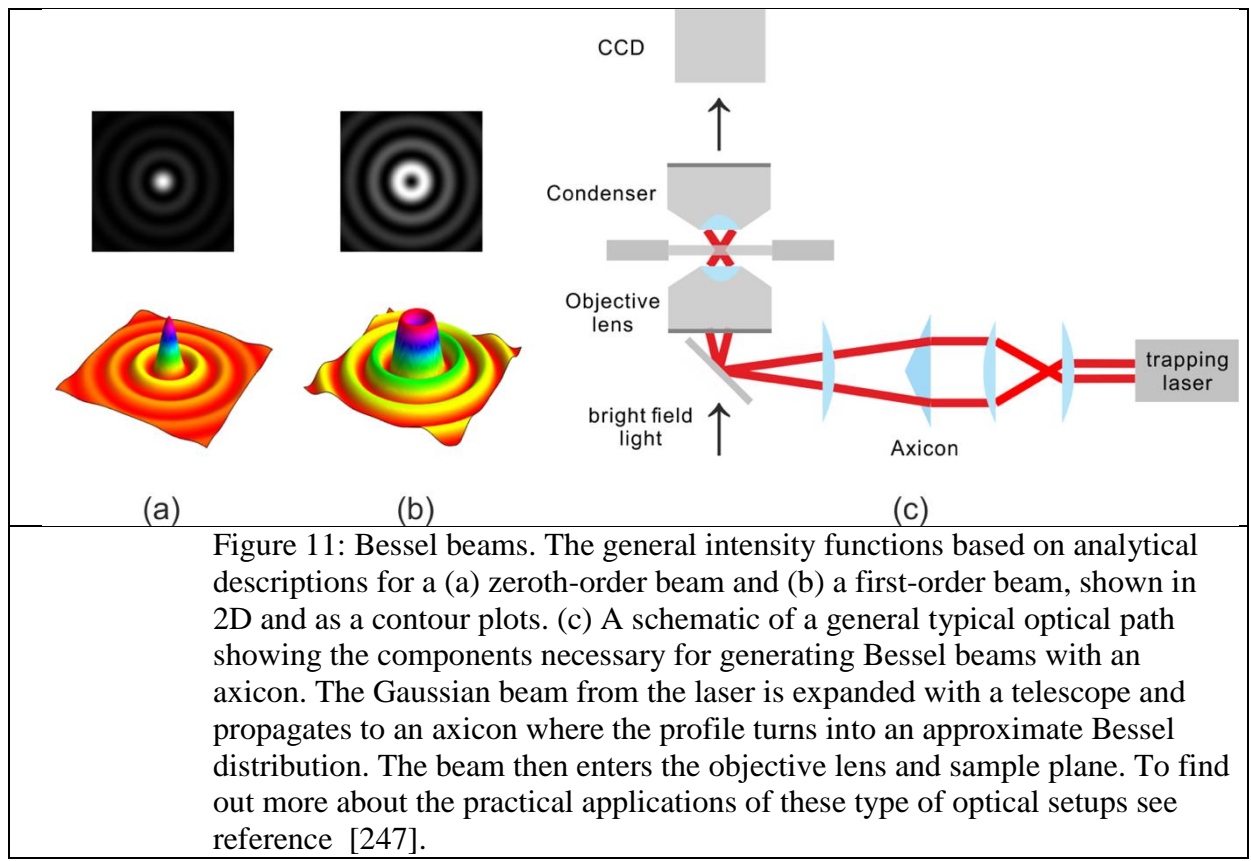

Figure 11: Bessel beams. The general intensity functions based on analytical descriptions for a (a) zeroth-order beam and (b) a first-order beam, shown in 2D and as a contour plots. (c) A schematic of a general typical optical path showing the components necessary for generating Bessel beams with an axicon. The Gaussian beam from the laser is expanded with a telescope and propagates to an axicon where the profile turns into an approximate Bessel distribution. The beam then enters the objective lens and sample plane. To find out more about the practical applications of these type of optical setups see reference [247].

Like the generation of Laguerre-Gaussian beams, there are a few ways to generate a Bessel beam. The simplest method to create a zeroth-order Bessel beam is by passing a Gaussian beam through an annular aperture and then a positive lens a focal length away [246], since mathematically the zeroth-order Bessel beam is the Fourier transform of a ring. A more efficient way (not losing all the light blocked by the annular aperture) is by passing a Gaussian beam through an axicon, which is a conical prism [248]. Higher-order beams are possible with the light source being Laguerre-Gaussian of specific orders and indices [249]. Figure 11 (c) shows the optical diagram for using an axicon to generate a zeroth-order Bessel beam. A caveat with this method is that the high requirement in alignment of the axicon with the optical axis is crucial to the quality of the generated beam. Holograms [250] and SLM [251] can of course be used as well with the advantage of more flexibility both in the order of Bessel beam generated and, in the case of SLM, the real-time control of the phase pattern.

The non-diffractive and self-reconstructive properties can be harnessed to increase the workspace of the trap [252]: to achieve trapping a Gaussian beam needs to be sharply focused to a diffraction-limited tight spot with low divergence downstream of the trap, whereas a Bessel beam eliminates diffractive divergence and thus traps along a long distance along the optical axis, sometimes even forming a channel for particle flow. In multi-trap setups where holographic traps are generated to manipulate multiple beads the lack of diffractive divergence from a Bessel beam allows the trap centre to incorporate multiple beads along the optical axis; when the beads are stacked along this axis, the beads at the front diffract and scatter the light, so efficient beam reconstruction is required to trap the beads at the back. Garcés-Chávez et al. [253] used a single Bessel beam to create multiple traps along the optical axis. The traps are as far apart as 3mm, allowing statistics gathering in high throughput assays as well as optically driven nanostructures. Another area where Bessel beams have found application is optical pulling force (OPF) traps [254,255] in which particles such as polystyrene beads are pulled towards the light source instead of away from it. When more photons are forward scattered by the particle than backward, a negative force is applied to the particle due to the conservation of momentum. Non-paraxial Bessel beams can be manipulated to achieve this. Higher-order Bessel beams have orbital angular momentum so can be used for rotating objects (see section 3.1.4).

3.1.4. Optical torque trap

Similar to the transfer of linear momentum from light to the trapped object to translate it, transfer of spin or angular momentum can rotate the object [256]. There are many ways to utilise angular momentum of light, some of which have been developed recently. For completeness, we will also briefly mention some of the older methods.

The simplest approach to rotation with light is trapping two points on an object with two separate traps and revolving the traps around each other. But, more commonly Bessel, Laguerre-Gaussian or other beams that contain orbital angular momentum are used to impart angular momentum on probe particles leading to their rotation. These are constant-torque, rather than constant-rotation traps. Alternatively, linearly polarised light can rotate a birefringent probe by a defined angular displacement. The extraordinary axis of the nanofabricated non-spherical probe is perpendicular to the trapping beam [257].

3.1.5. Holographic optical tweezers

Holographic optical tweezers (HOT) make use of holographic light modulation to create a single beam intensity profile so that it can trap large numbers of micro-particles in 3D with independent and dynamic control over each individual trap. The modulation of light can be achieved with nanofabricated diffractive optical elements [258] or SLMs [259,260]. Soon after its inception, HOT were capable of manipulating hundreds of micro-particles [261] dynamically. Apart from high throughput screening, HOT has applications in macromolecular sorting [262], or arranging materials into 3D structures as a new nanofabrication technique [263,264]. Compared to traditional tweezers, HOT provides some flexibility and adaptability. But, one of the advantages of traditional tweezers is the possibility of measure forces directly. There is no such capability in HOT indeed it is not clear that even force-distance curves are yet measurable in HOT.

3.1.6. Lab-on-a-chip optical tweezers

A lab-on-a-chip (LOC) is a small device with a characteristic length scale from millimetres to a few centimetres but has the ability, in principle, to function experimentally as a full scale laboratory instrument. Microfluidics technologies have been particularly valuable in catalysing the development of LOC devices. LOC is often automated and is capable of high throughput screening. LOC optical tweezers do not have the space to create a local laser intensity maximum with a high NA objective lens. Instead, they use designed dielectric nanostructures to shape light and create trapping potentials for single micro-particles. Light fields in this type of LOC OT are fixed due to the fixed design of the nanostructure [265,266].

### 3.2. Magnetic force techniques

Magnetic forces can be used to manipulate micron sized magnetic particles, which can be utilized in single-molecule biophysics. The forces measured and calculated depend on the type of magnetism used. Ferromagnetism or super-paramagnetism are the standard choices. The magnetic forces can be generated from the interaction between a magnetic field vector, **B**, and the magnetic dipole moment vector **m** of a magnetic particle. The magnetic force vector **f** is proportional to, and in the direction of, the **B** field gradient [267], whereas the torque vector **τ** aligns and scales with the field itself [268]:

| $f = \nabla(m \cdot B)$ | (11) |
|---|---|
| $\tau = B \times m$ | (12) |

Instruments that use these principles to manipulate magnetised micro-objects are collectively called magnetic tweezers (MT).

#### 3.2.1. Magnetic tweezers (MT)

MT have similarities to OT, though there are features which are distinctive and unique to MT. The equilibrium position for the magnetic bead is the local gradient maximum, which is inside the magnet (or electromagnet) so the magnetic micro-particle, typically a bead, can never reach the equilibrium position. Strictly speaking, the magnetic forces do not 'trap' the bead but either fix the bead in position by feedback loops that constantly adjust the field or by balancing the forces on a bead from a tethered biological molecule– see figures 12 (a) and (b) where the DNA molecule pulls the bead downwards and the magnets pull it upwards. The strength of the force can be adjusted by moving the magnets closer to or further from the sample (see figure 12(c)). Rotating permanent magnets will result in rotation of the magnetic bead. Such rotation may also be achieved by varying the electrical current in electromagnets, which can also apply horizontal forces with certain pole configurations. This has been demonstrated using six poles recently [269].

Since there is considerable flexibility in the MT design, especially those which are electromagnetic in origin, the magnetic forces can easily reach biologically relevant values of tens of pN. To put this into context, the muscle protein titin and other related proteins contain Ig and Fn domains which unfold with a reasonably high probability at applied force of 20 to 30 pN [79,80,222–224,270,271], and double-stranded DNA undergoes an overstretch transition at around 65 pN [272]. The minimum of applicable force is just as important since it determines whether the smallest biological forces can be measured. In the case of MT, this value can be as low as Brownian forces. The ability to easily and efficiently apply external torque to a magnetic bead is the key strength of MT (see figure 12 (d)). Compared to optical tweezers, the manufacturing and implementation of MT for rotation is more robust and requires less effort in regards to fine optical alignment in particular. Biological values of torque are in the range of a few pN·nm up to a few tens of pN.nm. Double stranded DNA separates when a ~9 pN·nm torque is applied, for example in molecular machines which unravel the two strands of the DNA double helix [273]. The ATP synthase $F_1$ motor, along with the associated membrane integrated Fo rotary motor, generates a torque ~40 pN·nm to couple electrochemical potential energy across a phospholipid bilayer membrane to mechanical rotation of these two machines, remarkably transforming the rotation into chemical potential energy locked into newly manufactured molecules of ATP, which some biologist describe as the universal 'fuel' in all living cells (though note that this statement is not strictly true e.g. GTP in eukaryotes or $H^+$ ions with prokaryotes are also used as fuels). That being said, there is still a challenge in providing very high magnetic field gradients for very large forces to be applied in magnetic tweezers.

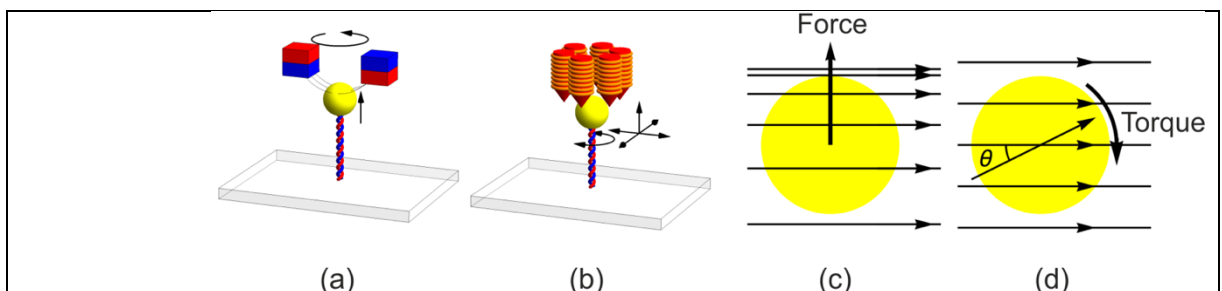

Figure 12. Schematic diagrams of magnetic tweezers configurations and force and torque generation. Here, we are assuming ferromagnetism in the figure with a permanent magnetisation. It should be noted that the big problem is then to calibrate the tweezers due to hysteresis effects, which is why people often use super-paramagnetic spheres instead. (a) Typical permanent-magnet MT. Double-stranded DNA is tethered to the coverslip at the bottom and a paramagnetic bead at the top. The force always points upward. (b) Electromagnet with six poles with the capability of applying horizontal forces. (c) Magnetic force due to the field gradient points towards regions of higher field. (d) In a uniform B field the force is zero, but the misalignment of the magnetisation of the magnetic bead from the background ***B*** field gives rise to a torque.

3.2.2.     3D control of magnetic traps

The MT in figure 12 (a) can apply force in only the positive *z* direction and torque along the *z* axis. The magnets are held by a motorised arm that can move the magnets closer to or further away from the sample along the vertical axis. Also, the arm can rotate the pair around the same axis. MT of various other translational and rotational degrees of freedom have been designed to meet the needs of biological molecule manipulation. The MT in figure 12 (b) [269] has poles at six corners above the bead so can apply radial force in addition to upward pulling. Theoretically four poles can apply radial force as well; Huang et al. designed such an MT [274], but in addition, they had a matching four poles underneath the sample so full 3D translation and rotation was possible.

Spatial constraints are a major design consideration for MT: a typical commercially available light microscope which uses a high numerical aperture objective lens for high resolution imaging, as is most relevant to single-molecule biophysics, has limited space available between the objective lens and the sample, and the sample and the condenser lens. These microscopes are therefore not ideal, as they stand, for implementing many of the bulkier MT designs. Whereas the setup in figure 12(b) can be incorporated into a commercial microscope, Huang's design needs to be incorporated into a bespoke microscope. Miniature MT [275,276], like the above electromagnetic tweezers in capabilities of manipulation of degrees of freedom, but small enough to fit into the flow cell, have been designed. These nanofabricated MT completely avoid the tight-space problem (see section 3.2.4 for more details.)

It is sometimes desirable to simply harness just the rotational capability of MT. Helmholtz coils are the simplest geometry to achieve a near-uniform magnetic field. Two pairs arranged perpendicular to each other can rotate the bead along one axis [277–280] whereas three pairs can rotate along all three directions [281]. Although one pair cannot rotate the bead by itself, it has found application in oscillating the beads [75]. Also, ultimately the impedance of electromagnetic coils limits dynamic measurements with this type of magnetic tweezers.

Clever methods to rotate magnet pairs without the capability of rotation along that axis have been devised. For example, the magnet pair can be rotated in the same way as in figure 12 (a) but then the arm holding the magnets can be moved sideways so that the molecule attached to the magnetic bead lies in the transverse plane. Loenhout et al. [282] used such an approach to supercoil a DNA molecule and extend it in the transverse plane for fluorescence imaging.

### 3.2.3. Torque measurement

A problem that challenges the traditional MT (such as the two shown in figure 12) is the large torsional stiffness. $B$-field gradients capable of applying forces whose magnitude is physiologically relevant require $B$-field strengths that can exert torques at many orders of magnitude above both biologically relevant values and measurable values. The reason that high-stiffness MT cannot be calibrated for torsional stiffness is that the angle between the bead magnetisation direction and the $B$-field is measured. The torsional stiffness is inversely proportional to the angle. But, the angle resulting from either Brownian rotation or the torque from a biological molecule decreases with stronger field to the extent that it is below measurement resolution. A transversely orientated rod can be added to the bead to amplify the angular displacement [283]. See figure 13 (a). Other clever geometries involving the combination of a cylindrically symmetric torque-less magnet to pull the bead and a small side magnet to apply force have been devised [284] (See figure 13 (b)). Similar designs with the rotation controlled by Helmholtz coils are created to reduce mechanical vibration and increase control precision [280]. The top magnet has even been replaced by optical tweezers for holding the bead in position [277–279]. Also, rotation faster than the angular response of the bead has been utilised to rotate the bead at a speed non-linearly dependent on the B-field rotation frequency [269].

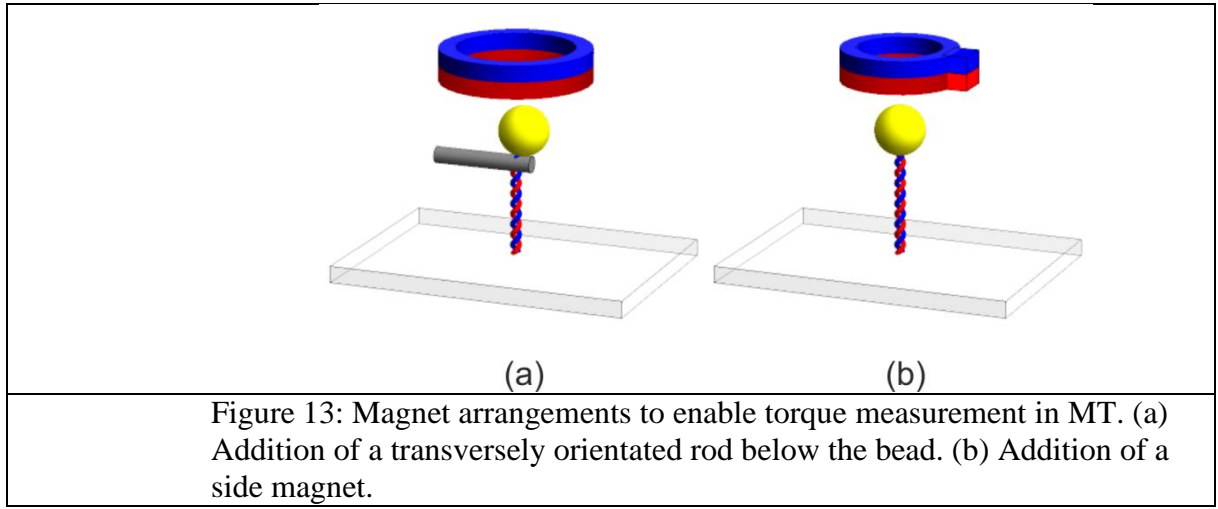

Figure 13: Magnet arrangements to enable torque measurement in MT. (a) Addition of a transversely orientated rod below the bead. (b) Addition of a side magnet.

3.2.4. Nanofabricated magnetic tweezers

Traditional magnetic tweezers whose profiles are in the range of centimetres to tens of centimetres suffer from spatial constraints imposed by the microscope. This is why magnetic tweezers are often built around home-made microscopes rather than commercially available ones [269,274,284,285]. But even in the former case, MT design still must make compromises in versatility to accommodate the positioning of microscope components. On the other hand, miniature MT that are lithographically embedded in a flow cell seamlessly integrate with any existing microscopes. Figure 14 (a) shows an example of nanofabricated MT with six poles in the same plane (the yellow structures) and a ring of coil (or 'ring trapper') lying flat in the centre (red) designed by Chiou *et al* [275]. The poles can apply forces along in two dimensions in the plane of the magnets while the ring provides vertical attractive force. Once trapped, the superparamagnetic or ferromagnetic bead can also be rotated. The proximity of the poles to the biological sample can be on the order of 100s of micrometres (at least one order of magnitude smaller than traditional MT). Since the *B* field scales with $1/r^2$, a much smaller current can be used to generate comparable levels of *B* field at the sample. Indeed, since magnetic forces scale with the gradient of the *B* field, and adjacent pole pieces in nanofabricated MT are so close that the gradient is huge compared to traditional MT, a smaller *B* field is needed to apply the same force on the bead. A similar design by Fisher *et al.*[276] shown in figure 14 (b) has every other pole raised. The placement of three poles above and three below the biological sample allows full translational and rotational control along all spatial directions. Thus, the need for the ring trapper is removed and the control of the *B* field is more versatile. Note though, flat pole pieces are not necessarily good, since they have a small field gradient. Often people using pointed pole pieces.

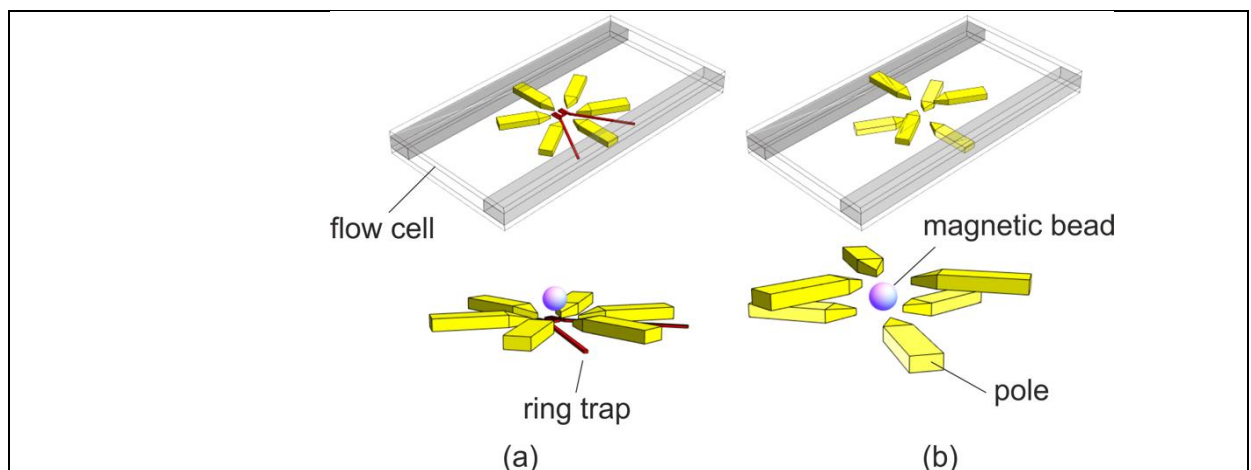

Figure 14. This shows a general schematic of two possible miniature electromagnetic tweezers designs. The top panels show overviews with the flow cells and the lower panels are zoomed-in figures of the poles with magnetic beads featured. (a) Here the hypothetical MT comprises six poles in the same plane, each of which is an independently controlled coil with a pole piece in the centre. There is also a circle of wire in the centre to provide a vertical force. The wires of the coils are not shown for clarity. For further information on how an MT device similar to this shown was implemented in practice see reference [275]. (b) Here, the putative MT has three poles that are in one level and three are raised relative to each other, to allow full translational and rotational control of the magnetic bead. For further information on how an MT device similar to the schematic shown was implemented in practice see reference [276].

The disadvantages of nanofabricated MT include the fact that the MT is fixed with respect to the biological samples so it is not possible to navigate the flow cell with the microscope stage keeping the $B$ field centred at the optical axis of the objective lens. The big advantage though is that small coils have small inductances, which mean you can do faster experiments. Chiou's design includes a small patch of gold on the slide above the ring tapper so thiol-functionalised magnetic beads, once moving near the gold surface, can attach to it. Also, the flow cell can be reused so the space inside the flow cell must be cleaned before use. The placement of coils in contact with the flow cell means that resistive heat dissipates directly into the flow cell. Also, the cross-sectional area of the coils is 2 to 4 orders of magnitude smaller than large MTs, significantly increasing resistive heat generation at the same current level. This poses a maximum on the applicable current, somewhat offsetting the strengthening of the $B$ field due to the closeness of the poles.

3.2.5. <u>High throughput devices</u>

The heterogeneity in molecular behaviours and conformations, coupled to the existence of probabilistically rare events, call for a statistical strategy for the interpretation of data for the study of single molecules. Investigating multiple single molecules in parallel has advantages in (i) saving time and (ii) automatically setting the same experimental conditions across a single population of many molecules. Magnetic tweezers (excluding micro-MTs introduced in section 3.2.4 above) can be considered to be intrinsically high throughput. This is because the ***B*** field is relatively uniform over a relatively large volume in these devices. For example, the length scale over which this uniformity extends can be several mm, which can potentially encapsulate several biological molecules in the sample under study.

In the case of ATP synthase, this molecular machine spans a volume whose effective diameter is ~10 nm, and in the case of DNA topology studies, the DNA molecules under study are extended no more than ∼15 μm along the long axis of the DNA in most cases. These length scales are significantly smaller than the mm length scale range for uniformity in the ***B*** field. The magnetic field formed by either permanent magnets or electromagnets has a field variation of only a few percent over the mm range. The limit of the useable ***B*** field for single-molecule biophysics experiments is normally set by the camera detector field of view in conjunction with how densely biological molecules are actually positioned in the sample and the numerical aperture of the objective lens used (see figure 15(a)). Early attempts at multiplexed experiments used a low magnification low numerical aperture objective lens, and thus a large field of view [286], to magnetically trap 34 DNA-tethered beads which could be monitored in the same camera detector field of view simultaneously. Demagnification in their setup introduced inaccuracy in bead position measurement so the researchers had to develop methods to mitigate the error statistically by tracking multiple beads.

The magnetic field has negligible variance over the extent of the field of view so any force and rotation applied to the beads are constant over all beads, given that all beads have nearly the same value of magnetisation. Vlaminck et al. [287] made key improvements in the positioning of samples on coverslip surfaces so that they could track ~450 DNA-bead tethers to yield data on 357 molecules from a single experiment (see figure 15(b)). Instead of letting DNA randomly immobilise to the coverslip surface, Vlaminck et al. micro-printed the molecules on the coverslip in a 2D patterned array.

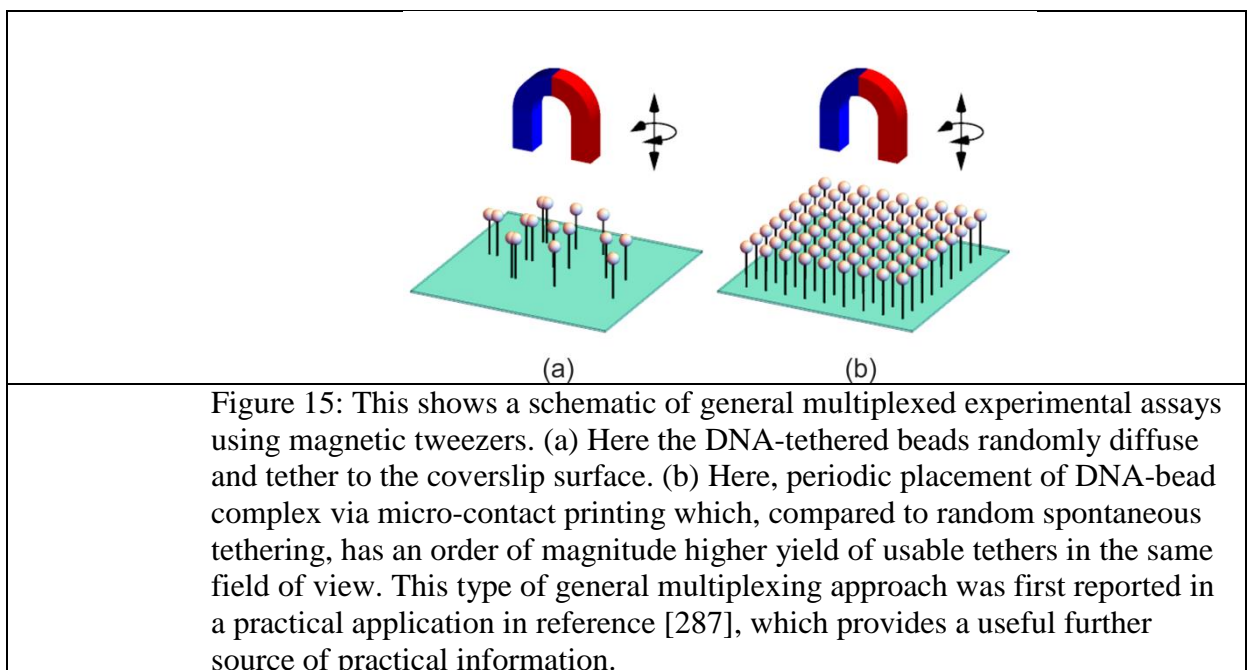

Figure 15: This shows a schematic of general multiplexed experimental assays using magnetic tweezers. (a) Here the DNA-tethered beads randomly diffuse and tether to the coverslip surface. (b) Here, periodic placement of DNA-bead complex via micro-contact printing which, compared to random spontaneous tethering, has an order of magnitude higher yield of usable tethers in the same field of view. This type of general multiplexing approach was first reported in a practical application in reference [287], which provides a useful further source of practical information.

High throughput MTs have been used to differentiate DNAs that are annealed and ligated [288].

### 3.2.6. Hybrid devices

OT have been added to MT as an independent source of force [277–279,289]. This also conveniently solves the large-torsional-stiffness problem and in some cases the combination allows easy measurement of torsional stiffness of the device. Figure 16 shows a DNA supercoiling assay where the magnets introduce twists into the DNA and the OT presses the bead downwards. Upon release of the OT, the downward force disappears and the plectoneme unravels [289]. The authors used this setup to investigate fast structural transitions of single DNA molecules; only with an independent source of force is it possible to suddenly remove the force and thereby explore out of thermal equilibrium behaviour.

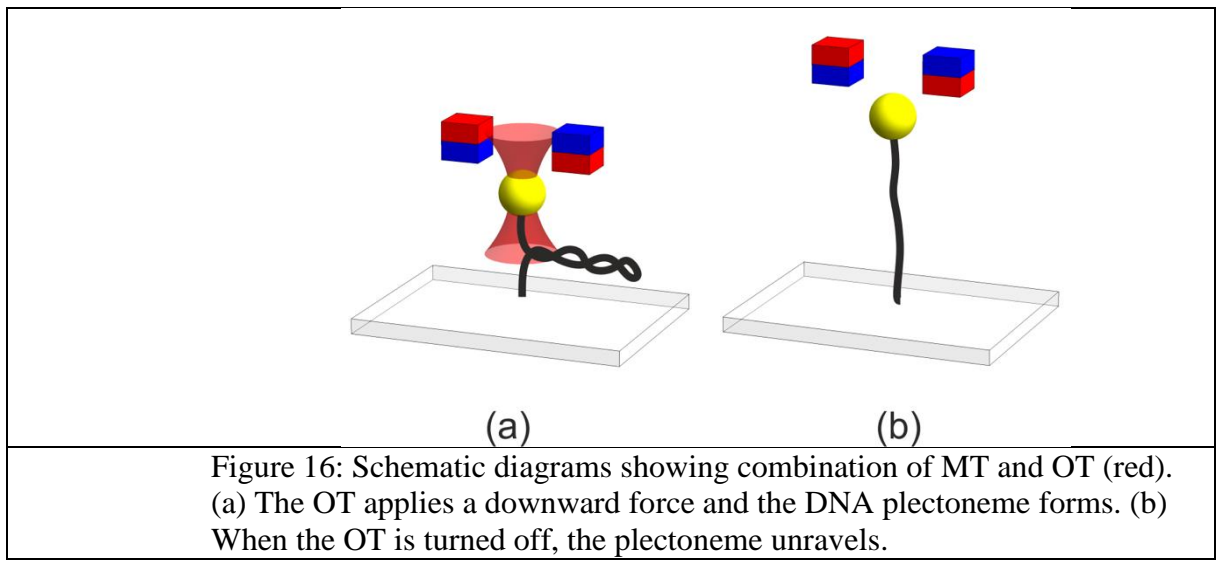

Figure 16: Schematic diagrams showing combination of MT and OT (red). (a) The OT applies a downward force and the DNA plectoneme forms. (b) When the OT is turned off, the plectoneme unravels.

### 3.3. Acoustic force to manipulate single molecules

Acoustic waves create air pressure gradients which can be used to trap micro-particles. Sound waves are longitudinal oscillations whose peaks are local medium (air or liquid) pressure maxima, also called pressure nodes, and troughs are local minima, or pressure antinodes. Micro-particles will be forced to either nodes or antinodes. If two sound wave transducers (known as interdigital transducers) are placed opposite each other, they generate two travelling waves that superimpose to form a standing wave. The positions of the nodes and antinodes are thus fixed and the particle fixed in 1D. Adjusting the wavelength and phase of the two travelling waves results in controlled movement of the positions of the nodes and antinodes.

Shi et al. [290] focussed a stream of micro-particles into a thin file with acoustic waves. Placing two pairs of interdigital transducers at right angles traps the particles in 3D – trapping in the vertical dimension comes from the flow of liquid at the nodes/antinodes that balance out gravitation, buoyancy and viscous drag – a setup named acoustic tweezers (AT) [291]. Among the advantages that AT offer are: the absence of heating as energy density is ~10 million times lower than that of optical tweezers, innate high throughput capabilities due to the multiple nodes and antinodes present in the chamber [292], biocompatibility between sound waves and biological molecules and, as an LOC device, convenience and ease of use. AT have even been applied to the selective control of micro-particles [293,294] as well as to trapping *in vivo* [295]. With the rise of acoustic holograms [296], we can expect the arrival of holographic AT capable of more dexterous shaping of the traps.

3.4. Surface probe methods

There are more than 20 different surface probe methods of potential relevance to single-molecule biophysics. These approaches utilise some physical readout of force interaction between a probe and a surface to map out the surface's topographical features. The most useful surface probe technique in terms single-molecule biophysics is atomic force microscopy or AFM.

3.4.1. Atomic force microscopy (AFM)

Atomic force microscopy uses an AFM tip attached to a thin, compliant microscopic metal cantilever to probe the surface of a sample, which can comprise soft biological matter. A laser beam shines onto the back of the cantilever, which reflects the beam to a photodiode to measure the beam deflection. An AFM tip of microscopic length scale but with a nanoscale radius of curvature is fixed to the cantilever. This tip is normally silicon based such as silicon nitride, and interacts with the sample through mainly Van der Waals attraction and Coulomb repulsion, resulting in bending depending on the distance of the sample from the tip [297]. The resolution of AFM is routinely on the Ångström level. AFM is applied to both imaging and force measurements. In imaging, the cantilever performs a 2D scan over the sample surface and builds the surface profile. However, until recent efforts to speed up AFM, the temporal resolution was low due to the mechanical nature of the imaging mode. In force measurement, the AFM tip moves vertically via a piezo electric actuator. The cantilever is tethered to the biological molecule so it can then be pulled/pushed to determine a force-displacement curve, which can reveal conformational changes of the biological sample [298].

AFM has been used in the detection and localization of single molecular recognition events [299], cell surface probing with molecular resolution [300], and in investigating the properties and activities of proteins [301–304] and nucleic acids [305]. Recent development in AFM include high speed AFM [306], and externally tuning the oscillator's response characteristics for robust position control [307]. The coating of cantilever tip has also been modified for higher resolution [308].

3.4.2. Electrostatic force microscopy

Electrostatic force microscopy has a mechanism similar to that of AFM except electrostatic forces are measured. A cantilever with a conductive tip at the sample end is used to measure the electrostatic force of a surface. A voltage is applied between the tip and the sample so the local charge distribution over the sample surface causes the tip to experience an attractive or repulsive force, which bends the cantilever. The extent of displacement is read out in the same way as that of the AFM, ie. with a laser beam reflecting off the cantilever surface onto a split photodiode. In the non-contact mode of operation, the cantilever is placed at a distance sufficient away from the sample surface that the force is always attractive. The cantilever is oscillated at its resonant frequency. The changing electric force varies the resonance frequency of the cantilever. Images are then formed by measuring the resonance frequency and phase. In the contact mode of operation, the tip-sample distance is kept constant. When an ac electrostatic modulation signal is applied to the tip, the cantilever sustains some level of vibration even though the tip is in contact with the sample surface. The surface electric potential, charge density and topography can be measured in this mode.

EFM has been applied to measuring surface hardness, surface potential, and charge distribution [309]. Applications in biological samples include the imaging of photosynthetic proteins [310] the visualisation of charge propagation along individual proteins [311], synthetic biological protein nanowires [312], electron transfer between microorganisms and minerals [313] as well as among microbial communities [314].

3.4.3. Cut-and-paste tools

Cut-and-Paste tools are used for the assembly of biomolecules at surfaces. These methods combine AFM with DNA hybridization to pick individual molecules from a depot chip and arrange them on a construction site one by one. Anchors and handles are composed of DNA, but alternatively a broad range of ligand receptor systems may be employed. Kufer et al. [315] used complementary DNA strands to pick up functionalised DNA oligomers with Ångström level precision. Each molecule transfer is monitored with fluorescence microscopy and force spectroscopy. They also used this method to deposit fluorophores in pre-defined patterns [316].

3.5. Electrical force manipulation tools

There are several biophysics techniques which utilize electrical forces, and many of these are relevant to single-molecule science, in particular including methods to rotate molecules and dielectrophoresis tools to translationally manipulate molecules.

3.5.1. Electrorotation

Electrorotation has many similarities to magnetic tweezers as a biophysical single-molecule rotation technique. A micro-particle with a permanent electric dipole moment is rotated by application of a 3D electric field generated from micro-electrodes built inside a flow cell. The physical principle of operation is that mobile electrical charges take a finite time to reorientate to follow an alternating electrical field and thus for a rapidly oscillating E-field there is a phase lag between the electrical dipole of a suitable particle which contains mobile surface electrical charges (for example, many typical chemically functionalized microbeads containing mobile surface ions) and the E-field. This phase lag establishes a torque between the external E-field vector and the dipole moment of the particle, thus resulting in rotation. Rowe et al. [57] applied rotation to bacterial flagellar motors via electrical charged microbeads, and measured speeds in excess of 1000Hz.

### 3.5.2. Anti-Brownian electrokinetic traps

Brownian motion of probe particles in solution poses challenges to the imaging of nanoscale objects. Free diffusion only leaves a time window at the millisecond level or less before the object moves out of observation volume. Anti-Brownian electrokinetic (ABEL) traps use electric field to confine the probe so it stays in position. Wang and Moerner [317] implemented ABEL to measure the diffusion coefficient and mobility of single trapped fluorescent proteins and oligomers.

# 4 *In silico* single-molecule biophysics

Much has been said in this review thus far about experimental methods in biophysics which operate at the single-molecule level. However, there are also several invaluable theoretical tools which utilise computational methods using biophysical modelling in particular to add insight into our understanding of biological process. The most important of these are molecular dynamics simulation approaches. Here we explore recent advances to the range of '*in silico*' methods for single-molecule biophysics, i.e. those which utilise intensive computational analysis.

## 4.1 Simulations of biological molecules

Computational simulations of the movement and interactions and dynamics of biological molecules have had a huge impact on our understanding of several fundamental biological processes. Here we make a brief overview of the basic simulation methods and review recent progress in this area.

### 4.1.1 Time and length scales

Simulations of biological molecules may be broadly but usefully classified as quantum mechanical, atomistic, or coarse-grained – divided along lines set by limitations in computer memory and processing power – and overlap between the three has historically been difficult although great developments have been seen in each class, shown pictorially in figure 17. Alongside divisions in the computational world, the *in silico* realm as a whole has in general operated at length and time scales orders of magnitude too small to be replicated in the laboratory. Nevertheless, increasingly sophisticated computational techniques are facilitating more and more computational experiments spanning multiple time and length scales in order to more fully elucidate biological phenomena, and recent developments in coarse-grained simulation software have made possible studies of experimentally tractable systems, while larger quantum mechanical calculations not only describe natural biological processes but throw light on designed systems which may be experimentally useful. Linking the two, atomistic molecular dynamics remains a workhorse of the biological physics community, and continued development in force fields and computational efficiency have enabled more complex systems to be within reach of this detailed level of investigation.

However, finite computational power leads inevitably to limits on what may be accomplished. In the coarse-grained regime, simulations of cellular environments have been reported with simulation times of 1 microsecond for a cubic simulation cell with edges of length 1 micron [318], while simulations of micron contour lengths of DNA are possible with modern coarse-grained simulation techniques [319]. Simulations of proteins with coarse-grained techniques have been performed over times in excess of milliseconds, allowing conformations and folding with long time scales to be elucidated.

Atomistic simulations of proteins are not far behind. With a specialised machine and high-performance algorithms, a simulation of a protein's folding and conformational dynamics were carried out, revealing conformational states with relaxation times orders of magnitude slower than those previously seen [320]. With traditional computers and widely available software, simulations of hundreds of base pairs of DNA have been seen and simulations of DNA over microsecond timescales have been reported.

*Ab initio* calculations of biological molecules have been of considerable interest for some time due to their superior accuracy, but the extent to which quantum mechanical effects are involved in biological processes is only recently becoming clear. Single point calculations of biological molecules have continued to grow in scale enough that a DNA sequencing device may be characterised using commercial density functional theory packages [321] while path integral simulations of a protein have been performed over 30ps [322], the increased computational demands tightly constraining the scale of the calculations. But, with ingenuity, exciting biological insight may be gleaned from comparatively modest time scales.

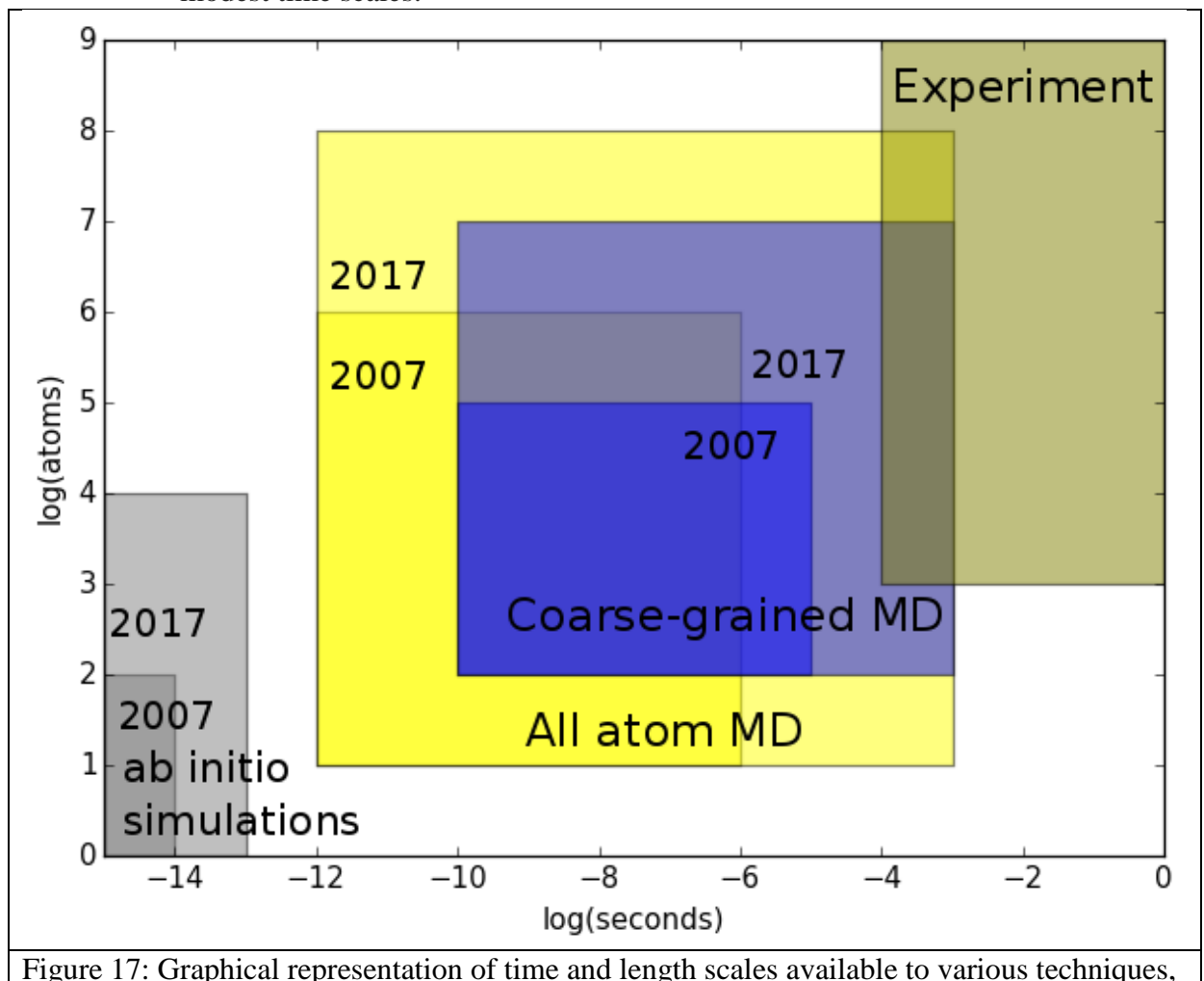

Figure 17: Graphical representation of time and length scales available to various techniques,

darker shades indicating the state of the field in 2007, lighter shades indicating the situation at the time of writing (in the year 2017). Grey: *ab initio* simulations; Yellow: atomistic molecular dynamics; Blue: coarse-grained DNA simulations; Olive: single-molecule experiments.

### 4.1.2 Atomistic molecular dynamics

For years the workhorse of computational biophysics, atomic scale molecular dynamics has enjoyed renewed success as computational power has increased. Rather than simulate individual biomolecules, it is possible now to explore complex systems of various interaction species, and to model full systems rather than choice subsystems. No longer limited to tens of nanoseconds, simulations of several hundred nanoseconds have become routine thanks to improved cluster computers, better exploitation of the power of general purpose graphics processing units (GPGPUs) to perform on-board rapid molecular simulations making use of the large number of computational cores, their specialised architecture, and more efficient computational techniques.

One of the most striking examples of this increase in scale across the board is recent investigations on an all-atom level of the full HIV capsid [323], a system composed of over 64 million atoms and several hundred component subsystems, which was simulated using the software package NAMD for 100 ns, providing a model which elucidates the details of subunit interactions in a way that experimental methods would struggle to do. Identification of the crucial sites in the capsid experimentally allows a greater focus when approaching simulations, and through targeting in this way it may be possible in future to use MD simulations as the basis of a bottom-up drug design methodology, so-called 'in silico drug design'.

Complex biological systems such as the virus capsid are the norm rather than the exception in nature, and the environment in which a biomolecule finds itself plays an active role in determining a molecule's structure and dynamics. The cellular cytoplasm is an archetypal demonstration of this. With thousands of molecular species present and competing for space and interacting with one another, the conditions in the living cell are strikingly different from those in many simulations, in which the number of simulated molecules is brought to a minimum for reasons of computational necessity. However, recent work has begun to explore the crowding effects proteins can have on one another within the living cell [324]. Focussing on proteins with high copy numbers, simulated cytoplasmic conditions have shown the effect that a high density of biomolecules may have on individual proteins' conformations and dynamical behaviour, and may lead to conformations unfamiliar from those found through the usual computational or experimental means.

How DNA acts under stress and torsion is of fundamental importance to the cell, with epigenetics studies suggesting that these responses may help to regulate gene expression and hence cellular function. Simulating these responses has historically been a tricky proposition - a long enough stretch of DNA is hard to simulate effectively, while applying forces consistent with those experienced due to biomolecules complicates simulation conditions substantially. One avenue which has been of interest to understand the mechanisms through which DNA relieves stress has been MD studies of the DNA minicircle. A DNA minicircle is like an enclosed circle of DNA, similar to a plasmid, but with far fewer nucleotide base pairs (only a few hundred), occurring naturally in mitochondria and so single-celled organisms such as trypanosomes. While the biological prevalence of the minicircle is disputed despite its presence in mitochondria organelles and trypanosome cells which cause sleeping sickness it does clearly offer a good test system to understand how an under- or over-twist of the DNA helix may be converted into tertiary structural parameters such as writhe, which has applications to the mechanics of DNA in the cell. Although the minicircles used are relatively large and the simulations necessarily span a relatively long time scale, the closed nature of the loop and the control which is available when introducing torsion to the DNA makes the minicircle a tractable means of DNA structure exploration when using modern machines and algorithms.

In order for biomolecules to interact and bind, they must often cross energy barriers. In the cell, the complex environment offers plenty of sources of the energy required to surmount barriers, often over timescales of milliseconds. Nevertheless, simulating this is outside the realm of possibility at present. Few simulations are performed with the crowding of the cellular environment, and the timescales required are too long to be accessible in most cases. It is therefore essential that the physicist find methods in order to reduce the energy barrier, the simulated time, or both.

In accelerated molecular dynamics (aMD), the energy landscape is altered as the simulation progresses such that the conformational space is fully explored over a short time. Changing the landscape irrevocably changes the physics of the simulation, and as a result it can be necessary to perform post hoc data manipulation. aMD offers access to effective milliseconds of simulation[325] over a much shorter simulation time. But as the authors note, the changed potential landscape makes time resolving events largely impossible, beyond indicating long- and short-timescale events. Meanwhile, temperature-accelerated molecular dynamics (TAMD) takes the opposite approach. Rather than modify the energy barriers to suit the energy available to the molecule, the temperature of the simulated molecule is increased so that it explores the potential energy landscape more quickly and easily, and is also capable of impressive predictive power at an all-atom level without biasing[326].

aMD and TAMD are both unbiased methods, in which the simulated molecules are free of any external forces or interference and can explore their conformational space as they please. For some processes, this is very effective, but not for all. In order to understand for example some unfolding or binding pathways, it is necessary to introduce an external force inducing the transition desired. This is known as steered molecular dynamics, named for the steering force which promotes a certain behaviour in the system. The steering force may be either a field which atoms move in (grid-steered MD), or a direct pulling force comparable to stretching studies done with AFM. This is a versatile technique which has been applied to problems as diverse as the effect of defects on DNA overstretching[327], motion of molecules through membranes[328], ssDNA self-assembly inside nanotubes[329], drug ligand binding affinities [330], and DNA repair[331]. Addition of this external force pushes the systems out of equilibrium, and for that reason if free energy calculations are desired, the usual thermodynamic integration should be replaced by a more complex approach based on Jarzynsky's equality[332].

Steered MD is not the only directed molecula dynamics technique. If the final structure is known, it can be used as a target for the system to evolve towards. Usually, a certain molecule or part of a molecule is selected to be targeted, and this subsystem experiences a harmonic force which pulls it towards the target geometry. This method is known unsurprisingly as targeted molecular dynamics, and can be used to study conformational transitions[333] as well as ligand binding pathways[334].

### 4.1.3  Coarse-grained Simulation

In recent years, there have been developed various highly successful approaches to the previously complex task of coarse-graining DNA such that long and large simulations may be performed. One prolific example of this is the free and open source package oxDNA. Developed at the University of Oxford, it set out to recreate physical parameters of DNA and RNA such as melting temperature and proclivity for single-stranded D/RNA to form helices, and represented each nucleic acid as just two particles – one acting as the backbone and the other as the base itself. Finding that the model could successfully model these behaviours of DNA, work was undertaken to put the DNA and RNA molecules in situations unfamiliar to the force field. It proved a success. Stretched DNA was observed with good agreement to experiment, and shone light on the rarely seen 'S-form' of DNA: as it could not be seen in the simulations, it is likely that smaller dynamics than those represented in oxDNA are responsible for the conformational change. The model proved to be so robust that DNA structure self-assembly can be seen, and the design of molecular machines like a DNA walker [335] or DNA tweezers has been accomplished, suggesting that in the future molecular machines for use in experiments may be designed in part through simulation – reducing time, cost, and complexity of the design process.

Despite being the most studied biomolecule, DNA is far from being the only one. Proteins and other structures within the cell mediate key processes and their behaviour is very often on too long a time scale for atomistic molecular dynamics to be practicable. Similarly, many processes happen on a relatively large time scale and in order to study them properly multiple simulation runs would be needed, once again taking atomistic simulation out of the question. One such process is the self-assembly of virus capsids, which vary in complexity greatly. Coarse-grained simulations studying this self-assembly have been performed which include key features of the process seen *in vitro*, and the behaviour may be related to the underlying physics of the potential energy landscape, a key feature in materials modelling [336].

Self-assembly as a theme is popular in coarse-grained computational biophysics, as it is an interesting phenomenon with applications ranging from nanotechnology to medicine. Important in cells is the self-assembly of the lipid bilayer around membrane proteins, which requires atomistic knowledge of the system and therefore spiralling computational cost as the complexity of the studied system increases. In 2008, a mechanism by which this could be coarse grained based on the protein structure was published [337]. The authors found that the self-assembled systems were in good agreement with experimental data, and the simulations were stable over a time scale of hundreds of nanoseconds, while the bilayer formed in tens of nanoseconds. Efficient coarse-grained approaches such as this will be of vital importance as larger and more complex systems are considered.

### 4.1.4 Quantum mechanics

Quantum mechanics principles underpin all physical processes in biological systems, but it is debatable whether one needs to address these principles in order to understand many biological processes. At one level one can argue that there are multiple 'trivial' quantum mechanical effects, such as the underlying physical principles behind chemical bonding. But, many of the 'spooky' quantum mechanical effects, such as quantum coherence effects over long length scales, might simply not be relevant to many biological processes which occur in relatively 'hot and wet' environments; the associated thermal energy scales are much larger than those involved involved in typical quantum mechanical energy transitions, and coupled to the effects of randomly moving water molecules bombarding biomolecules stochastically, quantum decoherence is, arguably, inevitable. Nevertheless, there is some evidence for quantum oscillations involved in photosynthesis, and for quantum tunnelling effects in electron transport proteins involved in oxidative phosphorylation.

Also, it has long been known that electron delocalisation has a great effect in real materials, and that fundamental physics such as the exclusion principle and electron band structures affect, for example, the means by which an electron may travel between pi orbitals on a DNA backbone. Few consider the quantum effects of the proton, a much less responsive and more massive particle. However, path integral molecular dynamics (PIMD) studies have been performed on biomolecules. In such studies, a proton is represented by a number of beads in a circle, joined by springs. By integrating modified equations of motion, an approximation to Feynman's path integrals may be recovered. An excellent example of the surprising behaviour which may be seen was published in work applying PIMD to an enzyme's binding site [322]. It was discovered that the delocalisation of a key proton drove vastly increased acidity at the active site, a finding which would not have been possible through any other means, and one which has implications for the workings of all biomolecular active sites.

More usual than PIMD simulations is the characterisation of ground electronic states for various biomolecular conformations. This type of work is now feasible to understand the energies and forces experienced by biomolecules as full *ab initio* detail is possible for systems of thousands of atoms, computationally expensive though it may be. One such study enabled calculation of the activation energy of an enzyme through description of the entire enzyme's electronic structure [338], made possible by the linearly scaling density functional theory implementation ONETEP. This kind of electronic structure characterisation may become a powerful tool in the future, as it may enable the properties and functions of biomolecules to be predicted and understood from the electronic distribution [339].

Entire useful systems which can be examined through fully quantum means also include those systems which may be the work of human endeavour. High throughput DNA sequencing is a highly valuable technology, both in the scientific and monetary sense, and design and characterisation of the probes which make it possible is a distinctly quantum mechanical task. Theoretical novel probe design has been highly popular in recent years [340–342], with the electronic properties of graphene and DNA used to demonstrate that the conductive behaviour would allow DNA sequencing though there are many engineering challenges. The scale of interest in graphene-based sequencing devices is such that after just six years of development, an in-depth review article was dedicated solely to them [343]. Here, graphene has been used primarily for its high structural rigidity which allow nanopores to be created across a single graphene sheet whose wall width can be as low as a single interplanar spacing of ~0.36nm, comparable to the stacking separation of a single base pair in a DNA double helix structure. In other words, there is the potential for single base pair precise sequence. The main issue with this approach is that graphene is also a very brittle material, which has thus far limited the manufacture of nanopores of consistent size.

Quantum mechanics, already underpinning the study of materials, seems poised to play a role in the exploration of biophysics, though at present there seems to be a gulf between experimentalists and theorists that needs to be bridged before significant progress can be made.

## 4.2 Determining the presence of single molecules

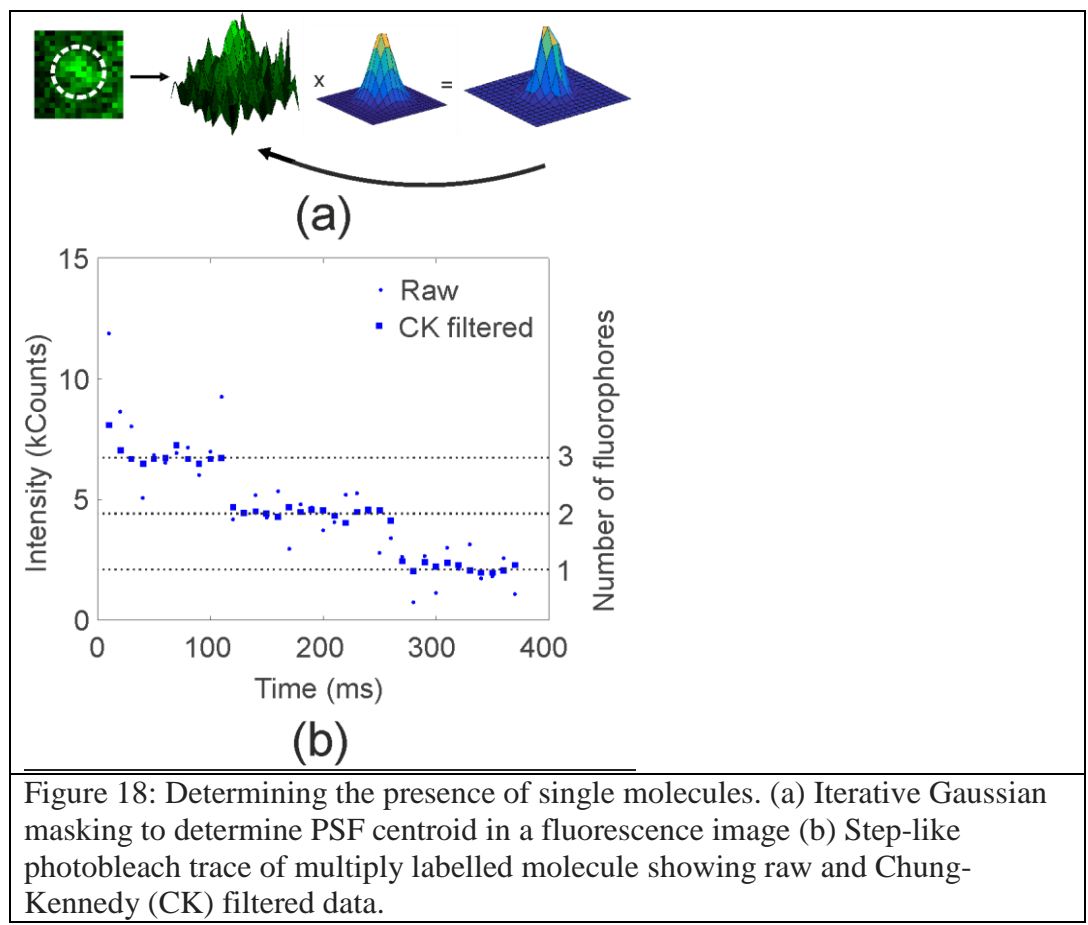

Figure 18: Determining the presence of single molecules. (a) Iterative Gaussian masking to determine PSF centroid in a fluorescence image (b) Step-like photobleach trace of multiply labelled molecule showing raw and Chung-Kennedy (CK) filtered data.

### 4.2.1 Tracking single molecules

Single-molecule emitters appear as diffraction limited PSFs. Often they are labelled with a single fluorophore and are very dim, particularly if a fluorescent protein is used. Thus, tracking at the single-molecule level is limited by noise in these instances, mainly shot noise in the detector but also from stray photons, not from the fluorescent emitter of interest. There can also be problems with the density of emitters; if PSFs are too close together this will adversely affect the spatial precision depending on the algorithm [344]. Single-molecule diffusion also deforms the PSF [345]. PSFs engineered for enabling the extraction of 3D information from a tracked particle require particularly good fitting algorithms to obtain accurate z information. Algorithms for particle tracking have been extensively reviewed recently previously [346] and quantitatively compared on simulated data [96].

Tracking a PSF is a statistical problem where a model for the PSF and noise is evaluated against the measured pixels at the detector, to determine the sub-pixel intensity centroid co-ordinates. A mathematically rigorous but relative simple to implement evaluation uses least squares, with the iterative Gaussian masking approach particularly fast (Figure 18 (a)) [60,347,348]. Models can also be evaluated using maximum likelihood estimation (MLE) [349] which is advantageous as it computes the likelihood of the observed data given the model and thus the precision and variance but is more computationally complex and requires a deterministic noise model. Other approaches do not require fitting such as pixel centroid determination (utilised by the popular QuickPALM software) [92], fast Fourier transform [350] or pixel triangulation [351]. Advantages of these single iteration approaches is that they can be implemented onto separate processors decoupled from a PC motherboard, e.g. on a camera CCD pixel array chipset itself, and so potentially are very fast and can operate in a high throughput multiplexed mode, but at the expense of sacrificing some level of localization precision and tracking flexibility.

These tracking techniques require sparse PSFs for precise localisations. Denser samples may require simultaneous localisation of multiple fluorophores. This has been implemented with MLE [352] and least squares - the DAOSTORM algorithm like many of these tracking techniques is borrowed from astrophysics [353]. The principle problem with these approaches is that models often tend to add more fluorophores than are present to reduce errors and so often semi-arbitrary thresholds must be set. Alternatively, one can use image estimation to try to determine the actual density of fluorophores in an image, usually by sub-arraying. There are several approaches [354–356], but the theoretical and practical limits have not yet been explored thoroughly. Once a track is determined there are several analytical tools available to investigate the mode of molecular diffusion, such as whether or not the behaviour is Brownian, anomalous (i.e. 'sub-diffusive'), directed or confined etc. [357], which calculate parameters such as the mean square displacement and compare this against probabilistic expectations from different diffusion models. These methods can be especially valuable in studying apparent changes of biomolecule localization dependent on different stages in the cell cycle [358] and complex molecular system involving tight-packing such as those in cell membranes which involve proteins that enable the transport of electrons [359–361].

4.2.2 <u>Counting single molecules</u>
Quantifying the number of molecules associated in a complex, the stoichiometry, is very valuable in helping us to understand biological function at a molecular scale. The main technique requires exciting all fluorophores in a molecular complex at once and imaging whilst simultaneously photobleaching. The intensity of complexes bleaches in a step-like fashion (figure 18 (b)). Quantifying the number of steps gives the stoichiometry, either by counting the actual number of steps [362] or, if many (typically more than six) fluorophores are present, by using the initial intensity [124,363,364]. Quantification using the decay rate is also possible [365]. These techniques have been used to determine the stoichiometry of the bacterial replication complex [366], a method which has also now been extended recently to yield also dynamic information regarding how the different components of the replication molecular machinery actually turnover with respect to time [367]. By extension to multi-colour microscopy, the structural maintenance of chromosome proteins used for remodelling DNA have also be investigated with these methods [123]. Recently these techniques have also been combined with image deconvolution to determine the total protein copy number of a transcription factor in yeast [368].

Other techniques have been developed for measuring stoichiometry, including utilising the information embodied in a brightness parameter in the autocorrelation fitting function of FCS measurements [369]. PALM can also generate stoichiometry estimates if molecules in a complex are sufficiently static over the image reconstruction time [370]. Care must be taken to account for over-counting the same molecule multiple times and under-counting by failing to detect molecules. The phenomenon of photon anti-bunching, using short pulsed excitation and the correlation function of emitted photons, can also generate stoichiometry estimates [371].

4.3 Artificial intelligence (AI)

For our discussion in the application of artificial intelligence (AI) in single-molecule biophysics, we focus on one approach of AI: using machine learning to perform tasks that are challenging for classical algorithms. Approaches include random decision forests [372], genetic algorithms [373], artificial neural networks [374], hidden Markov models [356,375,376], etc. Here in this review we put emphasis in particular on artificial neural networks, since these have achieved promising successes in recent years.

Artificial neural networks (ANNs), represented with the schematic diagrams in figure 19 (a), are algorithms inspired by the biological brain in information processing. ANNs are software-implemented neurons arranged in layers and each connected with neurons in adjacent layers, analogous to biological neurons which are connected via dendrites and axons to each other. The layers comprise input layer(s) that accept data and that pass them on to further hidden layer(s). The information is processed at each hidden layer before advancing to the next hidden layer. Finally, the computed result emerges from the output layer(s). Deep neural networks (DNNs) are the type of ANNs that have multiple hidden layers, more complex DNNs have a larger number of neurons per layer and more hidden/input/output layers which are needed to solve most real-life problems [377].

The connections between neurons are assigned dynamic weights. The values of weights reflect the extent to which the output of one neuron excites the next, with negative values corresponding to inhibitory effects on the downstream neurons rather than excitatory. The learning process in practice is the dynamic adjustment of these weights: the weights are modified with the addition of new learning materials to optimise performance and the final weight assignments after many training cycles optimise the network to satisfactorily perform the required tasks.

The neurons calculate a weighted sum of all incoming signals and run it through an activation function $K(x)$ as well as a bias function and fire a signal if the result surpasses a pre-set threshold. The purpose of the activation function is to rectify the output value by scaling and shifting it, i.e. an extra degree of freedom of the training parameters. Backpropagation is the modification method whereby the intended outcome is compared with the obtained outcome and the difference is used to evaluate the extent of modification of weights. This process starts at the output layer and computes backwards through the network. Figure 19 (b) shows the action of a neuron and (c) shows three examples of $K(x)$.

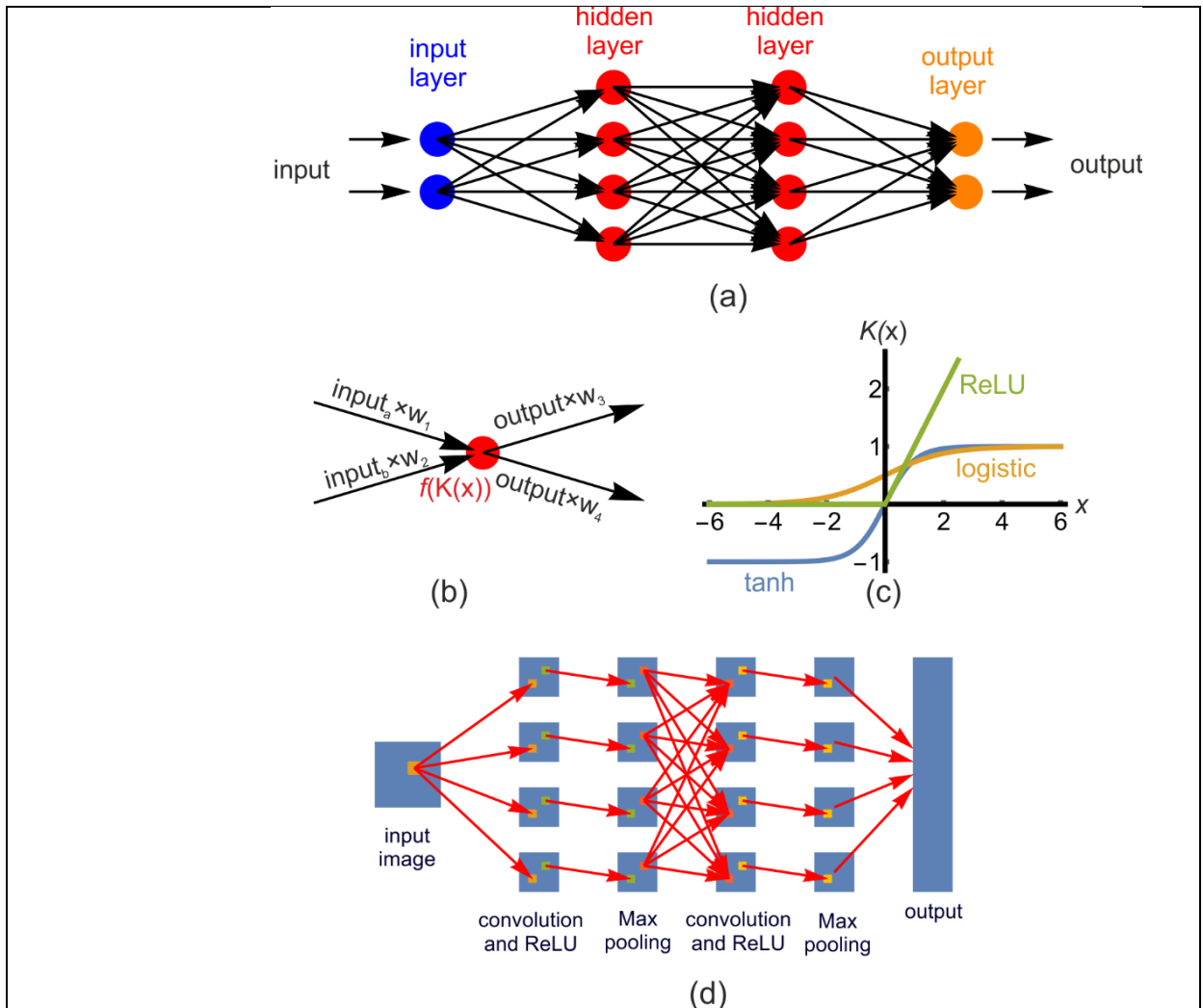

Figure 19. Schematic diagrams representing artificial neural networks. (a) A DNN with four layers of neurons. The disks represent neurons and the arrows represent connections and forward propagation. Input data enter the input neurons and proceed to deeper layers ('hidden layers') before being fed to and exiting the output nodes. (b) A node, or neuron (red disk), with two of its inputs and two of its outputs. The $w_i$ are weights. The neuron action is a summation of the weight inputs, followed by an activation function K, and finally multiplied by a bias (bias not shown in diagram). (c) Three examples of activation functions – tanh and logistic functions and rectified linear unit (ReLU). The ReLU activation function is simply $K(x) = \max(0, x)$. (d) Convolutional neural networks (ConvNets) showing convolution and pooling hidden layers.

As pointed out previously, the key advantage of DNNs-based algorithms is their ability to learn, or equivalently, the lack of the need to be explicitly programmed. Raw input data are supplied to the input layer of neurons. The next step, multiplying the input layer values with a weight, is a selection process to filter out data that are key to the task: after training, the weight assignments over the connections are such that data most relevant will have high weights and thus high influence over later layers of neurons and those with less relevance will have low weights. Using tagged data to train deep forward networks with backpropagation is called supervised learning [378]. A more automated learning is unsupervised learning [379,380] where only raw data are supplied and features are automatically detected. Only a small number of labelled inputs are then needed to introduce tags.

For image analysis, potentially relevant in the future to noisy single-molecule level data, convolutional neural networks (ConvNets) [381,382] have proved to be remarkably accurate and efficient. Instead of every neuron connected to all neurons in adjacent layers, ConvNets are only partially connected to boost performance thanks to the assumption that the input data are arrays (e.g. an image is a 2D array of pixels) and have certain properties. The hidden layers compose of two types: convolution and pooling layers. The former search for features at multiple sub-regions in the earlier layer and the latter pool the features together to make sense of the data. Like general DNNs, ConvNets layers increase in abstraction as the computation progresses into deeper layers. Microsoft used ConvNets with 30 layers to beat humans in image object recognition from the ImageNet 2012 classification dataset [383,384]. Figure 19 (d) schematically illustrates the architecture of ConvNets. The sub-regions of an input image are passed to neurons that are organised in feature maps. Pixel values are weighted via a filter bank, which is shared by all neurons in the same feature map. The next layer is a pooling layer where conjunctions of features are pooled together semantically. Pooling layers can reduce the dimensionality of the representation as input patches into adjacent neurons in the pooling layer are separated. The effectiveness of ConvNets is partially due to its utilisation of properties of image data, such as the fact that local pixels can be highly correlated. More detailed descriptions can be found in two recent 'deep learning' reviews [377,385].

### 4.3.1 Computer vision

In computer vision, deep learning hugely simplifies the image analysis process, potentially relevant in the future to single-molecule light microscopy methods, since it self-determines the representations needed for the task at hand [386], be it object identification, counting, localisation, etc., so that human programming for performing the task is neither detailed nor explicit. Traditional image processing algorithms perform a series of human-defined low-level tasks such as brightness thresholding or length matching that is both tedious and lacking of high-level understanding. Thus, AI is capable of much more general-purpose analysis while human input during training is as simple as supplying images with tagged objects. This is akin to acquainting young children with objects – there is no need to point out that a banana is long and curved with a yellow skin. Simply showing a banana and telling the child it is a banana suffices. Another powerful use of AI is the analysis of objects with complex and numerous features that cannot even be exhaustively listed by a human programmer, let along the subjectivity that humans introduce in representation selection.

Lempitsky et al. wrote a ConvNets based algorithm [387,388] that interactively counts the number of objects in an image, which similarly could have future utility for single-molecule image analysis. The human 'teaches' the package what counts as signal by clicking on several. The algorithm can then count objects in the whole frame, including ones that are in contact of or partially overlapping each other. Minimum effort from the user is required – there is no need to specify features such as intensity levels or signal sizes.

Another area of application is optical tomography. It reconstructs 3D volume images from 2D phase or intensity images at planes of different depths. With the application of tomography to increasingly complex systems, it becomes harder to find analytical solutions for image reconstruction. Multiple scattering of the emission light through inhomogeneous media only exacerbates the analysis problem. Kamilov et al. [389] demonstrated the reconstruction of phase voxels using an ANN learning model that avoids the need for explicit analytical reconstructions and intensive modelling of scattering, such as coupled dipole approximation. After training, the model not only reconstructs the 3D image but also maps the 3D index distribution of the medium.

Non-ANN based learning algorithms are less general in the sense that features need to be supplied to initialise the training. Nevertheless, it is still much easier to use in that the quantification of features are automatically explored. Wu and Rifkin [390,391] developed a machine learning approach to accurately distinguish single-molecule fluorescent spots from noise speckles. Their software, called Aro, uses supervised random forest classifiers to assess several features of local intensity maxima to estimate the probability of the spot being signal.

Here, no thresholding or any other whole-image manoeuvring is done. The key innovation was their use of statistics to enable the estimate of the error of image classification as well as the quality of the data. The user has the option to train Aro by labelling select spots as signal, noise or unsure. The results are compared to manual counting to yield a r-squared values of greater than 99%, on par with established non-learning approach FISH-Quant [392]. In contrast, threshold-picking method yields only 54%. Borgmann et al. [393] used classification algorithms including random forests, support vector machines, genetic programming, etc. to detect the Rhesus D type DEL phenotype, which is expressed in very low levels, among other more highly expressed phenotypes. The fluorescent antibodies, optics design and imaging are all standard. The resulting differentiation between phenotypes are largely unambiguous due to the high sensitivity of the algorithm to weak signals, which cannot be achieved with the previous practiced method of adsorption-elution.

In video-rate or faster fluorescence detection, where noisy background and low contrast pose difficulty in fluorophore detecting and tracking with traditional methods, such as live-cell movies, AI learning methods are also starting to show promise. Jiang et al. [394] used a Haar training classifier to detect and track particles and achieved a true positive rate of 98% for TIRF and 99% for epifluorescence of yellow fluorescent protein YFP engineered into African green monkey kidney cells.

4.3.2 Big data

Fast digital cameras, motorised and automated microscopes and multi-colour fluorophores in fluorescence imaging increasingly create large quantities of data generated in short spans of time and can make it impractical for humans to analyse or even for hard drives to store. AI can be trained to assess the quality of the data and discard low-value ones at or close to the production stage. Experiments limited by the rate at which humans visually examine the quality of data can now be conducted more efficiently.

Valentine and Woodhouse [395] trained ANNs to recognise high-quality discrete time series data with both high- and low-quality tagged data. No *a priori* assumptions were made about the data so although their focus was on seismic tomography, the method can be generalised. Huang and Murphy [396] wrote AdaBoost that can even generate classifiers of either neural network or decision tree types during training to localise fluorescently labelled protein and DNA *in vivo*.

4.3.3 Instrument design

Nanophotonics devices are structures designed to manipulate light to retrieve the high spatial resolution information of molecules beyond diffraction limit. The design of such devices is hindered by the tedious modelling of optical response given the structure, as well as the often near-impossible computation of a structure which has the desired optical response. This is precisely the class of problems where DNNs have decisive advantages both in the quality of results and in the avoidance of the direct understanding of the problem. Malkiel et al. [397] demonstrated the prediction of the nanophotonics structures only by their far-field response. They also showed the design of structures with DNNs given on-demand design of optical responses.

Genetic algorithms (GAs) are suited to instrument design where optimisation incorporates a large amount of parameter space, and is limited by case-specific constraints. The user only specifies the functionality and constraints while the design algorithm explores the whole parameter landscape in an evolution-style process to create the optimal design: these evolutionary approaches can appear at first glance indirect and over-engineered, nevertheless, they can perform complex, multi-dimensional calculations with high efficiency. Tehrani et al. [181] used GAs to design adaptive optics that accommodate aberrations from the inhomogeneity in the refractive index in a spatially extended sample. The large fluctuations in photon emission mean that traditional methods to optimise wave fronts are not feasible. Whole optical system design has also been optimised with GAs [398].

### 4.3.4 Molecular property prediction

Prediction of protein activity from structure or nucleic acid properties is another area where learning algorithms start to routinely outperform existing methods of choice. Quantitative structure–activity relationship (QSAR) predictions in the pharmaceutical industry compute the activity of on-target and off-target drugs. Ma et al. [399] showed that DNNs outperforms more traditional learning methods including random forest and support vector machine (SVM), etc. in making predictions on a set of large diverse QSAR data sets while reducing the computation time. QSARs are used in a range of engineering and scientific disciplines in the chemical and biological sciences as part of classification systems. A comprehensive review of deep learning in molecular behaviour prediction in drug discovery can be found in [400].

In protein binding, DeepBind [401] is cable of predicting the sequence specificities of nucleic acid binding proteins. Again, this is a deep learning model that achieves more accurate prediction than state-of-the-art methods. DeepMind can even train on *in vitro* data and test on *in vivo* data and still beat other programmes.

## 5 Future outlook and challenges

The exciting recent developments in single-molecule biophysics techniques have resulted in a suite of intriguing future challenges. Many challenges are being actively tackled right now. We discuss some of these future challenges, and their implications, in this section below.

### 5.1 Correlative single-molecule techniques

Much new insight may come from combining existing single-molecule techniques in the same investigations. Here we review recent progress is such correlative methods relevant to single-molecule biophysics.

#### 5.1.1 AFM/light microscopy

Atomic force microscopy (AFM) and fluorescence imaging have been combined for decades, but the advancement to AFM with single-molecule fluorescence detection is very recent. There are three main technical challenges to achieving this combined technique; the first of these is the limited space around an optical microscope in which to integrate an AFM, which requires a highly stable platform to achieve high resolution [402]. For thin, transparent specimens AFM can be integrated from above with fluorescence imaging from below (see figure 20 (a)). But, for thicker specimens there can be uncertainty as to whether the same molecule is being imaged in both modalities [403]. The second challenge is that AFM, in general, requires a high density of target molecules, due to the difficulty in directing the AFM tip detecting individual single molecules: this in direct contrast to single-molecule fluorescence microscopy in which fluorophores are ideally separated by more than the diffraction limit to be localised efficiently by tracking algorithms [403]. The third challenge is that stray light used for measuring the deflection of the AFM cantilever can interfere with the fluorescence detection, contributing to the background intensity detected.

The technique has been used on many samples, many using the TIRF geometry for single-molecule fluorescence microscopy [404], although this is limited due to the finite depth of field of fluorescence excitation that can be achieved (see for example [405]). Correlative AFM and super-resolution single-molecule fluorescence microscopy on structural elements in live mammalian cells has now been shown, but the two techniques are demonstrated sequentially; AFM first, then a buffer exchange, then super-resolution imaging [406]. AFM-FRET microscopy (see figure 20 (a), inset) allows mechanical changes to be monitored over time simultaneously by both techniques, for example the force required to enact a conformational change can be measured and corresponds to a loss of acceptor signal [403]. An alternative to labelling the protein being manipulated by AFM is to perform an AFM study on an enzyme that produces a fluorescent cleavage product, and monitor the level of fluorescence to measure the catalytic activity [407]. There are also powerful combinations between force spectroscopy with optical tweezers and fluorescence [408].

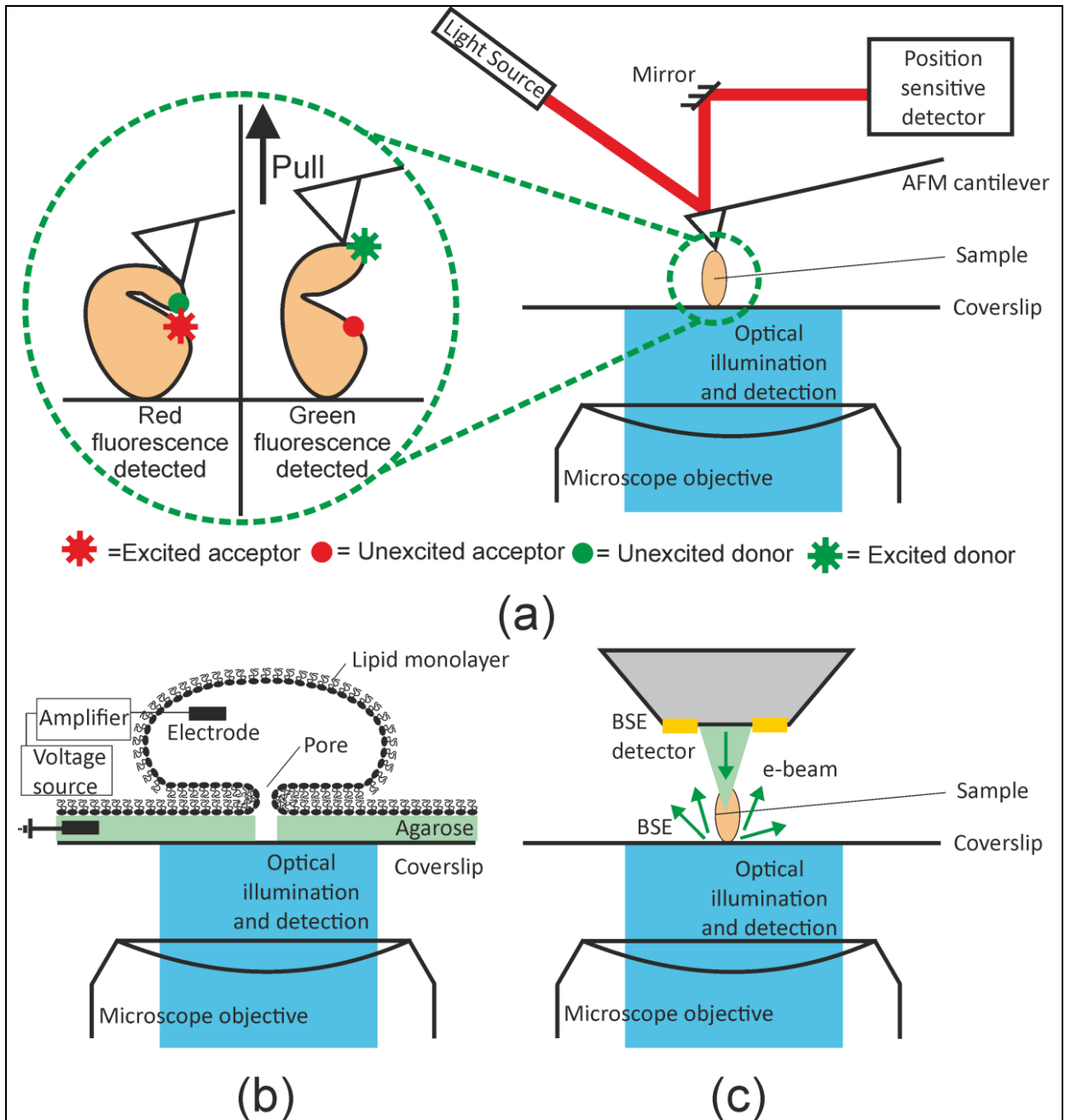

Fig 20: Correlative imaging techniques (A) AFM-FRET microscopy. Single-molecule fluorescence microscopy is performed in inverted mode, enabling top down AFM. As the target molecule is extended the FRET signal changes from the longer wavelength emission of the acceptor to the shorter wavelength emission of the acceptor molecule. (B) Cartoon of a hypothetical combined patch clamp and fluorescence experiment using a droplet interface bilayer. The droplet contains calcium and calcium sensitive fluorescent dye. The electrodes can measure voltage across the lipid bilayer area, all of which is illuminated with the laser. The laser illumination can be switched to TIRF microscopy to increase image contrast. (C) Cartoon of CLEM microscopy. Back scattered electrons (BSE) are detected to perform electron microscopy, and fluorescence microscopy is performed in inverted mode.

### 5.1.2 Patch clamp/single molecule imaging

One of the earliest correlative techniques, combined fluorescence microscopy and patch clamp experiments (figure 20 (b)), have been performed since the late 1990s [409–411], however early experiments struggled due to the mismatch in temporal resolution available with electrical and optical recording. With the increase in speed and sensitivity of emCCD cameras, coupled with other technological developments, the technique is undergoing a *renaissance*. The patch clamp can be combined with different single molecule fluorescence techniques such as FRET or TIRF microscopy to study different membrane transport processes.

The patch clamp combined with FRET imaging is well suited to studying molecular conformational changes of ion channels following ligand binding, combined with electrophysiological measurements. By placing the FRET donor and acceptor on the channel subunits, high FRET signal occurs when the channel is closed. States with low FRET signal but no current flow identify intermediate states between the channel opening and closing [412,413], supporting previous work with only electrophysiological measurements [414].

The early combined patch clamp and fluorescence techniques offered spatial information with single pore resolution, but until recently were almost exclusively applied to study calcium channels via fluorescence detection of calcium using TIRF microscopy on large *Xenopus* oocytes (egg cells from a large toad which is often used as a model organism). This was due to the requirement of large vesicles to apply a patch clamp, and the availability of calcium sensitive fluorescent dyes [409,415] compared to the lack of sensitive reporters for other physiologically relevant ions such as potassium. Calcium sensitive dyes such as calcium green-1 dextran and fluo-8 AM increase fluorescence emission flux on binding to calcium on the same time scale that calcium concentration changes due to gating events [416], enabling *in vivo* imaging of neuronal networks [417]. Recently, potassium channels have been studied *in vitro* using the dye Asante Potassium Green 4, although the dye's response limits the utility of these measurements – it takes on the order of seconds for dye to reach maximum signal after a millisecond time scale gating event [418].

Advances in the ability to make stable lipid bilayers [419,420] and droplet interface bilayers [421], coupled with improvements in microfabrication, are enabling the study of pore formation that does not involve channel proteins [422]. Electroporation is a process used extensively to transform DNA into cells [423], and also used for applications such as transdermal drug delivery [424]. Electroporation has been the subject of several theoretical studies [425,426], but these are largely unconfirmed by experimental evidence [422]. A recent study [422] uses a geometry similar to that shown in figure 20 (b), but with a microfabricated substrate which creates multiple droplets on one chip, enabling higher throughput. The researchers correlate the electrical data over an area with the fluorescence imaging data from each pore in the area, identifying the contributions of each pore.

### 5.1.3 Light and electron microscopy

Electron microscopy (EM) has advantages over optical microscopy. It has better resolution but lacks the specificity of fluorescence microscopy. Correlated light and electron microscopy (CLEM, figure 20 (c)) [427,428] can, in theory, achieve the best of both worlds but there are many challenges. Typically EM samples are chemically fixed, stained and embedded in plastic/wax for sectioning. Special procedures must be followed to preserve fluorescence of normal probes. Alternatively, cryo-EM can be used to flash freeze the sample and preserve fluorescence. Samples can be imaged separately on EM and optical microscopes using finder grids or fiducial markers. Recently several integrated CLEM instruments have become available which either move the sample between optical and electron imaging modes, preserving registration, or contain paraxial optical and electron imaging.

Scanning electron microscopy (SEM), which detects back scattered electrons, has been combined with STORM in a paraxial geometry to image antibody labelled NUP proteins in nuclear pore complexes. [429] SEM has also been combined with PALM and STED but using fiduciary markers and imaged separately to observe fluorescent protein fusions to histones, mitochondrial protein and presynaptic dense protein in the model flatworm *Caenorhabditis elegans* [430], Transmission electron microscopy (TEM) has also been combined with fluorescence using PALM to image the protein clatherin in a dried and stained cell membrane, again using fiducial markers to image separately [431]. A similar approach tagged mitochondria with the mEos2 protein and imaged using 3D interference PALM. The whole sample, including glass coverslip and slide was then sectioned using a focused ion beam and the sections imaged with TEM using fiducial markers to correlate [432].

## 5.2 Some specific challenges which are likely to emerge in single-molecule biophysics

Here we outline a few specific challenges relating to future single-molecule biophysics techniques, based on interpolating our current knowledge.

### 5.2.1 Challenges in statistical methods

Single-molecule methods that examine biological systems and processes one molecule at a time are inherently low throughput, and examining enough molecules to reach a statistically significant result can be extremely time consuming. There are several emerging strategies to increase molecular throughput, either by increasing the number of molecules that are observed at one time, or by increasing efficiency of observing molecules sequentially.

The drive to higher throughput super-resolution fluorescence imaging is in many cases being enabled by new microfluidics systems. Microfluidics systems allow the manipulation of fluid flow to move single particles, or change the conditions within a sample chamber. For example, in DNA curtains [433] [434] liquid flow over micro- or nano-fabricated substrates is used to produce multiple aligned DNA strands that can be used to study the interaction of proteins with DNA, with multiple proteins being recorded in a single field of view. Alternatively narrow channels can be used, with the single molecules to be studied moved through sequentially, allowing them to be fluorescently imaged [435] or manipulated (for example with optical tweezers), with less time between repeat experiments than in earlier experiments where the researcher had to find individual particles or wait for them to diffuse into view. Microfluidics can also be used to change the conditions around a cell and study the single-molecule responses within it [436,437].

Optical tweezers experiments can be parallelised using several different methods: acousto-optical deflectors (AODs), scanning mirrors and interference can be used to produce multiple traps via time-sharing of the laser beam. Holographic methods use spatial light modulators to create multiple optical traps in the same sample chamber, and can be used to create higher numbers of parallel traps [438,439] than time-sharing methods. Multiple beams can also be used to decrease acquisition times in super-resolution imaging, for example with the use of multiple STED 'donuts' to acquire images faster [99].

It is clear that high throughput methods are important for taking single molecule techniques to real world problems, such as medical diagnosis, but not all the techniques are currently available in high throughput versions.

5.2.2 Checking our assumptions
The first single-molecule experiments required a number of simplifying assumptions to evaluate parameters. As the complexity of experimental design increases we move further from the realm in which those assumptions are correct, and we must be mindful of the assumptions on which analysis techniques are based.

In most applications of fluorescence microscopy it is a basic assumption that the fluorescence dipole is free to rotate and produces a symmetric image during the time for a single exposure. But, when a fluorophore is bound to a molecule of interest it can become highly orientated, for example when bound to immobilised DNA, producing a non-symmetric image and resulting in an incorrect localisation [440]. This is a concern particularly when localisation precisions <10 nm are found and should be accounted for in experiments where the rotation of the fluorescence dipole is constrained.

Another example is found in many early examples of STORM analysis software, which are still in common usage today. These programs assume that each bright spot is a single fluorophore. As technological advances have driven higher-throughput STORM imaging it is now often the case that higher numbers of molecules are on in one diffraction-limited volume, and care must be taken to ensure the software used can recognise this to avoid mis-localisation.

These problems are not limited to light microscopy: in electron microscopy alignment and clustering algorithms are sometimes found to produce outcomes that strongly resemble initial guesses. Care must be taken to compare results to initial parameterisations, or new methods that check for this must be used [441].

The huge expansion in all areas of single-molecule biophysics has allowed the complexity of experiments performed to increase dramatically in the last ten years, but amongst this progress it is important to continue to check the underlying assumptions of our methods as we move to higher temporal resolution and systems with increased dynamic complexity.

5.2.3 Management of 'Big Data'
Approaches to manage big image data involve every aspect of data analysis. Good practice to keep compatibility between metadata, file formats, and open data interfaces is encouraged [442]. Cloud storage and processing facilities solve storing, sharing and intense processing problems that are too challenging for local machines [443]. Automated data selection and analysis tools [444,445] are increasingly indispensable to keep pace with data generation.

5.2.4. Use by non-specialists
The last decade has seen a huge growth in the number of available single-molecule techniques, and much effort has been expended to develop protocols for making samples, characterise the machines and their limits and develop analysis software. Much of this work has been carried out on test sample systems that are already well characterised – microtubules (figure 21) are a particularly notable example, with their regular structure of α and β-tubulin dimers.

The development of super-resolution techniques to increase the range and complexity of samples they are able to look at means that the tools are better placed than ever before to tackle difficult and interesting biological questions. Nevertheless, experimental setups capable of highly complex experiments involving multiple techniques become difficult to operate, such that as we move to techniques more capable of answering biological questions, we move further from experimental design that a non-specialist in the technology could operate. There has been a large effort in this area for light sheet microscopy with the OpenSPIM project [446,447], which has resulted in a high uptake of the technique amongst biologists. Also, there are now a range of commercial systems that can perform, for instance, STORM, but they are not in general as simple to customise as bespoke systems.

Coupled with the increased use of single-molecule technology by non-specialists there are now many examples of data analysis software, particularly for super-resolution microscopy, such as ThunderSTORM [94], which are user-friendly and designed to simplify analysis for non-specialists. Whilst these software often come with comprehensive manuals it is imperative that those using them understand the physical principles on which the technique works to avoid artefacts in images. Reviews of techniques aimed at non-specialists [448] aim to bridge this gap, but greater communication between developers and users is required in this regard.

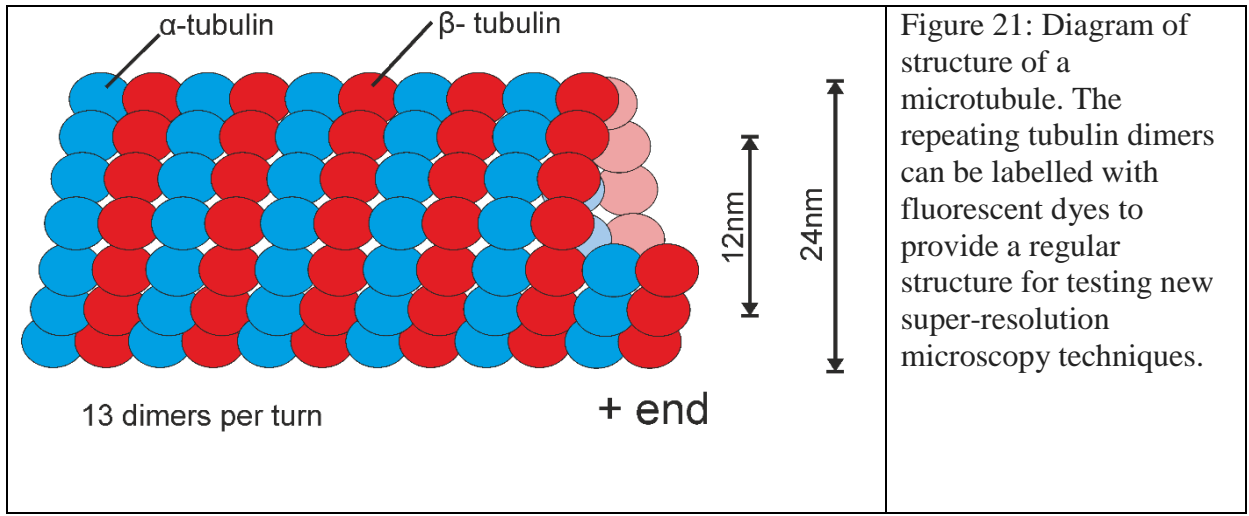

Figure 21: Diagram of structure of a microtubule. The repeating tubulin dimers can be labelled with fluorescent dyes to provide a regular structure for testing new super-resolution microscopy techniques.

### 5.3 Probing single molecules in populations of cells and tissues

For *in vivo* imaging, a range of issues need be considered: biocompatibility of fluorophores [449], the number of emitted photons and lifetime of fluorophores in a cellular environment [450], and noise due to scattering of inhomogeneous cellular content. Techniques such as *in vivo* FRET [451] and fluorescence imaging with near infrared light [452] have been developed to tackle these issues. Now proteins can be imaged in cells with nm level resolution [29]. Adaptive optics have also proved to be particularly valuable in correcting for the inhomogeneity in refractive index in a deep tissue sample. Transverse illumination methods such as selective plane illumination microscopy (SPIM) also known as 'light sheet' can result in the detection of less out of focal plane scattered light from deep tissue samples, to enable single-molecule detection in multicellular tissues up to a few hundred microns deep (i.e. a few tenths of a mm).

### 5.4 Personalized medicine

Personalized medicine is a medical model which caters healthcare specifically to an individual patient, as opposed to having to rely on generic treatments relevant to population level data, and so potentially targets more effective diagnosis and treatment with fewer side effects. The relevance to single-molecule biophysics lies in developments in miniaturized biosensing devices to enable smart diagnostics of health disorders such as LOC technologies discussed previously in this review, but also in the application of targeted treatment and cell delivery tool such as those of 'nanomedicine'.

Developments in microfluidics, surface chemistry tools, nanophotonics, and bioelectronics have all facilitated miniaturization of biosensing devices, designed to detect specific features in single biomolecules of biological samples. For example, the presence of particular types of cells through the detection of surface receptor complexes. Typically, these tools consist of a silicon based substrate which acts as a microscopic flow cell for detecting specific biomolecules in a sample, with synthetic arrangements of biological material bound to surfaces inside the flow cell, in a complex arrangement which often employs microfluidics to convey dissolved sample material, such as from blood, urine, and sputum etc., to detection zones inside the device. For detection of specific 'bio-markers' (labels specific to certain biomolecules types), a molecular 'surface pull-down' approach is often used: the surface of a detection zone is coated with a chemical reagent that binds specifically to one or more bio-markers.

Once immobilized the pulled down molecule can then be detected by a range of biophysical measurements. Fluorescence detection can be applied if the bio-marker can be fluorescently labelled, and to achieve fluorescence excitation these devices can utilize photonics properties of the silicon-based flow cell. Photonic waveguiding for example can enable excitation light to be guided to the detection region of the device, with photonic bandgap filtering used to separate fluorescence excitation from emission wavelengths. Microfabricated photonic surface geometries can generate evanescent excitation fields to increase the detection signal-to-noise- ratio by minimizing signal detection from unbound biomolecules.

Label-free detection LOC biosensors are also emerging, including interferometry surface plasmon resonance (SPR) methods, Raman spectroscopy, electrical impedance and ultrasensitive microscale quartz crystal microbalance (QCM) detectors that operate through detecting small changes in resonance frequency due to the surface binding of biomolecules. Also, microcantilevers similar to those in AFM imaging can be used for biomolecule detection, involving chemical functionalization of the cantilever using e.g. a specific antibody, which results in binding of a specific biomolecules as a sample solution is flowed across, detected again as changes to the resonance frequency.

Significant research developments have been made recently in the ability to efficiently and cheaply sequence single molecules of DNA. A promising new type of sequencing technology uses ion conductance measurements through engineered nanopores, either solid-state or manufactured from protein adapters. An applied electric field drives DNA translocation through a nanopore which then blocks off some of the ion flux as it passes through, but the extent to which this occurs depends on the specific nucleotide base pairs translocating through the pore due to size and shape differences, and thus the drop in detected ionic current is a molecular signature for the DNA sequence.

The use of nanomedicine is already emerging at the level of targeted drug binding. For example, pharmaceutical treatments which destroy specific cells such as those of cancers, such as radioactive nanoparticles coated with specific single-molecule antibody probes. 'Aptamers' (synthetic recognition molecules composed either of nucleic acids or more rarely peptides) have an important role here in being similar in evoking a minimal immune response compared to antibodies, thus fewer side-effects. Targeted binding can also be valuable for the visualization of disease in tissue e.g. antibody-tagged quantum dots can specifically bind to cancer tumours and assist in diagnosis.

Targeted drug delivery is also an important area of emerging development. These tools increase the specificity and efficacy of drugs actually being internalized by the cells in which they are designed to act, such as using delivery of certain drugs on the normal process of endocytosis by which many cells internalize biomolecules. Also, development of DNA origami devices involving synthetic 3D nanostructures made from DNA to acts as 'molecular cages' to permit the delivery of a variety of drugs deep into a cell while protecting the molecular cargo from cellular degradation processes before it can be released.

One of the most promising areas of nanomedicine research involves methods to facilitate tissue regeneration, achieved through 'biomimetic' materials. These can serve as replacements for damaged/diseased tissues and/or act as a growth template to permit stem cells to assemble in highly specific regions of space to facilitate generation of new tissues. Tissue replacement materials focus on mimicking the structural properties of healthy tissue, e.g. bone and teeth but also softer structural material such as collagen. Inorganic biomimetics have focused on using materials that can be synthesized in aqueous environments under physiological conditions that exhibit chemical and structural stability, particularly noble metals as well as metal oxide semiconductors and also chemically inert plastics which benefit from having a relatively low frictional drag while being relatively nonimmunogenic. Ceramics can also be used, and these inorganic surfaces are often pre-coated with short sequence peptides to encourage binding of cells from surrounding tissue. Much of the characterization of these regenerative medicine techniques is still at a bulk ensemble level in regards to biophysics, nevertheless, increasing use of single-molecule methods of electron microscopy and super-resolution microscopy are being used as quality control techniques for these materials, also coupled to methods using X-ray spectroscopy analysis,

6 Conclusions

In conclusion it is clear that techniques in single-molecule biophysics have outgrown the constraints of the pioneering techniques of structural biology and physiology. There is a distinct trajectory in moving towards methods which can render single-molecule precise information but still retain the functionality of biological processes under study. The future challenges stem as much from engineering as they do from human creativity. There is, for example, so much that these emerging techniques can *do*, but the key perhaps is to focus on really translating these into techniques that can *do good*.

Acknowledgements

Supported by the Biological Physical Sciences Institute (HM, ZZ), MRC (ML) (grant MR/K01580X/1), BBSRC (ML, AW) (grant BB/N006453/1), and EPSRC DTA PhD studentship (JS).

List of acronyms

| | |
|---|---|
| ABEL | Anti-Brownian electrokinetic (as in ABEL trap) |
| ADP | Adenosine diphosphate |
| AFM | Atomic force microscopy |
| AI | Artificial intelligence |
| ALEX | Alternating laser excitation |
| ANN | Artificial neural networks |
| AOD | Acousto-optical deflector |
| AT | Acoustic tweezers |
| ATP | Adenosine triphosphate |
| BALM | Binding activated localisation microscopy |
| BaLM | Bleaching/binding assisted localisation microscopy |
| BFM | Back focal plane |
| CLEM | Correlated light and electron microscopy |
| ConvNets | Convolutional neural networks |
| DNA | Deoxyribonucleic acid |
| DNN | Deep neural networks |
| EM | Electron microscopy |
| emCCD | Electron multiplying charge-coupled device |
| FCS | Fluorescence correlation spectroscopy |
| FLIP | Fluorescence loss in photobleaching |
| FLIPPER | Fluorescent indicator and peroxidase with precipitation for EM resolution |
| FND-Au | Fluorescent nanodiamond gold nanoparticle |
| FRAP | Fluorescence recovery after photobleaching |
| FRET | Förster resonance energy transfer |
| GA | Genetic algorithms |
| GPU | Graphics processing units |
| HOT | Holographic optical tweezers |
| iSCAT | Interferometric scattering microscopy |
| LC | Liquid crystal |
| LCOS-SLM | Liquid crystal-on-silicon spatial light modulator |
| LCP | Lipidic cubic phase |
| LED | Light emitting diode |
| LG | Laguerre-Gaussian |
| LOC | Lab-on-a-chip |
| MD | Molecular dynamics |
| MLE | Maximum likelihood estimation |
| MT | Magnetic tweezers |
| NA | Numerical aperture |
| NIR | Near infrared |
| NSOM | Near-field scanning microscopy |
| OT | Optical tweezers |
| OPF | Optical pulling force |
| PALM | Photoactivated localisation microscopy |
| PCR | Polymerase chain reaction |
| PDMS | Polymethylsiloxane |
| PIMD | Path integral molecular dynamics |
| PSF | Point spread function |
| QCM | Quartz crystal microbalance |
| QPD | Quadrant photodiode |
| QSAR | Quantitative structure–activity relationship |
| ReLU | Rectified linear unit |
| RESOLFT | reversible saturable optical linear fluorescene transitions |
| RNA | Ribonucleic acid |
| sCMOS | Scientific complementary metal oxide semiconductor |
| SEM | Scanning electron microscopy |
| SERS | Surface enhanced Raman spectroscopy |
| SFX | Serial femtosecond crystallography |
| SICM | Scanning ion conductance microscopy |
| SIM | Structured illumination microscopy |
| SLM | spatial light modulator |
| smFRET | single molecule FRET |
| SPAD | Single photon avalanche photodiode |
| SPR | Surface plasmon resonance |
| SSIM | Saturated structured illumination microscopy |
| STED | Stimulated emission depletion |
| STORM | Stochastic optical reconstruction microscopy |
| SVM | Support vector machine |
| TEM | Transmission electron microscopy |
| TERS | Tip enhanced Raman spectroscopy |
| TIRF | Total internal reflection fluorescence |
| UEM | Ultrafast electron microscopy |
| XFEL | X-ray free electron lasers |

References

[1]   Leake MC. The physics of life: one molecule at a time. Philos Trans R Soc Lond B Biol Sci 2013;368:20120248. doi:10.1098/rstb.2012.0248.

[2]   Cordes T, Moerner W, Orrit M, Sekatskii S, Faez S, Borri P, et al. Plasmonics, Tracking and Manipulating, and Living Cells: General discussion. Faraday Discuss 2015;184. doi:10.1039/c5fd90093j.

[3]   Harriman OLJ, Leake MC. Single molecule experimentation in biological physics: exploring the living component of soft condensed matter one molecule at a time. J Phys Condens Matter 2011;23:503101. doi:10.1088/0953-8984/23/50/503101.

[4]   Ritort F. Single-molecule experiments in biological physics: methods and applications. J Phys Condens Matter 2006;18:R531–83. doi:10.1088/0953-8984/18/32/R01.

[5] Lenn T, Leake MC. Experimental approaches for addressing fundamental biological questions in living, functioning cells with single molecule precision. Open Biol 2012;2:120090. doi:10.1098/rsob.120090.

[6] Leake MC. Analytical tools for single-molecule fluorescence imaging in cellulo. Phys Chem Chem Phys 2014;16:12635–47. doi:10.1039/c4cp00219a.

[7] Leake MC. Shining the spotlight on functional molecular complexes: The new science of single-molecule cell biology. Commun Integr Biol 2010;3:415–8. doi:10.4161/cib.3.5.12657.

[8] Leake MC. Single-Molecule Cellular Biophysics. Cambridge University Press; 2013.

[9] Hall CE. Method for the observation of macromolecules with the electron microscope illustrated with micrographs of DNA. J Biophys Biochem Cytol 1956;2:625–8.

[10] Williams RC, Wyckoff RWG. The Thickness of Electron Microscopic Objects. J Appl Phys 1944;15:712–6. doi:10.1063/1.1707376.

[11] Rotman B. Measurement of activity of single molecules of beta-D-galactosidase. Proc Natl Acad Sci U S A 1961;47:1981–91.

[12] Hirschfeld T. Optical microscopic observation of single small molecules. Appl Opt 1976;15:2965. doi:10.1364/AO.15.002965.

[13] Barak LS, Webb WW. Diffusion of low density lipoprotein-receptor complex on human fibroblasts. J Cell Biol 1982;95:846–52.

[14] Gelles J, Schnapp BJ, Sheetz MP. Tracking kinesin-driven movements with nanometre-scale precision. Nature 1988;331:450–3. doi:10.1038/331450a0.

[15] Schmidt T, Schütz GJ, Baumgartner W, Gruber HJ, Schindler H. Imaging of single molecule diffusion. Proc Natl Acad Sci U S A 1996;93:2926–9.

[16] Sako Y, Minoghchi S, Yanagida T. Single-molecule imaging of EGFR signalling on the surface of living cells. Nat Cell Biol 2000;2:168–72. doi:10.1038/35004044.

[17] Betzig E. Single Molecules, Cells, and Super-Resolution Optics (Nobel Lecture). Angew Chemie Int Ed 2015;54:8034–53. doi:10.1002/anie.201501003.

[18] Moerner WE. Single-Molecule Spectroscopy, Imaging, and Photocontrol: Foundations for Super-Resolution Microscopy (Nobel Lecture). Angew Chemie Int Ed 2015;54:8067–93. doi:10.1002/anie.201501949.

[19] Hell SW. Nanoscopy with Focused Light (Nobel Lecture). Angew Chemie Int Ed 2015;54:8054–66. doi:10.1002/anie.201504181.

[20] Wollman AJM, Nudd R, Hedlund EG, Leake MC. From Animaculum to single molecules: 300 years of the light microscope. Open Biol 2015;5:150019–150019. doi:10.1098/rsob.150019.

[21] Moerner WE, Kador L. Optical detection and spectroscopy of single molecules in a solid. Phys Rev Lett 1989;62:2535–8. doi:10.1103/PhysRevLett.62.2535.

[22] Orrit M, Bernard J. Single pentacene molecules detected by fluorescence excitation in a p -terphenyl crystal. Phys Rev Lett 1990;65:2716–9. doi:10.1103/PhysRevLett.65.2716.

[23] Ash EA, Nicholls G. Super-resolution Aperture Scanning Microscope. Nature 1972;237:510–2. doi:10.1038/237510a0.

[24] Betzig E, Isaacson M, Lewis A. Collection mode near-field scanning optical microscopy. Appl Phys Lett 1987;51:2088–90. doi:10.1063/1.98956.

[25] Betzig E, Chichester RJ, Lanni F, Taylor DL. Near-field fluorescence imaging of cytoskeletal actin. Bioimaging 1993;1:129–35. doi:10.1002/1361-6374(199309)1:3<129::AID-BIO1>3.0.CO;2-8.

[26] Betzig E, Chichester RJ. Single molecules observed by near-field scanning optical microscopy. Science 1993;262:1422–5. doi:10.1126/science.262.5138.1422.

[27] Hell SW. Improvement of lateral resolution in far-field fluorescence light microscopy

by using two-photon excitation with offset beams. Opt Commun 1994;106:19–24. doi:10.1016/0030-4018(94)90050-7.

[28] Betzig E. Proposed method for molecular optical imaging. Opt Lett 1995;20:237. doi:10.1364/OL.20.000237.

[29] Betzig E, Patterson GH, Sougrat R, Lindwasser OW, Olenych S, Bonifacino JS, et al. Imaging intracellular fluorescent proteins at nanometer resolution. Science 2006;313:1642–5. doi:10.1126/science.1127344.

[30] Okhonin V. A. METHOD OF INVESTIGATING SPECIMEN MICROSTRUCTURE. SU 1374922, 1986.

[31] Klar TA, Jakobs S, Dyba M, Egner A, Hell SW. Fluorescence microscopy with diffraction resolution barrier broken by stimulated emission. Proc Natl Acad Sci U S A 2000;97:8206–10. doi:10.1073/PNAS.97.15.8206.

[32] Capitanio M, Pavone FS. Interrogating Biology with Force: Single Molecule High-Resolution Measurements with Optical Tweezers. Biophys J 2013;105:1293–303. doi:10.1016/j.bpj.2013.08.007.

[33] Kapanidis AN, Strick T. Biology, one molecule at a time. Trends Biochem Sci 2009;34:234–43. doi:10.1016/j.tibs.2009.01.008.

[34] Bockelmann U. Single-molecule manipulation of nucleic acids. Curr Opin Struct Biol 2004;14:368–73. doi:10.1016/j.sbi.2004.03.016.

[35] Hu X, Li H. Force spectroscopy studies on protein-ligand interactions: A single protein mechanics perspective. FEBS Lett 2014;588:3613–20. doi:10.1016/j.febslet.2014.04.009.

[36] VALE R, REESE T, SHEETZ M. Identification of a novel force-generating protein, kinesin, involved in microtubule-based motility. Cell 1985;42:39–50. doi:10.1016/S0092-8674(85)80099-4.

[37] Block SM, Goldstein LSB, Schnapp BJ. Bead movement by single kinesin molecules studied with optical tweezers. Nature 1990;348:348–52. doi:10.1038/348348a0.

[38] Svoboda K, Block SM. Force and velocity measured for single kinesin molecules. Cell 1994;77:773–84. doi:10.1016/0092-8674(94)90060-4.

[39] Mehta AD, Cheney RE, Rock RS, Rief M, Spudich JA, Mooseker MS. Myosin-V is a processive actin-based motor. Nature 1999;400:590–3. doi:10.1038/23072.

[40] Greenleaf WJ, Woodside MT, Abbondanzieri EA, Block SM. Passive all-optical force clamp for high-resolution laser trapping. Phys Rev Lett 2005;95:208102. doi:10.1103/PhysRevLett.95.208102.

[41] Moffitt JR, Chemla YR, Izhaky D, Bustamante C. Differential detection of dual traps improves the spatial resolution of optical tweezers. Proc Natl Acad Sci U S A 2006;103:9006–11. doi:10.1073/pnas.0603342103.

[42] Gebhardt JCM, Bornschlogl T, Rief M. Full distance-resolved folding energy landscape of one single protein molecule. Proc Natl Acad Sci 2010;107:2013–8. doi:10.1073/pnas.0909854107.

[43] Ribezzi-Crivellari M, Huguet JM, Ritort F. Counter-propagating dual-trap optical tweezers based on linear momentum conservation. Rev Sci Instrum 2013;84:43104. doi:10.1063/1.4799289.

[44] Alder BJ, Wainwright TE. Phase Transition for a Hard Sphere System. J Chem Phys 1957;27:1208–9. doi:10.1063/1.1743957.

[45] Alder BJ, Wainwright TE. Studies in Molecular Dynamics. II. Behavior of a Small Number of Elastic Spheres. J Chem Phys 1960;33:1439–51. doi:10.1063/1.1731425.

[46] Levitt M, Lifson S. Refinement of protein conformations using a macromolecular energy minimization procedure. J Mol Biol 1969;46:269–79. doi:http://dx.doi.org/10.1016/0022-2836(69)90421-5.

[47] McCammon JA, Gelin BR, Karplus M. Dynamics of folded proteins. Nature 1977;267:585–90.
[48] McCammon JA, Gelin BR, Karplus M, Wolynes PG. The hinge-bending mode in lysozyme. Nature 1976;262:325–6.
[49] Jorgensen WL, Chandrasekhar J, Madura JD, Impey RW, Klein ML. Comparison of simple potential functions for simulating liquid water. J Chem Phys 1983;79:926–35. doi:10.1063/1.445869.
[50] Drew HR, Dickerson RE. Structure of a B-DNA dodecamer. J Mol Biol 1981;151:535–56. doi:http://dx.doi.org/10.1016/0022-2836(81)90009-7.
[51] Still WC, Tempczyk A, Hawley RC, Hendrickson T. Semianalytical treatment of solvation for molecular mechanics and dynamics. J Am Chem Soc 1990;112:6127–9. doi:10.1021/ja00172a038.
[52] Cheatham TEIII, Miller JL, Fox T, Darden TA, Kollman PA. Molecular Dynamics Simulations on Solvated Biomolecular Systems: The Particle Mesh Ewald Method Leads to Stable Trajectories of DNA, RNA, and Proteins. J Am Chem Soc 1995;117:4193–4. doi:10.1021/ja00119a045.
[53] Carloni P, Sprik M, Andreoni W. Key Steps of the cis-Platin-DNA Interaction: Density Functional Theory-Based Molecular Dynamics Simulations. J Phys Chem B 2000;104:823–35. doi:10.1021/jp992590x.
[54] Car R, Parrinello M. Unified Approach for Molecular Dynamics and Density-Functional Theory. Phys Rev Lett 1985;55:2471–4. doi:10.1103/PhysRevLett.55.2471.
[55] Marko JF, Siggia ED. Statistical mechanics of supercoiled DNA. Phys Rev E 1995;52:2912–38. doi:10.1103/PhysRevE.52.2912.
[56] Levitt M, Warshel A. Computer simulation of protein folding. Nature 1975;253:694–8.
[57] Rowe AD, Leake MC, Morgan H, Berry RM. Rapid rotation of micron and submicron dielectric particles measured using optical tweezers. J Mod Opt 2003;50. doi:10.1080/0950034031000069361.
[58] Sowa Y, Rowe AD, Leake MC, Yakushi T, Homma M, Ishijima A, et al. Direct observation of steps in rotation of the bacterial flagellar motor. Nature 2005;437:916–9. doi:10.1038/nature04003.
[59] Reid SW, Leake MC, Chandler JHJ, Lo C-J, Armitage JP, Berry RM. The maximum number of torque-generating units in the flagellar motor of Escherichia coli is at least 11. Proc Natl Acad Sci U S A 2006;103:8066–71. doi:10.1073/pnas.0509932103.
[60] Thompson RE, Larson DR, Webb WW. Precise nanometer localization analysis for individual fluorescent probes. Biophys J 2002;82:2775–83. doi:10.1016/S0006-3495(02)75618-X.
[61] Chung E, Kim D, Cui Y, Kim Y-H, So PTC. Two-Dimensional Standing Wave Total Internal Reflection Fluorescence Microscopy: Superresolution Imaging of Single Molecular and Biological Specimens. Biophys J 2007;93:1747–57. doi:10.1529/biophysj.106.097907.
[62] Engelbrecht CJ, Stelzer EH. Resolution enhancement in a light-sheet-based microscope (SPIM). Opt Lett 2006;31:1477. doi:10.1364/OL.31.001477.
[63] Plank M, Wadhams GH, Leake MC. Millisecond timescale slimfield imaging and automated quantification of single fluorescent protein molecules for use in probing complex biological processes. Integr Biol (Camb) 2009;1:602–12. doi:10.1039/b907837a.
[64] Grimm JB, English BP, Chen J, Slaughter JP, Zhang Z, Revyakin A, et al. A general method to improve fluorophores for live-cell and single-molecule microscopy. Nat Methods 2015;12:244–50. doi:10.1038/nmeth.3256.
[65] Grimm JB, Sung AJ, Legant WR, Hulamm P, Matlosz SM, Betzig E, et al.

Carbofluoresceins and Carborhodamines as Scaffolds for High-Contrast Fluorogenic Probes. ACS Chem Biol 2013;8:1303–10. doi:10.1021/cb4000822.

[66] Dempsey GT, Vaughan JC, Chen KH, Bates M, Zhuang X. Evaluation of fluorophores for optimal performance in localization-based super-resolution imaging. Nat Methods 2011;8:1027–36. doi:10.1038/nmeth.1768.

[67] Lukinavičius G, Reymond L, D'Este E, Masharina A, Göttfert F, Ta H, et al. Fluorogenic probes for live-cell imaging of the cytoskeleton. Nat Methods 2014;11:731–3. doi:10.1038/nmeth.2972.

[68] Chan KG, Streichan SJ, Trinh LA, Liebling M. Simultaneous Temporal Superresolution and Denoising for Cardiac Fluorescence Microscopy. IEEE Trans Comput Imaging 2016;2:348–58. doi:10.1109/TCI.2016.2579606.

[69] Mourabit I El, Rhabi M El, Hakim A, Laghrib A, Moreau E. A new denoising model for multi-frame super-resolution image reconstruction. Signal Processing 2017;132:51–65. doi:10.1016/j.sigpro.2016.09.014.

[70] Nugent-Glandorf L, Perkins TT. Measuring 0.1-nm motion in 1 ms in an optical microscope with differential back-focal-plane detection. Opt Lett 2004;29:2611–3. doi:10.1364/OL.29.002611.

[71] Lang MJ, Asbury CL, Shaevitz JW, Block SM. An automated two-dimensional optical force clamp for single molecule studies. Biophys J 2002;83:491–501. doi:10.1016/S0006-3495(02)75185-0.

[72] Abbondanzieri EA, Greenleaf WJ, Shaevitz JW, Landick R, Block SM. Direct observation of base-pair stepping by RNA polymerase. Nature 2005;438:460–5. doi:10.1038/nature04268.

[73] Liu R, Garcia-Manyes S, Sarkar A, Badilla CL, Fernández JM. Mechanical Characterization of Protein L in the Low-Force Regime by Electromagnetic Tweezers/Evanescent Nanometry. Biophys J 2009;96:3810–21. doi:10.1016/j.bpj.2009.01.043.

[74] Haber C, Wirtz D. Magnetic tweezers for DNA micromanipulation. Rev Sci Instrum 2000;71:4561. doi:10.1063/1.1326056.

[75] Trepat X, Grabulosa M, Buscemi L, Rico F, Fabry B, Fredberg JJ, et al. Oscillatory magnetic tweezers based on ferromagnetic beads and simple coaxial coils. Rev Sci Instrum 2003;74:4012–20. doi:10.1063/1.1599062.

[76] Neuman KC, Nagy A. Single-molecule force spectroscopy : optical tweezers , magnetic tweezers and atomic force microscopy 2008;5:491–505. doi:10.1038/NMETH.1218.

[77] Svoboda K, Schmidt CF, Schnapp BJ, Block SM. Direct observation of kinesin stepping by optical trapping interferometry. Nature 1993;365:721–7. doi:10.1038/365721a0.

[78] Shroff H, Reinhard BM, Siu M, Agarwal H, Spakowitz A, Liphardt J. Biocompatible force sensor with optical readout and dimensions of 6 nm3. Nano Lett 2005;5:1509–14.

[79] Leake MC, Wilson D, Bullard B, Simmons RM. The elasticity of single kettin molecules using a two-bead laser-tweezers assay. FEBS Lett 2003;535:55–60. doi:10.1016/S0014-5793(02)03857-7.

[80] Linke WA, Leake MC. Multiple sources of passive stress relaxation in muscle fibres. Phys Med Biol 2004;49:3613–27. doi:10.1088/0031-9155/49/16/009.

[81] Carter AR, King GM, Ulrich TA, Halsey W, Alchenberger D, Perkins TT. Stabilization of an optical microscope to 0.1 nm in three dimensions. Appl Opt 2007;46:421. doi:10.1364/AO.46.000421.

[82] Chiu S-W, Leake MC. Functioning nanomachines seen in real-time in living bacteria

using single-molecule and super-resolution fluorescence imaging. Int J Mol Sci 2011;12:2518–42. doi:10.3390/ijms12042518.

[83] Yildiz A, Forkey JN, McKinney S a, Ha T, Goldman YE, Selvin PR. Myosin V walks hand-over-hand: single fluorophore imaging with 1.5-nm localization. Science 2003;300:2061–5. doi:10.1126/science.1084398.

[84] Hell SW, Wichmann J. Breaking the diffraction resolution limit by stimulated emission: stimulated-emission-depletion fluorescence microscopy. Opt Lett 1994;19:780–2.

[85] Hess ST, Girirajan TPK, Mason MD. Ultra-high resolution imaging by fluorescence photoactivation localization microscopy. Biophys J 2006;91:4258–72. doi:10.1529/biophysj.106.091116.

[86] Rust MJ, Bates M, Zhuang X. Sub-diffraction-limit imaging by stochastic optical reconstruction microscopy (STORM). Nature 2006;3:793–5. doi:10.1038/NMETH929.

[87] Schoen I, Ries J, Klotzsch E, Ewers H, Vogel V. Binding-activated localization microscopy of DNA structures. Nano Lett 2011;11:4008–11. doi:10.1021/nl2025954.

[88] Burnette DT, Sengupta P, Dai Y, Lippincott-Schwartz J, Kachar B. Bleaching/blinking assisted localization microscopy for superresolution imaging using standard fluorescent molecules. Proc Natl Acad Sci U S A 2011;108:21081–6. doi:10.1073/pnas.1117430109.

[89] Bryan SJ, Burroughs NJ, Shevela D, Yu J, Rupprecht E, Liu L-N, et al. Localisation and interactions of the Vipp1 protein in cyanobacteria. Mol Microbiol 2014;94:1179–95. doi:10.1111/mmi.12826.

[90] Boettiger AN, Bintu B, Moffitt JR, Wang S, Beliveau BJ, Fudenberg G, et al. Super-resolution imaging reveals distinct chromatin folding for different epigenetic states. Nature 2016. doi:10.1038/nature16496.

[91] Liao Y, Li Y, Schroeder JW, Simmons LA, Biteen JS. Single-Molecule DNA Polymerase Dynamics at a Bacterial Replisome in Live Cells. Biophys J 2016;111:2562–9. doi:10.1016/j.bpj.2016.11.006.

[92] Henriques R, Lelek M, Fornasiero EF, Valtorta F, Zimmer C, Mhlanga MM. QuickPALM: 3D real-time photoactivation nanoscopy image processing in ImageJ. Nat Methods 2010;7:339–40. doi:10.1038/nmeth0510-339.

[93] Rees EJ, Erdelyi M, Schierle GSK, Knight AE, Kaminski CF. Elements of image processing in localization microscopy. J Opt 2013;15:94012. doi:10.1088/2040-8978/15/9/094012.

[94] Ovesny M, K i ek P, Borkovec J, Vindrych Z, Hagen GM. ThunderSTORM: a comprehensive ImageJ plug-in for PALM and STORM data analysis and super-resolution imaging. Bioinformatics 2014;30:2389–90. doi:10.1093/bioinformatics/btu202.

[95] Gustafsson N, Culley S, Ashdown G, Owen DM, Pereira PM, Henriques R. Fast live-cell conventional fluorophore nanoscopy with ImageJ through super-resolution radial fluctuations. Nat Commun 2016;7:12471. doi:10.1038/ncomms12471.

[96] Sage D, Kirshner H, Pengo T, Stuurman N, Min J, Manley S, et al. Quantitative evaluation of software packages for single-molecule localization microscopy. Nat Methods 2015. doi:10.1038/nmeth.3442.

[97] Hell SW, Dyba M, Jakobs S. Concepts for nanoscale resolution in fluorescence microscopy. Curr Opin Neurobiol 2004;14:599–609. doi:10.1016/j.conb.2004.08.015.

[98] Hofmann M, Eggeling C, Jakobs S, Hell SW. Breaking the diffraction barrier in fluorescence microscopy at low light intensities by using reversibly photoswitchable proteins. Proc Natl Acad Sci U S A 2005;102:17565–9. doi:10.1073/pnas.0506010102.

[99] Chmyrov A, Keller J, Grotjohann T, Ratz M, D'Este E, Jakobs S, et al. Nanoscopy

with more than 100,000 "doughnuts." Nat Methods 2013;10:737–40. doi:10.1038/nmeth.2556.

[100] Schnorrenberg S, Grotjohann T, Vorbrüggen G, Herzig A, Hell SW, Jakobs S, et al. In vivo super-resolution RESOLFT microscopy of Drosophila melanogaster. Elife 2016;5:1370–3. doi:10.7554/eLife.15567.

[101] Balzarotti F, Eilers Y, Gwosch KC, Gynnå AH, Westphal V, Stefani FD, et al. Nanometer resolution imaging and tracking of fluorescent molecules with minimal photon fluxes. Science (80- ) 2017;355.

[102] Juette MF, Gould TJ, Lessard MD, Mlodzianoski MJ, Nagpure BS, Bennett BT, et al. Three-dimensional sub-100 nm resolution fluorescence microscopy of thick samples. Nat Methods 2008;5:527–9. doi:10.1038/nmeth.1211.

[103] Kao HP, Verkman AS. Tracking of single fluorescent particles in three dimensions: use of cylindrical optics to encode particle position. Biophys J 1994;67:1291–300. doi:10.1016/S0006-3495(94)80601-0.

[104] Huang B, Wang W, Bates M, Zhuang X. Three-dimensional super-resolution imaging by stochastic optical reconstruction microscopy. Science 2008;319:810–3. doi:10.1126/science.1153529.

[105] Pavani SRP, Thompson MA, Biteen JS, Lord SJ, Liu N, Twieg RJ, et al. Three-dimensional, single-molecule fluorescence imaging beyond the diffraction limit by using a double-helix point spread function. Proc Natl Acad Sci U S A 2009;106:2995–9. doi:10.1073/pnas.0900245106.

[106] Badieirostami M, Lew MD, Thompson M a, Moerner WE. Three-dimensional localization precision of the double-helix point spread function versus astigmatism and biplane. Appl Phys Lett 2010;97:161103. doi:10.1063/1.3499652.

[107] Lew MD, Lee SF, Badieirostami M, Moerner WE. Corkscrew point spread function for far-field three-dimensional nanoscale localization of pointlike objects. Opt Lett 2011;36:202. doi:10.1364/OL.36.000202.

[108] Jia S, Vaughan JC, Zhuang X. Isotropic three-dimensional super-resolution imaging with a self-bending point spread function. Nat Photonics 2014;8:302–6. doi:10.1038/nphoton.2014.13.

[109] Shechtman Y, Sahl SJ, Backer AS, Moerner WE. Optimal Point Spread Function Design for 3D Imaging. Phys Rev Lett 2014;113:133902. doi:10.1103/PhysRevLett.113.133902.

[110] Forster T. Energiewanderung und Fluoreszenz. Naturwissenschaften 1946;33:166–75. doi:10.1007/BF00585226.

[111] Stryer L, Haugland RP. Energy transfer: a spectroscopic ruler. Proc Natl Acad Sci U S A 1967;58:719–26. doi:10.1146/annurev.bi.47.070178.004131.

[112] Hohlbein J, Craggs TD, Cordes T, Ha T, Ando R, Mizuno H, et al. Alternating-laser excitation: single-molecule FRET and beyond. Chem Soc Rev 2014;43:1156–71. doi:10.1039/C3CS60233H.

[113] Ha T, Enderle T, Ogletree DF, Chemla DS, Selvin PR, Weiss S. Probing the interaction between two single molecules: fluorescence resonance energy transfer between a single donor and a single acceptor. Proc Natl Acad Sci U S A 1996;93:6264–8.

[114] Kapanidis AN, Margeat E, Ho SO, Kortkhonjia E, Weiss S, Ebright RH. Initial Transcription by RNA Polymerase Proceeds Through a DNA-Scrunching Mechanism. Science (80- ) 2006;314.

[115] Andrecka J, Lewis R, Brückner F, Lehmann E, Cramer P, Michaelis J. Single-molecule tracking of mRNA exiting from RNA polymerase II. Proc Natl Acad Sci U S A 2008;105:135–40. doi:10.1073/pnas.0703815105.

[116] Lee J, Lee S, Ragunathan K, Joo C, Ha T, Hohng S. Single-Molecule Four-Color FRET. Angew Chemie Int Ed 2010;49:9922–5. doi:10.1002/anie.201005402.
[117] Kalinin S, Peulen T, Sindbert S, Rothwell PJ, Berger S, Restle T, et al. A toolkit and benchmark study for FRET-restrained high-precision structural modeling. Nat Methods 2012;9:1218–25. doi:10.1038/nmeth.2222.
[118] Llorente-Garcia I, Lenn T, Erhardt H, Harriman OL, Liu L-N, Robson A, et al. Single-molecule in vivo imaging of bacterial respiratory complexes indicates delocalized oxidative phosphorylation. Biochim Biophys Acta 2014;1837:811–24.
[119] Edidin M, Zagyansky Y, Lardner T. Measurement of membrane protein lateral diffusion in single cells. Science (80- ) 1976;191.
[120] Axelrod D, Ravdin P, Koppel DE, Schlessinger J, Webb WW, Elson EL, et al. Lateral motion of fluorescently labeled acetylcholine receptors in membranes of developing muscle fibers. Proc Natl Acad Sci U S A 1976;73:4594–8.
[121] Cole NB, Smith CL, Sciaky N, Terasaki M, Edidin M, Lippincott-Schwartz J. Diffusional Mobility of Golgi Proteins in Membranes of Living Cells. Science (80- ) 1996;273.
[122] Ellenberg J, Siggia ED, Moreira JE, Smith CL, Presley JF, Worman HJ, et al. Nuclear Membrane Dynamics and Reassembly in Living Cells: Targeting of an Inner Nuclear Membrane Protein in Interphase and Mitosis. J Cell Biol 1997;138.
[123] Badrinarayanan A, Reyes-Lamothe R, Uphoff S, Leake MC, Sherratt DJ. In vivo architecture and action of bacterial structural maintenance of chromosome proteins. Science 2012;338:528–31. doi:10.1126/science.1227126.
[124] Leake MC, Chandler JH, Wadhams GH, Bai F, Berry RM, Armitage JP. Stoichiometry and turnover in single, functioning membrane protein complexes. Nature 2006;443:355–8. doi:10.1038/nature05135.
[125] Mudumbi KC, Schirmer EC, Yang W. Single-point single-molecule FRAP distinguishes inner and outer nuclear membrane protein distribution. Nat Commun 2016;7:12562. doi:10.1038/ncomms12562.
[126] Gustafsson MG. Surpassing the lateral resolution limit by a factor of two using structured illumination microscopy. J Microsc 2000;198:82–7.
[127] Gustafsson MGL. Nonlinear structured-illumination microscopy: wide-field fluorescence imaging with theoretically unlimited resolution. Proc Natl Acad Sci U S A 2005;102:13081–6. doi:10.1073/pnas.0406877102.
[128] Rego EH, Shao L, Macklin JJ, Winoto L, Johansson G a, Kamps-Hughes N, et al. Nonlinear structured-illumination microscopy with a photoswitchable protein reveals cellular structures at 50-nm resolution. Proc Natl Acad Sci U S A 2012;109:E135-43. doi:10.1073/pnas.1107547108.
[129] York AG, Chandris P, Nogare DD, Head J, Wawrzusin P, Fischer RS, et al. Instant super-resolution imaging in live cells and embryos via analog image processing. Nat Methods 2013. doi:10.1038/nmeth.2687.
[130] Winter PW, York AG, Nogare DD, Ingaramo M, Christensen R, Chitnis A, et al. Two-photon instant structured illumination microscopy improves the depth penetration of super-resolution imaging in thick scattering samples. Optica 2014;1:181. doi:10.1364/OPTICA.1.000181.
[131] Ströhl F, Kaminski CF. Frontiers in structured illumination microscopy. Optica 2016;3:667. doi:10.1364/OPTICA.3.000667.
[132] Li D, Shao L, Chen B-C, Zhang X, Zhang M, Moses B, et al. Extended-resolution structured illumination imaging of endocytic and cytoskeletal dynamics. Science (80- ) 2015;349:aab3500-aab3500. doi:10.1126/science.aab3500.
[133] Ries J, Schwille P. Fluorescence correlation spectroscopy. BioEssays 2012;34:361–8.

doi:10.1002/bies.201100111.
[134] Ehrenberg M, Rigler R. Rotational brownian motion and fluorescence intensify fluctuations. Chem Phys 1974;4:390–401. doi:10.1016/0301-0104(74)85005-6.
[135] Magde D, Elson E, Webb WW. Thermodynamic Fluctuations in a Reacting System—Measurement by Fluorescence Correlation Spectroscopy. Phys Rev Lett 1972;29:705–8. doi:10.1103/PhysRevLett.29.705.
[136] Krichevsky O, Bonnet G. Fluorescence correlation spectroscopy: the technique and its applications. Reports Prog Phys 2002;65:251–97. doi:10.1088/0034-4885/65/2/203.
[137] Schwille P, Meyer-Almes FJ, Rigler R. Dual-color fluorescence cross-correlation spectroscopy for multicomponent diffusional analysis in solution. Biophys J 1997;72:1878–86. doi:10.1016/S0006-3495(97)78833-7.
[138] Schwille P, Haupts U, Maiti S, Webb WW. Molecular Dynamics in Living Cells Observed by Fluorescence Correlation Spectroscopy with One- and Two-Photon Excitation. Biophys J 1999;77:2251–65. doi:10.1016/S0006-3495(99)77065-7.
[139] Gowrishankar K, Ghosh S, Saha S, C. R, Mayor S, Rao M. Active Remodeling of Cortical Actin Regulates Spatiotemporal Organization of Cell Surface Molecules. Cell 2012;149:1353–67. doi:10.1016/j.cell.2012.05.008.
[140] Krieger JW, Singh AP, Bag N, Garbe CS, Saunders TE, Langowski J, et al. Imaging fluorescence (cross-) correlation spectroscopy in live cells and organisms. Nat Protoc 2015;10:1948–74. doi:10.1038/nprot.2015.100.
[141] Honigmann A, Mueller V, Ta H, Schoenle A, Sezgin E, Hell SW, et al. Scanning STED-FCS reveals spatiotemporal heterogeneity of lipid interaction in the plasma membrane of living cells. Nat Commun 2014;5:5412. doi:10.1038/ncomms6412.
[142] Vicidomini G, Ta H, Honigmann A, Mueller V, Clausen MP, Waithe D, et al. STED-FLCS: An Advanced Tool to Reveal Spatiotemporal Heterogeneity of Molecular Membrane Dynamics. Nano Lett 2015;15:5912–8. doi:10.1021/acs.nanolett.5b02001.
[143] Andrade DM, Clausen MP, Keller J, Mueller V, Wu C, Bear JE, et al. Cortical actin networks induce spatio-temporal confinement of phospholipids in the plasma membrane – a minimally invasive investigation by STED-FCS. Sci Rep 2015;5:11454. doi:10.1038/srep11454.
[144] Eggeling C. Super-resolution optical microscopy of lipid plasma membrane dynamics. Essays Biochem 2015;57:69–80. doi:10.1042/bse0570069.
[145] Benda A, Ma Y, Gaus K. Self-Calibrated Line-Scan STED-FCS to Quantify Lipid Dynamics in Model and Cell Membranes. Biophys J 2015;108:596–609. doi:10.1016/j.bpj.2014.12.007.
[146] Leutenegger M, Ringemann C, Lasser T, Hell SW, Eggeling C. Fluorescence correlation spectroscopy with a total internal reflection fluorescence STED microscope (TIRF-STED-FCS). Opt Express 2012;20:5243. doi:10.1364/OE.20.005243.
[147] Wachsmuth M, Conrad C, Bulkescher J, Koch B, Mahen R, Isokane M, et al. High-throughput fluorescence correlation spectroscopy enables analysis of proteome dynamics in living cells. Nat Biotechnol 2015;33:384–9. doi:10.1038/nbt.3146.
[148] Lindfors K, Kalkbrenner T, Stoller P, Sandoghdar V. Detection and spectroscopy of gold nanoparticles using supercontinuum white light confocal microscopy. Phys Rev Lett 2004;93:37401. doi:10.1103/PhysRevLett.93.037401.
[149] de Wit G, Danial JSH, Kukura P, Wallace MI. Dynamic label-free imaging of lipid nanodomains. Proc Natl Acad Sci U S A 2015;112:12299–303. doi:10.1073/pnas.1508483112.
[150] Nie S, Emory SR. Probing Single Molecules and Single Nanoparticles by Surface-Enhanced Raman Scattering. Science (80- ) 1997;275.
[151] Treffer R, Böhme R, Deckert-Gaudig T, Lau K, Tiede S, Lin X, et al. Advances in

TERS (tip-enhanced Raman scattering) for biochemical applications. Biochem Soc Trans 2012;40:609–14. doi:10.1042/BST20120033.

[152] Pieczonka NPW, Moula G, Aroca RF. SERRS for Single-Molecule Detection of Dye-Labeled Phospholipids in Langmuir−Blodgett Monolayers. Langmuir 2009;25:11261–4. doi:10.1021/la902486w.

[153] Taylor RW, Benz F, Sigle DO, Bowman RW, Bao P, Roth JS, et al. Watching individual molecules flex within lipid membranes using SERS. Sci Rep 2014;4. doi:10.1038/srep05940.

[154] Nenninger A, Mastroianni G, Robson A, Lenn T, Xue Q, Leake MC, et al. Independent mobility of proteins and lipids in the plasma membrane of Escherichia coli. Mol Microbiol 2014;92:1142–53. doi:10.1111/mmi.12619.

[155] Domenici F, Bizzarri AR, Cannistraro S. Surface-enhanced Raman scattering detection of wild-type and mutant p53 proteins at very low concentration in human serum. Anal Biochem 2012;421:9–15. doi:10.1016/j.ab.2011.10.010.

[156] Davidson MW, Campbell RE. Engineered fluorescent proteins: innovations and applications. Nat Methods 2009;6:713–7. doi:10.1038/nmeth1009-713.

[157] Cranfill PJ, Sell BR, Baird MA, Allen JR, Lavagnino Z, de Gruiter HM, et al. Quantitative assessment of fluorescent proteins. Nat Methods 2016;13:557–62. doi:10.1038/nmeth.3891.

[158] Fernández-Suárez M, Ting AY. Fluorescent probes for super-resolution imaging in living cells. Nat Rev Mol Cell Biol 2008;9:929–43. doi:10.1038/nrm2531.

[159] Terai T, Nagano T. Small-molecule fluorophores and fluorescent probes for bioimaging. Pflügers Arch - Eur J Physiol 2013;465:347–59. doi:10.1007/s00424-013-1234-z.

[160] Ngo JT, Adams SR, Deerinck TJ, Boassa D, Rodriguez-Rivera F, Palida SF, et al. Click-EM for imaging metabolically tagged nonprotein biomolecules. Nat Chem Biol 2016;12:459–65. doi:10.1038/nchembio.2076.

[161] Kuipers J, Ham TJ van, Kalicharan RD, Veenstra-Algra A, Sjollema KA, Dijk F, et al. FLIPPER, a combinatorial probe for correlated live imaging and electron microscopy, allows identification and quantitative analysis of various cells and organelles. Cell Tissue Res 2015;360:61. doi:10.1007/s00441-015-2142-7.

[162] Liu W, Naydenov B, Chakrabortty S, Wuensch B, Hübner K, Ritz S, et al. Fluorescent Nanodiamond–Gold Hybrid Particles for Multimodal Optical and Electron Microscopy Cellular Imaging. Nano Lett 2016;16:6236–44. doi:10.1021/acs.nanolett.6b02456.

[163] Streets AM, Huang Y. Microfluidics for biological measurements with single-molecule resolution. Curr Opin Biotechnol 2014;25:69–77. doi:10.1016/j.copbio.2013.08.013.

[164] Wollman AJM, Miller H, Foster S, Leake MC. An automated image analysis framework for segmentation and division plane detection of single live Staphylococcus aureus cells which can operate at millisecond sampling time scales using bespoke Slimfield microscopy. Phys Biol 2016;13:55002. doi:10.1088/1478-3975/13/5/055002.

[165] Wollman AJM, Sanchez-Cano C, Carstairs HMJ, Cross R a., Turberfield AJ. Transport and self-organization across different length scales powered by motor proteins and programmed by DNA. Nat Nanotechnol 2014;9:44–7. doi:10.1038/nnano.2013.230.

[166] Gambin Y, VanDelinder V, Ferreon ACM, Lemke EA, Groisman A, Deniz AA. Visualizing a one-way protein encounter complex by ultrafast single-molecule mixing. Nat Methods 2011;8:239–41. doi:10.1038/nmeth.1568.

[167] Guo MT, Rotem A, Heyman JA, Weitz DA, Quake SR, Samuel BS, et al. Droplet microfluidics for high-throughput biological assays. Lab Chip 2012;12:2146. doi:10.1039/c2lc21147e.

[168] N. Reginald Beer †,‡, Benjamin J. Hindson †, Elizabeth K. Wheeler †, Sara B. Hall †, Klint A. Rose †, Ian M. Kennedy ‡ and, et al. On-Chip, Real-Time, Single-Copy Polymerase Chain Reaction in Picoliter Droplets 2007. doi:10.1021/AC701809W.

[169] Unger MA, Chou H-P, Thorsen T, Scherer A, Quake SR. Monolithic Microfabricated Valves and Pumps by Multilayer Soft Lithography. Science (80- ) 2000;288.

[170] Kim S, Streets AM, Lin RR, Quake SR, Weiss S, Majumdar DS. High-throughput single-molecule optofluidic analysis. Nat Methods 2011;8:242–5. doi:10.1038/nmeth.1569.

[171] Sarkar SK. The total internal reflection fluorescence microscope. Single Mol. Biophys. Poisson Process Approach to Stat. Mech., IOP Publishing; 2016. doi:10.1088/978-1-6817-4116-1ch2.

[172] Huang Z-L, Zhu H, Long F, Ma H, Qin L, Liu Y, et al. Localization-based super-resolution microscopy with an sCMOS camera. Opt Express 2011;19:19156. doi:10.1364/OE.19.019156.

[173] Long F, Zeng S, Huang Z-L. Localization-based super-resolution microscopy with an sCMOS camera Part II: Experimental methodology for comparing sCMOS with EMCCD cameras. Opt Express 2012;20:17741. doi:10.1364/OE.20.017741.

[174] Singh AP, Krieger JW, Buchholz J, Charbon E, Langowski J, Wohland T. The performance of 2D array detectors for light sheet based fluorescence correlation spectroscopy. Opt Express 2013;21:8652. doi:10.1364/OE.21.008652.

[175] Jradi K, Pellion D, Ginhac D. Design, characterization and analysis of a 0.35 μm CMOS SPAD. Sensors (Basel) 2014;14:22773–84. doi:10.3390/s141222773.

[176] Rocca FM Della, Nedbal J, Tyndall D, Krstajić N, Li DD-U, Ameer-Beg SM, et al. Real-time fluorescence lifetime actuation for cell sorting using a CMOS SPAD silicon photomultiplier. Opt Lett 2016;41:673–6.

[177] Acconcia G, Cominelli A, Rech I, Ghioni M. High-efficiency integrated readout circuit for single photon avalanche diode arrays in fluorescence lifetime imaging. Rev Sci Instrum 2016;87:113110. doi:10.1063/1.4968199.

[178] Keller PJ, Schmidt AD, Wittbrodt J, Stelzer EHK. Reconstruction of Zebrafish Early Embryonic Development by Scanned Light Sheet Microscopy. Science (80- ) 2008;322.

[179] Bormuth V, Howard J, Schaffer E. LED illumination for video-enhanced DIC imaging of single microtubules. J Microsc 2007;226:1–5. doi:10.1111/j.1365-2818.2007.01756.x.

[180] Thornton KL, Findlay RC, Walrad PB, Wilson LG. Investigating the Swimming of Microbial Pathogens Using Digital Holography, Springer International Publishing; 2016, p. 17–32. doi:10.1007/978-3-319-32189-9_3.

[181] Tehrani KF, Xu J, Zhang Y, Shen P, Kner P. Adaptive optics stochastic optical reconstruction microscopy (AO-STORM) using a genetic algorithm. Opt Express 2015;23:13677. doi:10.1364/OE.23.013677.

[182] Gould TJ, Burke D, Bewersdorf J, Booth MJ. Adaptive optics enables 3D STED microscopy in aberrating specimens. Opt Express 2012;20:20998. doi:10.1364/OE.20.020998.

[183] Izeddin I, El Beheiry M, Andilla J, Ciepielewski D, Darzacq X, Dahan M. PSF shaping using adaptive optics for three-dimensional single-molecule super-resolution imaging and tracking. Opt Express 2012;20:4957. doi:10.1364/OE.20.004957.

[184] Chapman HN, Fromme P, Barty A, White TA, Kirian RA, Aquila A, et al. Femtosecond X-ray protein nanocrystallography. Nature 2011;470:73–7. doi:10.1038/nature09750.

[185] Weierstall U, et al DJ. Lipidic cubic phase injector facilitates membrane protein serial

femtosecond crystallography. Nat Comms 2014;5.

[186] Kierspel T, Wiese J, Mullins T, Robinson J, Aquila A, Barty A, et al. Strongly aligned gas-phase molecules at free-electron lasers. J Phys B At Mol Opt Phys 2015;48:204002.

[187] Sawaya MR, Cascio D, Gingery M, Rodriguez J, Goldschmidt L, Colletier J-P, et al. Protein crystal structure obtained at 2.9 Å resolution from injecting bacterial cells into an X-ray free-electron laser beam. Proc Natl Acad Sci 2014;111:12769–74. doi:10.1073/pnas.1413456111.

[188] Morgan JLW, Strumillo J, Zimmer J. Crystallographic snapshot of cellulose synthesis and membrane translocation. Nature 2013;493:181–7.

[189] Flannigan DJ, Barwick B, Zewail AH. Biological imaging with 4D ultrafast electron microscopy. PNAS 2010;107:9933–7.

[190] Shu X, Lev-Ram V, Deerinck TJ, Qi Y, Ramko EB, Davidson MW, et al. A Genetically Encoded Tag for Correlated Light and Electron Microscopy of Intact Cells, Tissues, and Organisms. PLOS Biol 2011.

[191] de Jongea N, Peckysb DB, Kremersa GJ, Pistona DW. Electron microscopy of whole cells in liquid with nanometer resolution. PNAS 2008;106:2159–64.

[192] Sato K, Pellegrino M, Nakagawa T, Nakagawa T, Vosshall LB, Touhara K. Insect olfactory receptors are heteromeric ligand-gated ion channels. Nature 2008;452:1002–6. doi:10.1038/nature06850.

[193] Voigt N, Li N, Wang Q, Wang W, Trafford AW, Abu-Taha I, et al. Enhanced Sarcoplasmic Reticulum $Ca^{2+}$ Leak and Increased $Na^+$-$Ca^{2+}$ Exchanger Function Underlie Delayed Afterdepolarizations in Patients With Chronic Atrial Fibrillation. Circulation 2012;125:2059–70. doi:10.1161/CIRCULATIONAHA.111.067306.

[194] Machtens J-P, Kortzak D, Lansche C, Leinenweber A, Kilian P, Begemann B, et al. Mechanisms of Anion Conduction by Coupled Glutamate Transporters. Cell 2015;160:542–53. doi:10.1016/j.cell.2014.12.035.

[195] Feng Y, Zhang Y, Ying C, Wang D, Du C. Nanopore-based Fourth-generation DNA Sequencing Technology. Genomics Proteomics Bioinformatics 2015;13:4–16. doi:10.1016/j.gpb.2015.01.009.

[196] Storm AJ, Chen JH, Ling XS, Zandbergen HW, Dekker C. Fabrication of solid-state nanopores with single-nanometre precision. Nat Mater 2003;2:537–40. doi:10.1038/nmat941.

[197] Garaj S, Hubbard W, Reina A, Kong J, Branton D, Golovchenko JA. Graphene as a subnanometre trans-electrode membrane. Nature 2010;467:190–3. doi:10.1038/nature09379.

[198] Cherf GM, Lieberman KR, Rashid H, Lam CE, Karplus K, Akeson M. Automated forward and reverse ratcheting of DNA in a nanopore at 5-Å precision. Nat Biotechnol 2012;30:344–8. doi:10.1038/nbt.2147.

[199] Bell NAW, Engst CR, Ablay M, Divitini G, Ducati C, Liedl T, et al. DNA Origami Nanopores. Nano Lett 2012;12:512–7. doi:10.1021/nl204098n.

[200] Kasianowicz JJ, Brandin E, Branton D, Deamer DW. Characterization of individual polynucleotide molecules using a membrane channel. Proc Natl Acad Sci 1996;93:13770–3.

[201] Mikheyev AS, Tin MMY. A first look at the Oxford Nanopore MinION sequencer. Mol Ecol Resour 2014;14:1097–102. doi:10.1111/1755-0998.12324.

[202] Wang Y, Zheng D, Tan Q, Wang MX, Gu L-Q. Nanopore-based detection of circulating microRNAs in lung cancer patients. Nat Nanotechnol 2011;6:668–74. doi:10.1038/nnano.2011.147.

[203] Bayley H, Movileanu L, Howorka S, Braha O. Detecting protein analytes that

modulate transmembrane movement of a polymer chain within a single protein pore. Nat Biotechnol 2000;18:1091–5. doi:10.1038/80295.

[204] Hornblower B, Coombs A, Whitaker RD, Kolomeisky A, Picone SJ, Meller A, et al. Single-molecule analysis of DNA-protein complexes using nanopores. Nat Methods 2007;4:315. doi:10.1038/nmeth1021.

[205] Talaga DS, Li J. Single-Molecule Protein Unfolding in Solid State Nanopores. J Am Chem Soc 2009;131:9287–97. doi:10.1021/ja901088b.

[206] Hansma PK, Drake B, Marti O, Gould SA, Prater CB. The scanning ion-conductance microscope. Science 1989;243:641–3.

[207] Klenerman D, Korchev YE, Davis SJ. Imaging and characterisation of the surface of live cells. Curr Opin Chem Biol 2011;15:696–703. doi:10.1016/j.cbpa.2011.04.001.

[208] Kremer F. Dielectric spectroscopy - Yesterday, today and tomorrow. J. Non. Cryst. Solids, vol. 305, 2002, p. 1–9. doi:10.1016/S0022-3093(02)01083-9.

[209] Müller F, Ferreira CA, Azambuja DS, Alemán C, Armelin E. Measuring the Proton Conductivity of Ion-Exchange Membranes Using Electrochemical Impedance Spectroscopy and Through-Plane Cell. J Phys Chem B 2014;118:1102–12. doi:10.1021/jp409675z.

[210] Christensen BJ, Coverdale T, Olson RA, Ford SJ, Garboczi EJ, Jennings HM, et al. Impedance Spectroscopy of Hydrating Cement-Based Materials: Measurement, Interpretation, and Application. J Am Ceram Soc 1994;77:2789–804. doi:10.1111/j.1151-2916.1994.tb04507.x.

[211] Russell E V, Israeloff NE. Direct observation of molecular cooperativity near the glass transition. Nature 2000;408:695–8. doi:10.1038/35047037.

[212] Grenier K, Dubuc D, Chen T, Artis F, Chretiennot T, Poupot M, et al. Recent advances in microwave-based dielectric spectroscopy at the cellular level for cancer investigations. IEEE Trans Microw Theory Tech 2013;61:2023–30. doi:10.1109/TMTT.2013.2255885.

[213] Artis F, Chen T, Chretiennot T, Fournie J-J, Poupot M, Dubuc D, et al. Microwaving Biological Cells: Intracellular Analysis with Microwave Dielectric Spectroscopy. IEEE Microw Mag 2015;16:87–96. doi:10.1109/MMM.2015.2393997.

[214] Maalouf R, Fournier-Wirth C, Coste J, Chebib H, Saïkali Y, Vittori O, et al. Label-free detection of bacteria by electrochemical impedance spectroscopy: Comparison to surface plasmon resonance. Anal Chem 2007;79:4879–86. doi:10.1021/ac070085n.

[215] Soley A, Lecina M, Gámez X, Cairó JJ, Riu P, Rosell X, et al. On-line monitoring of yeast cell growth by impedance spectroscopy. J Biotechnol 2005;118:398–405. doi:10.1016/j.jbiotec.2005.05.022.

[216] Leroy J, Dalmay C, Landoulsi A, Hjeij F, M??lin C, Bessette B, et al. Microfluidic biosensors for microwave dielectric spectroscopy. Sensors Actuators, A Phys 2015;229:172–81. doi:10.1016/j.sna.2015.04.002.

[217] Simmons RM, Finer JT, Warrick HM, Kralik B, Chu S, Spudich JA. Force on single actin filaments in a motility assay measured with an optical trap. Adv Exp Med Biol 1993;332:331–7.

[218] Block SM, Goldstein LSB, Schnapp BJ. Bead movement by single kinesis molecules studied with optical tweezers. Nature 1990;348:348–52.

[219] Lipfert J, Koster DA, Vilfan ID, Hage S, Dekker NH. Single-molecule magnetic tweezers studies of type IB topoisomerases. Methods Mol Biol 2009;582:71–89. doi:10.1007/978-1-60761-340-4_7.

[220] Charvin G, Strick TR, Bensimon D, Croquette V. Tracking topoisomerase activity at the single-molecule level. Annu Rev Biophys Biomol Struct 2005;34:201–19. doi:10.1146/annurev.biophys.34.040204.144433.

[221] Itoh H, Takahashi A, Adachi K, Noji H. Mechanically driven ATP synthesis by F1 - ATPase. Nature 2004;427:465–8. doi:10.1038/nature02229.1.

[222] Bullard B, Garcia T, Benes V, Leake MC, Linke WA, Oberhauser AF. The molecular elasticity of the insect flight muscle proteins projectin and kettin. Proc Natl Acad Sci U S A 2006;103:4451–6. doi:10.1073/pnas.0509016103.

[223] Bullard B, Ferguson C, Minajeva A, Leake MC, Gautel M, Labeit D, et al. Association of the Chaperone αβ-crystallin with Titin in Heart Muscle. J Biol Chem 2004;279. doi:10.1074/jbc.M307473200.

[224] Leake MC, Grützner A, Krüger M, Linke WA. Mechanical properties of cardiac titin's N2B-region by single-molecule atomic force spectroscopy. J Struct Biol 2006;155:263–72. doi:10.1016/j.jsb.2006.02.017.

[225] Neuman KC, Block SM. Optical trapping. Rev Sci Instrum 2004;75:2787–809. doi:10.1063/1.1785844.

[226] Tanase M, Biais N, Sheetz M. Magnetic Tweezers in Cell Biology. Methods Cell Biol 2007;83:473–93. doi:10.1016/S0091-679X(07)83020-2.

[227] Evans E, Ritchie K, Merkel R. Sensitive force technique to probe molecular adhesion and structural linkages at biological interfaces. Biophys J 1995;68:2580–7. doi:10.1016/S0006-3495(95)80441-8.

[228] Kim S, Blainey PC, Schroeder CM, Xie XS. Multiplexed single-molecule assay for enzymatic activity on flow-stretched DNA. Nat Methods 2007;4:397. doi:10.1038/nmeth1037.

[229] Zlatanova J, Lindsay SM, Leuba SH. Single molecule force spectroscopy in biology using the atomic force microscope. Prog Biophys Mol Biol 2000;74:37–61. doi:10.1016/S0079-6107(00)00014-6.

[230] Leake MC. Biophysics : tools and techniques. CRC Press; 2016.

[231] Smith SB, Finzi L, Bustamante C. Direct mechanical measurements of the elasticity of single DNA molecules by using magnetic beads. Science 1992;258:1122–6.

[232] Cluzel P, Lebrun A, Heller C, Lavery R, Viovy JL, Chatenay D, et al. DNA: an extensible molecule. Science 1996;271:792–4.

[233] Florin EL, Moy VT, Gaub HE. Adhesion forces between individual ligand-receptor pairs. Science 1994;264:415–7.

[234] Noji H, Yasuda R, Yoshida M, Kinosita K. Direct observation of the rotation of F1-ATPase. Nature 1997;386:299–302. doi:10.1038/386299a0.

[235] Ashkin A. Forces of a single-beam gradient laser trap on a dielectric sphere in the ray optics regime. Biophys J 1992;61:569–82. doi:10.1016/S0006-3495(92)81860-X.

[236] Nieminen TA, Loke VLY, Stilgoe AB, Knöner G, Brańczyk AM, Heckenberg NR, et al. Optical tweezers computational toolbox. J Opt A Pure Appl Opt 2007;9:S196–203. doi:10.1088/1464-4258/9/8/S12.

[237] Gittes F, Schmidt CF. Interference model for back-focal-plane displacement detection in optical tweezers. Opt Lett 1998;23:7. doi:10.1364/OL.23.000007.

[238] Sato S, Ishigure M, Inaba H. Optical trapping and rotational manipulation of microscopic particles and biological cells using higher-order mode Nd:YAG laser beams. Electron Lett 1991;27:1831. doi:10.1049/el:19911138.

[239] Allen L, Beijersbergen MW, Spreeuw RJC, Woerdman JP. Orbital angular momentum of light and the transformation of Laguerre-Gaussian laser modes. Phys Rev A 1992;45:8185–9. doi:10.1103/PhysRevA.45.8185.

[240] Abramochkin E, Volostnikov V. Beam transformations and nontransformed beams. Opt Commun 1991;83:123–35. doi:10.1016/0030-4018(91)90534-K.

[241] Tamm C, Weiss CO. Bistability and optical switching of spatial patterns in a laser. J Opt Soc Am B 1990;7:1034. doi:10.1364/JOSAB.7.001034.

[242] Chu S-C, Chen Y-T, Tsai K-F, Otsuka K. Generation of high-order Hermite-Gaussian modes in end-pumped solid-state lasers for square vortex array laser beam generation. Opt Express 2012;20:7128–41. doi:10.1364/OE.20.007128.

[243] Arlt J, Dholakia K, Allen L, Padgett MJ. The production of multiringed Laguerre–Gaussian modes by computer-generated holograms. J Mod Opt 1998;45:1231–7. doi:10.1080/09500349808230913.

[244] Turnbull GA, Robertson DA, Smith GM, Allen L, Padgett MJ. The generation of free-space Laguerre-Gaussian modes at millimetre-wave frequencies by use of a spiral phaseplate. Opt Commun 1996;127:183–8. doi:10.1016/0030-4018(96)00070-3.

[245] Matsumoto N, Ando T, Inoue T, Ohtake Y, Fukuchi N, Hara T. Generation of high-quality higher-order Laguerre-Gaussian beams using liquid-crystal-on-silicon spatial light modulators. J Opt Soc Am A Opt Image Sci Vis 2008;25:1642–51. doi:10.1364/JOSAA.25.001642.

[246] Durnin J, Miceli JJ, Eberly JH. Diffraction-free beams. Phys Rev Lett 1987;58:1499–501. doi:10.1103/PhysRevLett.58.1499.

[247] McGloin D, Dholakia K. Bessel beams: Diffraction in a new light. Contemp Phys 2005;46:15–28. doi:10.1080/0010751042000275259.

[248] Herman RM, Wiggins TA. Production and uses of diffractionless beams. J Opt Soc Am A 1991;8:932. doi:10.1364/JOSAA.8.000932.

[249] Arlt J, Dholakia K. Generation of high-order Bessel beams by use of an axicon. Opt Commun 2000;177:297–301. doi:10.1016/S0030-4018(00)00572-1.

[250] Vasara A, Turunen J, Friberg AT. Realization of general nondiffracting beams with computer-generated holograms. J Opt Soc Am A 1989;6:1748. doi:10.1364/JOSAA.6.001748.

[251] Chattrapiban N, Rogers EA, Cofield D, Hill, III WT, Roy R. Generation of nondiffracting Bessel beams by use of a spatial light modulator. Opt Lett 2003;28:2183. doi:10.1364/OL.28.002183.

[252] Sokolovskii GS, Dudelev V V, Losev SN, Soboleva KK, Deryagin AG, Kuchinskii VI, et al. Optical trapping with Bessel beams generated from semiconductor lasers. J Phys Conf Ser 2014;572:12039. doi:10.1088/1742-6596/572/1/012039.

[253] Garcés-Chávez V, McGloin D, Melville H, Sibbett W, Dholakia K. Simultaneous micromanipulation in multiple planes using a self-reconstructing light beam. Nature 2002;419:145–7. doi:10.1038/nature01007.

[254] Chen J, Ng J, Lin Z, Chan CT. Optical pulling force. Nat Photonics 2011;5:531–4. doi:10.1038/nphoton.2011.153.

[255] Novitsky A, Qiu C-W, Wang H. Single Gradientless Light Beam Drags Particles as Tractor Beams. Phys Rev Lett 2011;107:203601. doi:10.1103/PhysRevLett.107.203601.

[256] Poynting J. The wave motion of a revolving shaft, and a suggestion as to the angular momentum in a beam of circularly polarised light. Proc R Soc London Ser A, 1909.

[257] Deufel C, Forth S, Simmons CR, Dejgosha S, Wang MD. Nanofabricated quartz cylinders for angular trapping: DNA supercoiling torque detection. Nat Methods 2007;4:223–5. doi:10.1038/nmeth1013.

[258] Grier DG. Colloids: A surprisingly attractive couple. Nature 1998;393:621–3. doi:10.1038/31340.

[259] Reicherter M, Haist T, Wagemann EU, Tiziani HJ. Optical particle trapping with computer-generated holograms written on a liquid-crystal display. Opt Lett 1999;24:608. doi:10.1364/OL.24.000608.

[260] Liesener J, Reicherter M, Haist T, Tiziani HJ. Multi-functional optical tweezers using computer-generated holograms. Opt Commun 2000;185:77–82. doi:10.1016/S0030-

4018(00)00990-1.

[261] Curtis JE, Koss BA, Grier DG. Dynamic holographic optical tweezers. Opt Commun 2002;207:169–75. doi:10.1016/S0030-4018(02)01524-9.

[262] Roxworthy BJ, Ko KD, Kumar A, Fung KH, Chow EKC, Liu GL, et al. Application of Plasmonic Bowtie Nanoantenna Arrays for Optical Trapping, Stacking, and Sorting. Nano Lett 2012;12:796–801. doi:10.1021/nl203811q.

[263] Korda P, Spalding GC, Dufresne ER, Grier DG. Nanofabrication with holographic optical tweezers. Rev Sci Instrum 2002;73:1956–7.

[264] Agarwal R, Ladavac K, Roichman Y, Yu G, Lieber CM, Grier DG. Manipulation and assembly of nanowires with holographic optical traps. Opt Express 2005;13:8906. doi:10.1364/OPEX.13.008906.

[265] Cai H, Poon AW. Optical manipulation and transport of microparticles on silicon nitride microring-resonator-based add–drop devices. Opt Lett 2010;35:2855. doi:10.1364/OL.35.002855.

[266] Juan ML, Righini M, Quidant R. Plasmon nano-optical tweezers. Nat Photonics 2011;5:349–56. doi:10.1038/nphoton.2011.56.

[267] Shevkoplyas SS, Siegel AC, Westervelt RM, Prentiss MG, Whitesides GM. The force acting on a superparamagnetic bead due to an applied magnetic field. Lab Chip 2007;7:1294–302. doi:10.1039/b705045c.

[268] Mosconi F, Allemand JF, Croquette V. Soft magnetic tweezers: A proof of principle. Rev Sci Instrum 2011;82:34302. doi:10.1063/1.3531959.

[269] Gosse C, Croquette V. Magnetic tweezers: micromanipulation and force measurement at the molecular level. Biophys J 2002;82:3314–29. doi:10.1016/S0006-3495(02)75672-5.

[270] Kellermayer MS, Smith SB, Granzier HL, Bustamante C. Folding-unfolding transitions in single titin molecules characterized with laser tweezers. Science (80- ) 1997;276:1112–6. doi:10.1126/science.276.5315.1112.

[271] Leake MC, Wilson D, Gautel M, Simmons RM. The elasticity of single titin molecules using a two-bead optical tweezers assay. Biophys J 2004;87:1112–35. doi:10.1529/biophysj.103.033571.

[272] Smith SB, Cui Y, Bustamante C. Overstretching B-DNA: The Elastic Response of Individual Double-Stranded and Single-Stranded DNA Molecules. Science (80- ) 1996;271:795–9. doi:10.1126/science.271.5250.795.

[273] Strick TR, Bensimon D, Croquette V. Micro-mechanical measurement of the torsional modulus of DNA. Genetica 1999;106:57–62. doi:10.1023/a:1003772626927.

[274] Huang H, Dong CY, Kwon H-S, Sutin JD, Kamm RD, So PTC. Three-Dimensional Cellular Deformation Analysis with a Two-Photon Magnetic Manipulator Workstation. Biophys J 2002;82:2211–23. doi:10.1016/S0006-3495(02)75567-7.

[275] Chiou C-H, Huang Y-Y, Chiang M-H, Lee H-H, Lee G-B. New magnetic tweezers for investigation of the mechanical properties of single DNA molecules. Nanotechnology 2006;17:1217–24. doi:10.1088/0957-4484/17/5/009.

[276] Fisher JK, Cribb J, Desai K V, Vicci L, Wilde B, Keller K, et al. Thin-foil magnetic force system for high-numerical-aperture microscopy. Rev Sci Instrum 2006;77:nihms8302. doi:10.1063/1.2166509.

[277] Romano G, Sacconi L, Capitanio M, Pavone FS. Force and torque measurements using magnetic micro beads for single molecule biophysics. Opt Commun 2003;215:323–31. doi:10.1016/S0030-4018(02)02247-2.

[278] Claudet C, Bednar J. Magneto-optical tweezers built around an inverted microscope. Appl Opt 2005;44:3454–7. doi:10.1364/AO.44.003454.

[279] Zhou Z, Miller H, Wollman A, Leake M. Developing a New Biophysical Tool to

Combine Magneto-Optical Tweezers with Super-Resolution Fluorescence Microscopy. Photonics 2015;2:758–72. doi:10.3390/photonics2030758.

[280] Janssen XJA, Lipfert J, Jager T, Daudey R, Beekman J, Dekker NH. Electromagnetic Torque Tweezers: A Versatile Approach for Measurement of Single-Molecule Twist and Torque. Nano Lett 2012;12:3634–9. doi:10.1021/nl301330h.

[281] Schuerle S, Erni S, Flink M, Kratochvil BE, Nelson BJ. Three-Dimensional Magnetic Manipulation of Micro- and Nanostructures for Applications in Life Sciences. IEEE Trans Magn 2013;49:321–30. doi:10.1109/TMAG.2012.2224693.

[282] van Loenhout MTJ, de Grunt M V., Dekker C. Dynamics of DNA Supercoils. Science (80- ) 2012;338:94–7. doi:10.1126/science.1225810.

[283] Celedon A, Nodelman IM, Wildt B, Dewan R, Searson P, Wirtz D, et al. Magnetic Tweezers Measurement of Single Molecule Torque. Nano Lett 2009;9:1720–5. doi:10.1021/nl900631w.

[284] Lipfert J, Kerssemakers JWJ, Jager T, Dekker NH. Magnetic torque tweezers: measuring torsional stiffness in DNA and RecA-DNA filaments. Nat Methods 2010;7:977–80. doi:10.1038/nmeth.1520.

[285] Strick TR, Allemand J-F, Bensimon D, Bensimon A, Croquette V. The Elasticity of a Single Supercoiled DNA Molecule. Science (80- ) 1996;271:1835–7. doi:10.1126/science.271.5257.1835.

[286] Ribeck N, Saleh OA. Multiplexed single-molecule measurements with magnetic tweezers. Rev Sci Instrum 2008;79:94301. doi:10.1063/1.2981687.

[287] De Vlaminck I, Henighan T, van Loenhout MTJ, Pfeiffer I, Huijts J, Kerssemakers JWJ, et al. Highly Parallel Magnetic Tweezers by Targeted DNA Tethering. Nano Lett 2011;11:5489–93. doi:10.1021/nl203299e.

[288] Assi F, Jenks R, Yang J, Love C, Prentiss M. Massively parallel adhesion and reactivity measurements using simple and inexpensive magnetic tweezers. J Appl Phys 2002;92:5584–6. doi:10.1063/1.1509086.

[289] Crut A, Koster DA, Seidel R, Wiggins CH, Dekker NH. Fast dynamics of supercoiled DNA revealed by single-molecule experiments. Proc Natl Acad Sci U S A 2007;104:11957–62. doi:10.1073/pnas.0700333104.

[290] Shi J, Mao X, Ahmed D, Colletti A, Huang TJ. Focusing microparticles in a microfluidic channel with standing surface acoustic waves (SSAW). Lab Chip 2008;8:221–3. doi:10.1039/b716321e.

[291] Guo F, Mao Z, Chen Y, Xie Z, Lata JP, Li P, et al. Three-dimensional manipulation of single cells using surface acoustic waves. Proc Natl Acad Sci 2016;113:1522–7. doi:10.1073/pnas.1524813113.

[292] Shi J, Ahmed D, Mao X, Lin S-CS, Lawit A, Huang TJ. Acoustic tweezers: patterning cells and microparticles using standing surface acoustic waves (SSAW). Lab Chip 2009;9:2890. doi:10.1039/b910595f.

[293] Riaud A, Baudoin M, Bou Matar O, Becerra L, Thomas J-L. Selective Manipulation of Microscopic Particles with Precursor Swirling Rayleigh Waves. Phys Rev Appl 2017;7:24007. doi:10.1103/PhysRevApplied.7.024007.

[294] Courtney CRP, Demore CEM, Wu H, Grinenko A, Wilcox PD, Cochran S, et al. Independent trapping and manipulation of microparticles using dexterous acoustic tweezers. Appl Phys Lett 2014;104:154103. doi:10.1063/1.4870489.

[295] Galanzha EI, Viegas MG, Malinsky TI, Melerzanov A V, Juratli MA, Sarimollaoglu M, et al. In vivo acoustic and photoacoustic focusing of circulating cells. Sci Rep 2016;6:21531. doi:10.1038/srep21531.

[296] Melde K, Mark AG, Qiu T, Fischer P. Holograms for acoustics. Nature 2016;537:518–22. doi:10.1038/nature19755.

[297]  Binnig G, Quate CF, Gerber C. Atomic Force Microscope. Phys Rev Lett 1986;56:930–3. doi:10.1103/PhysRevLett.56.930.
[298]  Cappella B. Physical Principles of Force–Distance Curves by Atomic Force Microscopy, Springer International Publishing; 2016, p. 3–66. doi:10.1007/978-3-319-29459-9_1.
[299]  Hinterdorfer P, Dufrêne YF. Detection and localization of single molecular recognition events using atomic force microscopy. Nat Methods 2006;3:347–55. doi:10.1038/nmeth871.
[300]  Muller DJ, Helenius J, Alsteens D, Dufrene YF. Force probing surfaces of living cells to molecular resolution. Nat Chem Biol 2009;5:383–90. doi:10.1038/nchembio.181.
[301]  Rief M, Gautel M, Oesterhelt F, Fernandez JM, Gaub HE. Reversible Unfolgind of Individual Titin Immunoglobulin Domains by AFM. Science (80- ) 1997;276:1109–12. doi:10.1126/science.276.5315.1109.
[302]  Ido S, Kimiya H, Kobayashi K, Kominami H, Matsushige K, Yamada H. Immunoactive two-dimensional self-assembly of monoclonal antibodies in aqueous solution revealed by atomic force microscopy. Nat Mater 2014;13:264–70. doi:10.1038/nmat3847.
[303]  Uchihashi T, Iino R, Ando T, Noji H. High-speed atomic force microscopy reveals rotary catalysis of rotorless $F_1$-ATPase. Science 2011;333:755–8. doi:10.1126/science.1205510.
[304]  Vidal CMP, Zhu W, Manohar S, Aydin B, Keiderling TA, Messersmith PB, et al. Collagen-collagen interactions mediated by plant-derived proanthocyanidins: A spectroscopic and atomic force microscopy study. Acta Biomater 2016;41:110–8. doi:10.1016/j.actbio.2016.05.026.
[305]  Ido S, Kimura K, Oyabu N, Kobayashi K, Tsukada M, Matsushige K, et al. Beyond the Helix Pitch: Direct Visualization of Native DNA in Aqueous Solution. ACS Nano 2013;7:1817–22. doi:10.1021/nn400071n.
[306]  Ando T, Uchihashi T, Kodera N. High-speed AFM and applications to biomolecular systems. Annu Rev Biophys 2013;42:393–414. doi:10.1146/annurev-biophys-083012-130324.
[307]  Dagdeviren OE, Götzen J, Hölscher H, Altman EI, Schwarz UD. Robust high-resolution imaging and quantitative force measurement with tuned-oscillator atomic force microscopy. Nanotechnology 2016;27:65703. doi:10.1088/0957-4484/27/6/065703.
[308]  Ellner M, Pavliček N, Pou P, Schuler B, Moll N, Meyer G, et al. The Electric Field of CO Tips and Its Relevance for Atomic Force Microscopy. Nano Lett 2016;16:1974–80. doi:10.1021/acs.nanolett.5b05251.
[309]  Hong JW, Park S, Khim ZG. Measurement of hardness, surface potential, and charge distribution with dynamic contact mode electrostatic force microscope. Rev Sci Instrum 1999;70:1735. doi:10.1063/1.1149660.
[310]  Plumeré N. Single molecules: a protein in the spotlight. Nat Nanotechnol 2012;7:616–7.
[311]  Malvankar NS, Yalcin SE, Tuominen MT, Lovley DR. Visualization of charge propagation along individual pili proteins using ambient electrostatic force microscopy. Nat Nanotechnol 2014;9:1012–7. doi:10.1038/nnano.2014.236.
[312]  Tan Y, Adhikari RY, Malvankar NS, Pi S, Ward JE, Woodard TL, et al. Synthetic Biological Protein Nanowires with High Conductivity. Small 2016:4481–5. doi:10.1002/smll.201601112.
[313]  Shi L, Dong H, Reguera G, Beyenal H, Lu A, Liu J, et al. Extracellular electron transfer mechanisms between microorganisms and minerals. Nat Rev Microbiol

[314] Lovley DR. Happy together: microbial communities that hook up to swap electrons. ISME J 2016;11:1–10. doi:10.1038/ismej.2016.136.

[315] Kufer SK, Puchner EM, Gumpp H, Liedl T, Gaub HE. Single-Molecule Cut-and-Paste Surface Assembly. Science (80- ) 2008;319:594–6. doi:10.1126/science.1151424.

[316] Kufer SK, Strackharn M, Stahl SW, Gumpp H, Puchner EM, Gaub HE. Optically monitoring the mechanical assembly of single molecules. Nat Nanotechnol 2009;4:45–9. doi:10.1038/nnano.2008.333.

[317] Wang Q, Moerner WE. An Adaptive Anti-Brownian Electrokinetic Trap with Real-Time Information on Single-Molecule Diffusivity and Mobility. ACS Nano 2011;5:5792–9. doi:10.1021/nn2014968.

[318] Ridgway D, Broderick G, Lopez-Campistrous A, Ru'aini M, Winter P, Hamilton M, et al. Coarse-Grained Molecular Simulation of Diffusion and Reaction Kinetics in a Crowded Virtual Cytoplasm. Biophys J 2008;94:3748–59.

[319] Knotts IV TAK, Rathore N, Schwartz DC, de Pablo JJ. A coarse grain model for DNA. J Chem Phys 2007;126.

[320] Shaw DE, Maragakis P, Lindorff-Larsen K, Piana S, Dror RO, Eastwood MP, et al. Atomic-Level Characterization of the Structural Dynamics of Proteins. Science (80- ) 2010;330:341–6.

[321] Nelson T, Zhang B, Prezhdo O V. Detection of Nucleic Acids with Graphene Nanopores: Ab Initio Characterization of a Novel Sequencing Device. Nano Lett 2010;10:3237–42.

[322] Wang L, Fried SD, Boxer SG, Markland TE. Quantum Delocalization of Protons in the Hydrogen-Bond Network of an Enzyme Active Site. PNAS 2014;111:18454–9.

[323] Zhao G, Perilla JR, Yufenyuy EL, Meng X, Chen B, Ning J, et al. Mature HIV-1 capsid structure by cryo-electron microscopy and all-atom molecular dynamics. Nature 2013;497:643–6.

[324] Mori T, Miyashita N, Im W, Feig M, Sugita Y. Molecular dynamics simulations of biological membranes and membrane proteins using enhanced conformational sampling algorithms. Biochim Biophys Acta - Biomembr 2016;1858:1635–51. doi:10.1016/j.bbamem.2015.12.032.

[325] Pierce LCT, Salomon-Ferrer R, Augusto F de Oliveira C, McCammon JA, Walker RC. Routine Access to Millisecond Time Scale Events with Accelerated Molecular Dynamics. J Chem Theory Comput 2012;8:2997–3002. doi:10.1021/ct300284c.

[326] Abrams CF, Vanden-Eijnden E. Large-scale conformational sampling of proteins using temperature-accelerated molecular dynamics. Proc Natl Acad Sci 2010;107:4961–6. doi:10.1073/pnas.0914540107.

[327] Li H, Gisler T. Overstretching of a 30 bp DNA duplex studied with steered molecular dynamics simulation: effects of structural defects on structure and force-extension relation. Eur Phys J E Soft Matter 2009;30:325–32. doi:10.1140/epje/i2009-10524-5.

[328] Wells DB, Abramkina V, Aksimentiev A. Exploring transmembrane transport through alpha-hemolysin with grid-steered molecular dynamics. J Chem Phys 2007;127:125101. doi:10.1063/1.2770738.

[329] Cheng C-L, Zhao G-J. Steered molecular dynamics simulation study on dynamic self-assembly of single-stranded DNA with double-walled carbon nanotube and graphene. Nanoscale 2012;4:2301–5. doi:10.1039/c2nr12112c.

[330] Mai BK, Viet MH, Li MS. Top leads for swine influenza A/H1N1 virus revealed by steered molecular dynamics approach. J Chem Inf Model 2010;50:2236–47. doi:10.1021/ci100346s.

[331] Qi Y, Spong MC, Nam K, Banerjee A, Jiralerspong S, Karplus M, et al. Encounter and

extrusion of an intrahelical lesion by a DNA repair enzyme. Nature 2009;462:762–6. doi:10.1038/nature08561.

[332] Park S, Khalili-Araghi F, Tajkhorshid E, Schulten K. Free energy calculation from steered molecular dynamics simulations using Jarzynski's equality. J Chem Phys 2003;119:3559–66. doi:10.1063/1.1590311.

[333] Krüger P, Verheyden S, Declerck PJ, Engelborghs Y. Extending the capabilities of targeted molecular dynamics: simulation of a large conformational transition in plasminogen activator inhibitor 1. Protein Sci 2001;10:798–808. doi:10.1110/ps.40401.

[334] Cheng X, Wang H, Grant B, Sine SM, McCammon JA. Targeted molecular dynamics study of C-loop closure and channel gating in nicotinic receptors. PLoS Comput Biol 2006;2:e134. doi:10.1371/journal.pcbi.0020134.

[335] Ouldridge TE, Hoare RL, Louis AA, Doye JPK, Bath J, Turberfield AJ. Optimizing DNA Nanotechnology through Coarse-Grained Modeling: A Two-Footed DNA Walker. ACS Nano 2013;7:2479–90.

[336] Johnston IG, Louis AA, Doye JPK. Modelling the self-assembly of virus capsids. J Phys Condens Matter 2010;22.

[337] Scott KA, Bond PJ, Ivetac A, Chetwynd AP, Khalid S, Sansom MSP. Coarse-Grained MD Simulations of Membrane Protein-Bilayer Self-Assembly. Structure 2008;16:621–30.

[338] Lever G, Cole DJ, Lonsdale R, Ranaghan KE, Wales DJ, Mulholland AJ, et al. Large-Scale Density Functional Theory Transition State Searching in Enzymes. Phys Chem Lett 2014;5:3614–9.

[339] Matta CF. Modeling biophysical and biological properties from the characteristics of the molecular electron density, electron localization and delocalization matrices, and the electrostatic potential. J Comp Chem 2014;35:1165–98.

[340] Min SK, Kim WY, Cho Y, Kim KS. Fast DNA sequencing with a graphene-based nanochannel device. Nat Nanotechnol 2011;6:162–5.

[341] Wells DB, Belkin M, Comer J, Aksimentiev A. Assessing Graphene Nanopores for Sequencing DNA. Nano Lett 2012;12:4117–23.

[342] Prasongkit J, Grigoriev A, Pathak B, Ahuja R, Scheicher RH. Transverse Conductance of DNA Nucleotides in a Graphene Nanogap from First Principles. Nano Lett 2011;11:1941–5.

[343] Heerema SJ, Dekker C. Graphene nanodevices for DNA sequencing. Nat Nanotechnol 2016;11:127–36.

[344] Small AR. Theoretical Limits on Errors and Acquisition Rates in Localizing Switchable Fluorophores. vol. 96. 2009. doi:10.1016/j.bpj.2008.11.001.

[345] Deschout H, Neyts K, Braeckmans K. The influence of movement on the localization precision of sub-resolution particles in fluorescence microscopy. J Biophotonics 2012;5:97–109. doi:10.1002/jbio.201100078.

[346] Small A, Stahlheber S. Fluorophore localization algorithms for super-resolution microscopy. Nat Methods 2014;11:267–79. doi:10.1038/nmeth.2844.

[347] Wollman AJM, Miller H, Zhou Z, Leake MC. Probing DNA interactions with proteins using a single-molecule toolbox: inside the cell, in a test tube, and in a computer. Biochem Soc Trans 2015;43:139–45.

[348] Miller H, Zhou Z, Wollman AJM, Leake MC. Superresolution imaging of single DNA molecules using stochastic photoblinking of minor groove and intercalating dyes. Methods 2015;88:81–8. doi:10.1016/j.ymeth.2015.01.010.

[349] Smith CS, Joseph N, Rieger B, Lidke KA. Fast, single-molecule localization that achieves theoretically minimum uncertainty. Nat Methods 2010;7:373–5.

doi:10.1038/nmeth.1449.

[350] Yu B, Chen D, Qu J, Niu H. Fast Fourier domain localization algorithm of a single molecule with nanometer precision. Opt Lett 2011;36:4317. doi:10.1364/OL.36.004317.

[351] Andersson S. Localization of a fluorescent source without numerical fitting. Opt Express 2008;16:18714. doi:10.1364/OE.16.018714.

[352] Chao J, Ram S, Ward ES, Ober RJ. Ultrahigh accuracy imaging modality for super-localization microscopy. Nat Methods 2013;10:335–8. doi:10.1038/nmeth.2396.

[353] Holden SJ, Uphoff S, Kapanidis AN. DAOSTORM: an algorithm for high- density super-resolution microscopy. Nat Methods 2011;8:279–80. doi:10.1038/nmeth0411-279.

[354] Mukamel EA, Babcock H, Zhuang X. Statistical Deconvolution for Superresolution Fluorescence Microscopy. Biophys J 2012;102:2391–400. doi:10.1016/j.bpj.2012.03.070.

[355] Zhu L, Zhang W, Elnatan D, Huang B. Faster STORM using compressed sensing. Nat Methods 2012;9:721–3. doi:10.1038/nmeth.1978.

[356] Cox S, Rosten E, Monypenny J, Jovanovic-Talisman T, Burnette DT, Lippincott-Schwartz J, et al. Bayesian localization microscopy reveals nanoscale podosome dynamics. Nat Methods 2011;9:195–200. doi:10.1038/nmeth.1812.

[357] Robson A, Burrage K, Leake MC. Inferring diffusion in single live cells at the single-molecule level. Philos Trans R Soc Lond B Biol Sci 2013;368:20120029. doi:10.1098/rstb.2012.0029.

[358] Chiu S-W, Roberts MAJ, Leake MC, Armitage JP. Positioning of chemosensory proteins and FtsZ through the Rhodobacter sphaeroides cell cycle. Mol Microbiol 2013;90:322–37. doi:10.1111/mmi.12366.

[359] Lenn T, Leake MC, Mullineaux CW. Clustering and dynamics of cytochrome bd-I complexes in the Escherichia coli plasma membrane in vivo. Mol Microbiol 2008;70:1397–407. doi:10.1111/j.1365-2958.2008.06486.x.

[360] Lenn T, Leake MC. Single-molecule studies of the dynamics and interactions of bacterial OXPHOS complexes. Biochim Biophys Acta - Bioenerg 2016;1857. doi:10.1016/j.bbabio.2015.10.008.

[361] Lenn T, Leake MC, Mullineaux CW. Are Escherichia coli OXPHOS complexes concentrated in specialized zones within the plasma membrane? Biochem Soc Trans 2008;36:1032–6. doi:10.1042/BST0361032.

[362] Ulbrich MH, Isacoff EY. Subunit counting in membrane-bound proteins. Nat Methods 2007;4:319. doi:10.1038/nmeth1024.

[363] Leake MC, Greene NP, Godun RM, Granjon T, Buchanan G, Chen S, et al. Variable stoichiometry of the TatA component of the twin-arginine protein transport system observed by in vivo single-molecule imaging. Proc Natl Acad Sci U S A 2008;105:15376–81. doi:10.1073/pnas.0806338105.

[364] Delalez NJ, Wadhams GH, Rosser G, Xue Q, Brown MT, Dobbie IM, et al. Signal-dependent turnover of the bacterial flagellar switch protein FliM. Proc Natl Acad Sci U S A 2010;107:11347–51. doi:10.1073/pnas.1000284107.

[365] Liesche C, Grußmayer KS, Ludwig M, Wörz S, Rohr K, Herten D-P, et al. Automated Analysis of Single-Molecule Photobleaching Data by Statistical Modeling of Spot Populations. Biophys J 2015;109:2352–62. doi:10.1016/j.bpj.2015.10.035.

[366] Reyes-Lamothe R, Sherratt DJ, Leake MC. Stoichiometry and architecture of active DNA replication machinery in Escherichia coli. Science 2010;328:498–501. doi:10.1126/science.1185757.

[367] Beattie TR, Kapadia N, Nicolas E, Uphoff S, Wollman AJ, Leake MC, et al. Frequent

exchange of the DNA polymerase during bacterial chromosome replication. Elife 2017;6:e21763. doi:10.7554/eLife.21763.

[368] Wollman AJM, Leake MC. Millisecond single-molecule localization microscopy combined with convolution analysis and automated image segmentation to determine protein concentrations in complexly structured, functional cells, one cell at a time. Faraday Discuss 2015;184:401–24. doi:10.1039/c5fd00077g.

[369] Chen Y, Johnson J, Macdonald P, Wu B, Mueller JD. Chapter 16 - Observing Protein Interactions and Their Stoichiometry in Living Cells by Brightness Analysis of Fluorescence Fluctuation Experiments. Single Mol Tools Fluoresc Based Approaches, Part A n.d.;472:345–63. doi:10.1016/S0076-6879(10)72026-7.

[370] Lee S-H, Shin JY, Lee A, Bustamante C. Counting single photoactivatable fluorescent molecules by photoactivated localization microscopy (PALM). Proc Natl Acad Sci U S A 2012;109:17436–41. doi:10.1073/pnas.1215175109.

[371] Kurz A, Schmied JJ, Grußmayer KS, Holzmeister P, Tinnefeld P, Herten D-P. Counting Fluorescent Dye Molecules on DNA Origami by Means of Photon Statistics. Small 2013;9:4061–8. doi:10.1002/smll.201300619.

[372] Tin Kam Ho. Random decision forests. Proc. 3rd Int. Conf. Doc. Anal. Recognit., vol. 1, IEEE Comput. Soc. Press; n.d., p. 278–82. doi:10.1109/ICDAR.1995.598994.

[373] Mitchell M. An introduction to genetic algorithms. MIT press; 1998.

[374] Yegnanarayana B. Artificial neural networks. PHI Learning Pvt. Ltd.; 2009.

[375] Eddy SR. Hidden Markov models. Curr Opin Struct Biol 1996;6:361–5. doi:S0959-440X(96)80056-X [pii].

[376] Andrec M, Levy RM, Talaga DS. Direct Determination of Kinetic Rates from Single-Molecule Photon Arrival Trajectories Using Hidden Markov Models. J Phys Chem A 2003;107:7454–64. doi:10.1021/jp035514+.

[377] LeCun Y, Bengio Y, Hinton G. Deep learning. Nature 2015;521:436–44. doi:10.1038/nature14539.

[378] Rumelhart DE, Hinton GE, Williams RJ. Learning representations by back-propagating errors. Nature 1986;323:533–6. doi:10.1038/323533a0.

[379] Hinton GE, Osindero S, Teh Y-W. A Fast Learning Algorithm for Deep Belief Nets. Neural Comput 2006;18:1527–54. doi:10.1162/neco.2006.18.7.1527.

[380] Bengio Y, Lamblin P, Popovici D, Larochelle H. Greedy Layer-Wise Training of Deep Networks. Adv Neural Inf Process Syst 2007;19:153. doi:citeulike-article-id:4640046.

[381] LeCun Y, Boser B, Denker J, Henderson D. Handwritten digit recognition with a back-propagation network. Adv Neural Inf Process Syst 1990:396–404.

[382] Lecun Y, Bottou L, Bengio Y, Haffner P. Gradient-based learning applied to document recognition. Proc IEEE 1998;86:2278–324. doi:10.1109/5.726791.

[383] He K, Zhang X, Ren S, Sun J. Delving Deep into Rectifiers: Surpassing Human-Level Performance on ImageNet Classification 2015:1026–34.

[384] Russakovsky O, Deng J, Su H, Krause J, Satheesh S, Ma S, et al. ImageNet Large Scale Visual Recognition Challenge. Int J Comput Vis 2015;115:211–52. doi:10.1007/s11263-015-0816-y.

[385] Schmidhuber J. Deep learning in neural networks: An overview. Neural Networks 2015;61:85–117. doi:10.1016/j.neunet.2014.09.003.

[386] Bengio Y, Courville A, Vincent P. Representation Learning: A Review and New Perspectives. IEEE Trans Pattern Anal Mach Intell 2013;35:1798–828. doi:10.1109/TPAMI.2013.50.

[387] Lempitsky V, Zisserman A. Learning To Count Objects in Images 2010:1324–32.

[388] Arteta C, Lempitsky V, Noble JA, Zisserman A. Interactive Object Counting, Springer, Cham; 2014, p. 504–18. doi:10.1007/978-3-319-10578-9_33.

[389] Kamilov US, Papadopoulos IN, Shoreh MH, Goy A, Vonesch C, Unser M, et al. Learning approach to optical tomography. Optica 2015;2:517. doi:10.1364/OPTICA.2.000517.

[390] Wu AC-Y, Rifkin SA. Aro: a machine learning approach to identifying single molecules and estimating classification error in fluorescence microscopy images. BMC Bioinformatics 2015;16:102. doi:10.1186/s12859-015-0534-z.

[391] Wu AC-Y, Rifkin SA. Erratum to:'Aro: a machine learning approach to identifying single molecules and estimating classification error in fluorescence microscopy images'. BMC Bioinformatics 2016;17:207.

[392] Mueller F, Senecal A, Tantale K, Marie-Nelly H, Ly N, Collin O, et al. FISH-quant: automatic counting of transcripts in 3D FISH images. Nat Methods 2013;10:277–8. doi:10.1038/nmeth.2406.

[393] Borgmann DM, Mayr S, Polin H, Schaller S, Dorfer V, Obritzberger L, et al. Single Molecule Fluorescence Microscopy and Machine Learning for Rhesus D Antigen Classification. Sci Rep 2016;6:32317. doi:10.1038/srep32317.

[394] Jiang S, Zhou X, Kirchhausen T, Wong STC. Detection of molecular particles in live cells via machine learning. Cytom Part A 2007;71A:563–75. doi:10.1002/cyto.a.20404.

[395] Valentine AP, Woodhouse JH. Approaches to automated data selection for global seismic tomography. Geophys J Int 2010;182:1001–12. doi:10.1111/j.1365-246X.2010.04658.x.

[396] Chen X, Velliste M, Weinstein S, Jarvik JW, Murphy RF, Krajewski S, et al. Location proteomics: building subcellular location trees from high-resolution 3D fluorescence microscope images of randomly tagged proteins. In: Nicolau D V., Enderlein J, Leif RC, Farkas DL, editors. BMC Bioinforma. 2004 51, vol. 4962, BioMed Central; 2003, p. 298. doi:10.1117/12.477899.

[397] Malkiel I, Nagler A, Mrejen M, Arieli U, Wolf L, Suchowski H. Deep Learning for Design and Retrieval of Nano-photonic Structures. arXiv Prepr arXiv170207949 2017.

[398] Hesam N, Nezhad M. Optical system optimization using genetic algorithms. Technische Universitelt Delft, 2014.

[399] Ma J, Sheridan RP, Liaw A, Dahl GE, Svetnik V. Deep Neural Nets as a Method for Quantitative Structure–Activity Relationships. J Chem Inf Model 2015;55:263–74. doi:10.1021/ci500747n.

[400] Gawehn E, Hiss JA, Schneider G. Deep Learning in Drug Discovery. Mol Inform 2016;35:3–14. doi:10.1002/minf.201501008.

[401] Alipanahi B, Delong A, Weirauch MT, Frey BJ. Predicting the sequence specificities of DNA- and RNA-binding proteins by deep learning. Nat Biotechnol 2015;33:831–8. doi:10.1038/nbt.3300.

[402] Geisse NA. AFM and combined optical techniques. Mater Today 2009;12:40–5. doi:10.1016/S1369-7021(09)70201-9.

[403] He Y, Lu M, Cao J, Lu HP. Manipulating Protein Conformations by Single-Molecule AFM-FRET Nanoscopy. ACS Nano 2012;6:1221–9. doi:10.1021/nn2038669.

[404] Dobbie IM, Robson A, Delalez N, Leake MC. Visualizing single molecular complexes in vivo using advanced fluorescence microscopy. J Vis Exp 2009:1508. doi:10.3791/1508.

[405] Kellermayer MSZ, Karsai Á, Kengyel A, Nagy A, Bianco P, Huber T, et al. Spatially and Temporally Synchronized Atomic Force and Total Internal Reflection Fluorescence Microscopy for Imaging and Manipulating Cells and Biomolecules. Biophys J 2006;91:2665–77. doi:10.1529/biophysj.106.085456.

[406] Odermatt PD, Shivanandan A, Deschout H, Jankele R, Nievergelt AP, Feletti L, et al.

High-Resolution Correlative Microscopy: Bridging the Gap between Single Molecule Localization Microscopy and Atomic Force Microscopy. Nano Lett 2015;15:4896–904. doi:10.1021/acs.nanolett.5b00572.

[407] Gumpp H, Puchner EM, Zimmermann JL, Gerland U, Gaub HE, Blank K. Triggering Enzymatic Activity with Force. Nano Lett 2009;9:3290–5. doi:10.1021/nl9015705.

[408] Comstock MJ, Ha T, Chemla YR. Ultrahigh-resolution optical trap with single-fluorophore sensitivity. Nat Methods 2011;8:335–40. doi:10.1038/nmeth.1574.

[409] Zou H, Lifshitz LM, Tuft RA, Fogarty KE, Singer JJ. Imaging Ca2+ Entering the Cytoplasm through a Single Opening of a Plasma Membrane Cation Channel. J Gen Physiol 1999;114.

[410] Demuro A, Parker I. Imaging the Activity and Localization of Single Voltage-Gated Ca2+ Channels by Total Internal Reflection Fluorescence Microscopy. Biophys J 2004;86:3250–9. doi:10.1016/S0006-3495(04)74373-8.

[411] Kusch J, Zifarelli G. Patch-clamp fluorometry: electrophysiology meets fluorescence. Biophys J 2014;106:1250–7. doi:10.1016/j.bpj.2014.02.006.

[412] Sasmal DK, Lu HP. Single-Molecule Patch-Clamp FRET Microscopy Studies of NMDA Receptor Ion Channel Dynamics in Living Cells: Revealing the Multiple Conformational States Associated with a Channel at Its Electrical Off State. J Am Chem Soc 2014;136:12998–3005. doi:10.1021/ja506231j.

[413] Sasmal DK, Yadav R, Lu HP. Single-Molecule Patch-Clamp FRET Anisotropy Microscopy Studies of NMDA Receptor Ion Channel Activation and Deactivation under Agonist Ligand Binding in Living Cells. J Am Chem Soc 2016;138:8789–801. doi:10.1021/jacs.6b03496.

[414] Traynelis SF, Wollmuth LP, McBain CJ, Menniti FS, Vance KM, Ogden KK, et al. Glutamate Receptor Ion Channels: Structure, Regulation, and Function. Pharmacol Rev 2010;62.

[415] Kao JP. Practical aspects of measuring [Ca2+] with fluorescent indicators. Methods Cell Biol 1994;40:155–81.

[416] Molokanova E, Savchenko A. Bright future of optical assays for ion channel drug discovery. Drug Discov Today 2008;13:14–22. doi:10.1016/j.drudis.2007.11.009.

[417] Stosiek C, Garaschuk O, Holthoff K, Konnerth A. In vivo two-photon calcium imaging of neuronal networks. Proc Natl Acad Sci U S A 2003;100:7319–24. doi:10.1073/pnas.1232232100.

[418] Szabo M, Wallace MI. Imaging potassium-flux through individual electropores in droplet interface bilayers. Biochim Biophys Acta - Biomembr 2016;1858:613–7. doi:10.1016/j.bbamem.2015.07.009.

[419] Atanasov V, Knorr N, Duran RS, Ingebrandt S, Offenhäusser A, Knoll W, et al. Membrane on a Chip: A Functional Tethered Lipid Bilayer Membrane on Silicon Oxide Surfaces. Biophys J 2005;89:1780–8. doi:10.1529/biophysj.105.061374.

[420] Tanaka M, Sackmann E. Polymer-supported membranes as models of the cell surface. Nature 2005;437:656–63. doi:10.1038/nature04164.

[421] Leptihn S, Castell OK, Cronin B, Lee E-H, Gross LCM, Marshall DP, et al. Constructing droplet interface bilayers from the contact of aqueous droplets in oil. Nat Protoc 2013;8:1048–57. doi:10.1038/nprot.2013.061.

[422] Sengel JT, Wallace MI. Imaging the dynamics of individual electropores. Proc Natl Acad Sci 2016;113:5281–6. doi:10.1073/pnas.1517437113.

[423] Dower WJ, Miller JF, Ragsdale CW. High efficiency transformation of E.coli by high voltage electroporation. Nucleic Acids Res 1988;16:6127–45. doi:10.1093/nar/16.13.6127.

[424] Schoellhammer CM, Blankschtein D, Langer R. Skin permeabilization for transdermal

drug delivery: recent advances and future prospects. Expert Opin Drug Deliv 2014;11:393–407. doi:10.1517/17425247.2014.875528.

[425] Tarek M. Membrane Electroporation: A Molecular Dynamics Simulation. Biophys J 2005;88:4045–53. doi:10.1529/biophysj.104.050617.

[426] Tieleman DP. The molecular basis of electroporation. BMC Biochem 2004;5:10. doi:10.1186/1471-2091-5-10.

[427] de Boer P, Hoogenboom JP, Giepmans BNG. Correlated light and electron microscopy: ultrastructure lights up! Nat Methods 2015;12:503–13. doi:10.1038/nmeth.3400.

[428] Schirra RT, Zhang P, Zhang P. Correlative fluorescence and electron microscopy. Curr Protoc Cytom 2014;70:12.36.1-10. doi:10.1002/0471142956.cy1236s70.

[429] Löschberger A, Franke C, Krohne G, van de Linde S, Sauer M. Correlative super-resolution fluorescence and electron microscopy of the nuclear pore complex with molecular resolution. J Cell Sci 2014;127.

[430] Watanabe S, Punge A, Hollopeter G, Willig KI, Hobson RJ, Davis MW, et al. Protein localization in electron micrographs using fluorescence nanoscopy. Nat Methods 2011;8:80–4. doi:10.1038/nmeth.1537.

[431] Sochacki KA, Shtengel G, van Engelenburg SB, Hess HF, Taraska JW. Correlative super-resolution fluorescence and metal-replica transmission electron microscopy. Nat Methods 2014;11:305–8. doi:10.1038/nmeth.2816.

[432] Kopek BG, Shtengel G, Xu CS, Clayton DA, Hess HF. Correlative 3D superresolution fluorescence and electron microscopy reveal the relationship of mitochondrial nucleoids to membranes. Proc Natl Acad Sci U S A 2012;109:6136–41. doi:10.1073/pnas.1121558109.

[433] Annette Granéli, Caitlyn C. Yeykal, Tekkatte Krishnamurthy Prasad A, Greene* EC. Organized Arrays of Individual DNA Molecules Tethered to Supported Lipid Bilayers 2005. doi:10.1021/LA051944A.

[434] Greene EC, Wind S, Fazio T, Gorman J, Visnapuu M-L. Chapter 14 – DNA Curtains for High-Throughput Single-Molecule Optical Imaging. Methods Enzymol., vol. 472, 2010, p. 293–315. doi:10.1016/S0076-6879(10)72006-1.

[435] Horrocks MH, Tosatto L, Dear AJ, Garcia GA, Iljina M, Cremades N, et al. Fast Flow Microfluidics and Single-Molecule Fluorescence for the Rapid Characterization of α-Synuclein Oligomers. Anal Chem 2015;87:8818–26. doi:10.1021/acs.analchem.5b01811.

[436] Eriksson E, Sott K, Lundqvist F, Sveningsson M, Scrimgeour J, Hanstorp D, et al. A microfluidic device for reversible environmental changes around single cells using optical tweezers for cell selection and positioning. Lab Chip 2010;10:617–25. doi:10.1039/B913587A.

[437] Gustavsson A-K, van Niekerk DD, Adiels CB, du Preez FB, Goksör M, Snoep JL. Sustained glycolytic oscillations in individual isolated yeast cells. FEBS J 2012;279:2837–47. doi:10.1111/j.1742-4658.2012.08639.x.

[438] Bowman RW, Padgett MJ. Optical trapping and binding. Reports Prog Phys 2013;76:26401. doi:10.1088/0034-4885/76/2/026401.

[439] Dholakia K, Čižmár T. Shaping the future of manipulation. Nat Photonics 2011;5:335–42. doi:10.1038/nphoton.2011.80.

[440] Backlund MP, Arbabi A, Petrov PN, Arbabi E, Saurabh S, Faraon A, et al. Removing orientation-induced localization biases in single-molecule microscopy using a broadband metasurface mask. Nat Photonics 2016;10:459–62. doi:10.1038/nphoton.2016.93.

[441] Yang Z, Fang J, Chittuluru J, Asturias FJ, Penczek PA. Iterative Stable Alignment and

Clustering of 2D Transmission Electron Microscope Images. vol. 20. 2012. doi:10.1016/j.str.2011.12.007.

[442] Cromey DW. Digital images are data: and should be treated as such. Methods Mol Biol 2013;931:1–27. doi:10.1007/978-1-62703-056-4_1.

[443] Hedvat C V. Digital microscopy past, present, and future. Arch Pathol Lab Med 2010;134:1666–70. doi:10.1043/2009-0579-RAR1.1.

[444] Selinummi J, Seppälä J, Yli-Harja O, Puhakka JA. Software for quantification of labeled bacteria from digital microscope images by automated image analysis. Biotechniques 2005;39:859–63. doi:10.2144/000112018.

[445] Xiaobo Zhou, Wong STC. Informatics challenges of high-throughput microscopy. IEEE Signal Process Mag 2006;23:63–72. doi:10.1109/MSP.2006.1628879.

[446] Pitrone PG, Schindelin J, Stuyvenberg L, Preibisch S, Weber M, Eliceiri KW, et al. OpenSPIM: an open-access light-sheet microscopy platform. Nat Methods 2013;10:598–9. doi:10.1038/nmeth.2507.

[447] Marx V. Microscopy: OpenSPIM 2.0. Nat Methods 2016;13:979–82. doi:10.1038/nmeth.4070.

[448] Allen JR, Ross ST, Davidson MW. Single molecule localization microscopy for superresolution. J Opt 2013;15:94001. doi:10.1088/2040-8978/15/9/094001.

[449] Leatherdale CA, Woo W-K, Mikulec F V., Bawendi MG, Sakurai Y, Libchaber A. In Vivo Imaging of Quantum Dots Encapsulated in Phospholipid Micelles. J Phys Chem 2002;106:7619. doi:10.1021/JP025698C.

[450] Harms GS, Cognet L, Lommerse PHM, Blab GA, Schmidt T. Autofluorescent Proteins in Single-Molecule Research: Applications to Live Cell Imaging Microscopy. Biophys J 2001;80:2396–408. doi:10.1016/S0006-3495(01)76209-1.

[451] Truong K, Ikura M. The use of FRET imaging microscopy to detect protein–protein interactions and protein conformational changes in vivo. Curr Opin Struct Biol 2001;11:573–8. doi:10.1016/S0959-440X(00)00249-9.

[452] Ntziachristos V, Bremer C, Weissleder R. Fluorescence imaging with near-infrared light: new technological advances that enable in vivo molecular imaging. Eur Radiol n.d.;13:195–208. doi:10.1007/S00330-002-1524-X.